**Dynamics and Clouds in Planetary Atmospheres from Telescopic Observations**

Agustín Sánchez-Lavega[1], Patrick Irwin[2], Antonio García Muñoz[3]

1 Departamento Física Aplicada, Escuela de Ingeniería de Bilbao, Universidad del País Vasco UPV/EHU, 48013 Bilbao, Spain
2 Department of Physics, University of Oxford, Oxford, UK.
3 Université Paris-Saclay, Université Paris-Cité, CEA, CNRS, AIM, 91191 Gif-sur-Yvette, France

Corresponding author: agustin.sanchez@ehu.eus

**Abstract**

This review presents an insight into our current knowledge of the atmospheres of the planets Venus, Mars, Jupiter, Saturn, Uranus and Neptune, the satellite Titan, and those of exoplanets. It deals with the thermal structure, aerosol properties (hazes and clouds, dust in the case of Mars), chemical composition, global winds and selected dynamical phenomena in these objects. Our understanding of atmospheres is greatly benefitting from the discovery in the last three decades of thousands of exoplanets. The exoplanet properties span a broad range of conditions, and it is fair to expect as much variety for their atmospheres. This complexity is driving unprecedented investigations of the atmospheres, where those of the solar systems bodies are the obvious reference. We are witnessing a significant transfer of knowledge in both directions between the investigations dedicated to Solar System and exoplanet atmospheres, and there are reasons to think that this exchange will intensity in the future. We identify and select a list of research subjects that can be conducted at optical and infrared wavelengths with future and currently available ground-based and space-based telescopes, but excluding those from the space missions to solar system bodies.

**Contents**

## 1. Introduction

The continuous technological development of ground-based telescopes with large collecting areas, optical instrumentation (adaptive optics for imaging and spectroscopy), and detectors with increasing capabilities at optical and infrared wavelengths, are enabling deeper and more detailed studies of planetary atmospheres of solar system bodies and exoplanets. In the solar system, while in-situ and spacecraft observations of planetary atmospheres provide unique insights, they are necessarily limited in time coverage, in response to rapid and transient events, and also in wavelength coverage and spectral resolution. In contrast, observations from ground-based telescopes, such as Keck, Gemini, Very Large Telescope (VLT), Gran Telescopio Canarias (GTC)… and from space-based telescopes such as the Hubble Space Telescope (HST), and more recently the James Webb Space Telescope (JWST), provide a long-term basis of observations with good wavelength coverage and resolution. Two good examples of the value of these combined studies are the Martian Global Dust Storm in 2018 (Guzewich et al., 2020) and the giant storm in Saturn in 2010 (Sánchez-Lavega et al. 2018c). The combined observations applied to these two examples, with complementary data sets and analysis, provided context, synergies, and a broader scope of the studied phenomenon. In addition, telescopes will be the only source of information for studies of solar system atmospheres for which no space missions are foreseen in the short term, such as Saturn, Uranus and Neptune. They will also provide support before and after the arrival of missions to Venus (2029 onwards), Jupiter (2031 onwards) and Titan (2034 onwards). Mars is the planet with abundant in-situ and orbital missions in progress, but nevertheless, telescopic observations offer interesting observational aspects as will be discussed in this paper.

Turning next our attention beyond the solar system, we know that each FGK-type star is on average orbited by one planet, typically of radius ≤$4R_\oplus$, i.e. less than four Earth radii (Kunimoto and Matthews, 2020). The occurrence rate rises to 2.5 planets/star when the statistics refer to M-dwarf hosts (Dressing and Charbonneau, 2015). The large number of suspected planets in the galaxy, the variety of environments in which they are found, and the broad range of masses and radii, $M_p$ and $R_p$, that they exhibit suggest that many types of atmospheres exist. Understanding the basics of an exoplanet atmosphere such as its composition or energy budget is key to putting other planet properties in perspective, which may in turn provide information on the planet conditions at formation or even on its habitability. The atmospheric characterization of exoplanets is a very active field. Spectra for about one hundred planets have been gathered by means of ground-based and space-borne telescopes (Madhusudhan, 2019; Changeat et al., 2022; Edwards et al., 2022). In space, HST and the Spitzer telescope have been the leading facilities for characterizing exoplanet atmospheres. Other space telescopes such as CoRoT (Convection Rotation and planetary Transits), Kepler, TESS (Transiting Exoplanet Survey Satellite) or CHEOPS (CHaracterising ExOPlanets Satellite), which were conceived in the exoplanet age but whose priority was not atmospheric characterization, have also contributed greatly (Esteves et al., 2015; Shporer, 2017; Lendl et al., 2020; von Essen et al., 2021). On the ground, numerous instruments, in particular high-resolution spectrographs, have been commissioned in recent years with the atmospheric characterization of exoplanets as a priority. In the immediate future, the long-awaited launch of JWST in December 2021 is already offering unprecedented insights for a diverse and large sample of exoplanets. Going from the investigation of a few atmospheres in the solar system to hundreds in our galactic vicinity will enable us to test atmospheric theory in new ways. It will also raise new questions on, for example, the interaction of the atmospheres with the planet interiors or the host stars (Strugarek et al., 2019; Kite & Schaefer, 2021). In the long run, the variety of observing programs dedicated to these and many other planets will help build the bigger picture of exoplanet atmospheres. In response, atmospheric models must therefore evolve on many different fronts in order to keep pace with the speed of observational discoveries.

## 2. Telescopic observations of solar system planetary atmospheres

Venus is a difficult planet to observe with telescopes due to its angular proximity to the Sun, with maximum elongation at quadrature $\sim 45°$, when it reaches a size $\sim 20$ arcsec. Its low axial tilt implies that we cannot see the poles. On Venus, the existence of a phase prevents us from viewing the entire disc in either its daytime or nighttime part, which limits the spatial coverage of dynamical phenomena. Large telescopes (including HST) rarely observe the planet due to Sun angular proximity (an exception is NASA Infrared Telescope Facility, IRTF), so smaller telescopes are the most commonly used. From Earth and Earth orbit, we are mostly limited to one geometry for Mars, with the Sun directly behind us at opposition when its angular diameter is at maximum in the range $\sim 14\text{-}24$ arcsec. Outside opposition, the phase angle must be taken into account when planning to observe the illuminated part of the disk. Because of the small tilt angle there is a limitation in the coverage of the polar regions. Jupiter and Saturn present large disc sizes to the Earth (~ 48 arcsec and 21 arcsec at opposition, respectively) and practically full illumination (phase angles < 12° and 6° respectively). The large size of Jupiter means that much can be done without adaptive optics by amateur and professional astronomers. Telescopic observations of Saturn's hemispheric visibility is constrained by the tilt of 26.7° of the rotation axis, and the rings and their shadows projection onto the disk. Twice in the Saturn's year of 29.4 Earth years, during the equinox when the Sun is directly over the equator, a ring plane crossing occurs. Next ring crossings will occur on 23 March and 6 May 2025, followed by those in 15 October 2038, and in 22 January, 1 April and 9 July 2039. In the coming years, the best visibility will be the southern hemisphere of the planet. Since no space mission is planned in the future to Saturn (the "Dragonfly" mission to Titan in 2034 is a lander to the satellite with no orbiter), the planet will be only targeted by ground-based telescopes. Titan, the major satellite of Saturn is one of the key observational targets for the large telescopes on Earth and in space due to its small size (~ 0.8 arcsec) and to the fact that the next space mission Dragonfly, will arrive to Titan in 2034 (Barnes et al. 2021). Finally, both Uranus and Neptune present small apparent disc sizes from Earth of $\sim 3.6$ and $\sim 2.4$ arcsec, respectively. Despite the prioritization of a Uranus mission as the next NASA Flagship, both planets are clear targets for large ground-based telescopes with AO and space telescopes (HST, JWST).

The study of images of the planets with ground-based telescopes is mainly constrained by the spatial resolution due to Earth's atmospheric turbulence ("seeing") and ultimately, by the telescope diffraction. However, image resolution and image sensitivity and contrast can be improved with techniques like adaptive optics (AO, operative in the visible and near infrared), telescope active optics and lucky imaging (LI; Mendikoa et al., 2016), or combinations of them. In AO image distortions due to seing is corrected for. This is done using a pick-off mirror and sensor to quantify the flatness of the wavefront from a nearby point source, and this information is then used to distort a mirror in the optical chain to correct out the turbulence distortion in real time. The technique of 'lucky imaging', pioneered by amateur astronomers, can in some cases greatly improve the spatial resolution. Here, multiple images are taken and then co-added in such a way that poor frames are discarded while good ones are co-aligned to correct for the 1st order effect of atmospheric turbulence, which causes the image to shift from side to side. Image stacking, derotation, deconvolution and other image processes provide exceptional telescopic views of the planets.

Figure 1a shows examples of telescope resolution according to their aperture for different wavelength ranges where high-resolution methods as LI and AO apply. The development of advanced instrumentation such as the MCAO Assisted Visible Imager and Spectrograph (MAVIS, McDermid et al 2020) will bring the spatial resolution between wavelengths 400 to 900 nm to 0.01-0.04 arcsec. Figure 1b shows the spatial resolution that can be achieved in the center of the planet disk (nadir view) for representative values of the angular resolution provided by the telescope, including the future Extremely Large Telescope (ELT, 40m diameter), TMT (Thirty Meter Telescope 30m) and GMT (Giant Magellan Telescope, 25.5 m). Advanced instrumentation with these large telescopes could provide spatial resolutions of ~ 5-12 km/px on Jupiter and 10-25 km/px on Saturn, for example. Accurate tracking of the moving planet in the sky, with velocities ranging from ~ 0.017 arcsec s⁻¹ for Mars to ~ 0.0016 arcsec s⁻¹ for Neptune, is an essential element for these studies with future large telescopes.

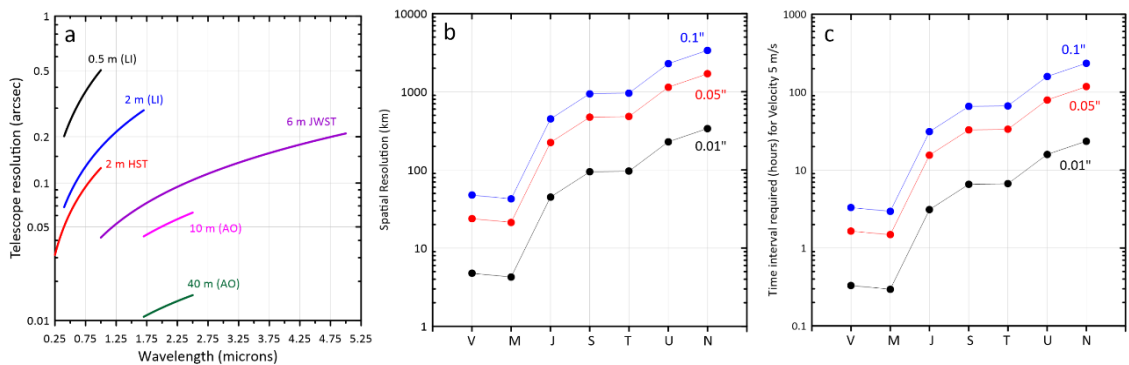

**Figure 1.** *(a) Telescope resolution as a function of wavelength for a wide range of apertures in meters (from a 0.5 m to 40 m). At short wavelengths the lucky imaging (LI) method puts the resolution close to the diffraction limit. At longer wavelengths Adaptive Optics (AO) methods are used. The Hubble Space Telescope (HST, 0.25 - 1 $\mu m$) and James Web Space Telescope (JWST, for the selected range 1 - 5 $\mu m$) are indicated; (b) Spatial resolution at the planet center for three cases of angular resolution as given in (a). Each planet is identified by its initials (T is for Titan); (c) Required time interval between two images to get, in each planet and according to the angular resolution, a minimum translation velocity for a feature of 5 m/s. We used the following angular size for the planets (in arcsec): Venus (40), Mars (25), Jupiter (50), Saturn (20), Titan (0.84), Uranus (3.5), Neptune (2.3).*

Modern telescopes, in addition to traditional imaging and long-slit spectroscopy instruments, often include Integral Field Unit Spectrometers, which can image planets in thousands of wavelengths simultaneously. Put another way, such instruments make observations where each pixel of the image is actually a complete spectrum covering a wide wavelength range at medium resolution, e.g. R=$\lambda/\Delta\lambda$ ~ 3000, where $\lambda$ is wavelength and $\Delta\lambda$ is the resolution element. Such observations are invaluable in discriminating between different cloud models, both in terms of wide spectral coverage, but also in terms of how the observed radiance varies with emission angle, leading to either limb-darkening or limb-brightening. Ground-based telescopes also allow for extremely high-resolution long-slit spectroscopy ($R \sim 30000$), which is sufficient to resolve individual molecular lines of atmospheric gases, allowing for greater selectivity against other constituents, but also allowing astronomers to probe for minor species and isotope ratios.

Images of planetary atmospheres provide information on the horizontal properties of the weather formations, their movement and their evolution over time. They also give information on the vertical properties by observing at different wavelengths sensitive to absorption and scattering by gases and aerosols at different altitudes. A fundamental aspect of studies of atmospheric dynamics from images are the temporal scales associated with the phenomena (e.g., time for the development and time for changes). For a weather system of horizontal size L with associated velocity V, the time scale is $\sim$ L/V, and similarly for vertical changes $\sim$ H/W (H vertical extent, W vertical velocity). For example, vigorous eruptive events related to convection can occur in very short time intervals of a few hours (H/W << L/V). These kinds of phenomena, as they are unpredictable, require a continuous monitoring of the planet in order to capture them from their early stages. This can only be done with a dedicated survey that monitors their occurrence and here amateur networks can nicely fulfil this requirement (Mousis et al., 2014).

Cloud tracking is the most commonly method used in planetary imaging to obtain the longitude and latitude drifts of a feature or cloud element and retrieve its horizontal velocity (u zonal, v meridional) in the rotating reference frame of the planet. For the terrestrial planets and Titan the motions are measured relative to the surface (the body rotation or sidereal period), but for the gas giants and icy planets, their radio-rotation period is the reference frame (Archinal et al., 2018). The rotation periods $\tau_{rot}$ range from -243.02 terrestrial days of Venus to 9.925 hr for Jupiter (in angular velocity from $\Omega = 2\pi/\tau_{rot} = 4.76\text{x}10^{-8}$ s$^{-1}$ for Venus to $2.8\text{x}10^{-5}$ s$^{-1}$ for Jupiter). For Saturn there is uncertainty in the radio-rotation (IAU adopted the Saturn Kilometric Radiation SKR to be 10.65622 h) and different alternative methods have been proposed to determine the rotation period of the planet ranging from 10.5333 h to 10.82 h (Carbary et al., 2018; see section 3.3.4).

The targeted feature should have enough contrast and size to be spatially resolved. The time interval between two images ($\Delta t$) and the telescope resolution ($\sigma$, assumed to be the minimum size of the resolved feature) can be used to first order estimate the wind velocity $V = \sigma/\Delta t$. Fig. 1c shows the time interval ($\Delta t$) required between two images of each planet to get a drift velocity of 5 ms$^{-1}$ according to the angular resolution provided by the telescope.

The feature's motion can be tracked manually (by visual identification) at different time intervals, fixing its trajectory to retrieve the mean and instantaneous velocity. Mapping the cloud field at two different time intervals can be used to get automatically the feature translations from brightness correlations allowing the retrieval of the velocity vector at each point in the map. This correlation velocimetry can be obtained from one-dimensional scans for flows essentially zonal (as in the giant planets), from two-dimensional scans and optical velocimetry methods (García-Melendo et al., 2001; Hueso et al., 2009; Asay-Davis et al., 2009; Liu et al., 2012; Sánchez-Lavega et al., 2019). From the horizontal wind field, the vorticity and the convergence or divergence of the flow (local, or averaged over an area) can be calculated. Additionally, if in the same area, clouds are observed and tracked in time at different altitudes, then the vertical shear of the wind velocity can be obtained, as we discuss below for each planet.

Another method that has been used successfully to measure wind speeds has been Doppler spectroscopy (visible, IR and mm spectra). At optical wavelengths, the method relies on the shift of solar lines reflected by the clouds (see, e.g., Machado et al., 2017 for Venus, and Gaulme et al., 2018 for Jupiter). In the case of the giant planets, the optical spectral range includes abundant absorption bands from the planet's atmosphere due ammonia (645 nm), methane (619, 724, 890 nm) and even hydrogen quadrupolar lines (e.g. 636.9, 815.3 nm). The height of the reflecting clouds in which these absorptions take place may differ from the cloud reflection heights sensed by the solar lines, so potentially they could be used to obtain velocities as a function of height. In the interferometric mm-range, winds have been measured by Doppler spectroscopy in all planets: Shah et al. (1991) and Clancy et al. (200'8) for Venus, Lellouch et al. (1991), Cavalié et al. (2008) and Moreno et al. (2009) for Mars, Cavalié et al. (2021) for Jupiter, Benmahi et al. (2022) for Saturn, Moreno et al. (2005), Lellouch et al. (2019) for Titan, Carrion-Gonzalez et al. (2023) for Neptune.

Wind velocities can be derived from temperature maps and the use of the thermal wind equation that relates the temperature gradients with the vertical shear in the wind velocity (Sánchez-Lavega, 2011; see section 3.1 and 3.3). The temperatures are retrieved from the inversion of thermal radiances and then the wind velocity is obtained after the vertical integration of the thermal wind equation. For zonal winds, the method uses as a reference the mean wind profile measured from cloud tracking. The assumed force balances are cyclostrophy (pressure gradient and centrifugal force) for Venus (Piccialli et al., 2008), geostrophy (pressure gradient and Corilis force) for Mars (Leovy, 2001), Jupiter (Simon-Miller et al., 2006), Saturn (Fletcher et al., 2018), Uranus (Sun et al., 1991) and Neptune (Tollefson et al., 2018), and gradient wind (pressure gradient, Coriolis and centrifugal force) for Titan (Achterberg et al., 2008). In the Equatorial region of rapidly rotating bodies, the divergence of the Coriolis force makes geostrophic balance innacurate and a different approach called the "Equatorial Thermal Wind Equation" (EQTWE) has been proposed by Marcus et al (2019). Additionally, when composition gradients are important, as for example occurs with methane in Uranus and Neptune, the thermal wind equation is also modified to include an additional term called "humidity winds" (Sun et al., 1991; Tollefson et al., 2018).

## 3. Solar system atmospheres: aerosols, chemical composition and temperature

In this section we present a review of the vertical structure (temperature and pressure profile), chemical compositions and aerosols (hazes, clouds and dust) in the atmospheres of the planets Venus, Mars, Jupiter, Saturn, Uranus, Neptune, and of the satellite Titan. We describe first the basic physics behind the atmospheric structure followed by the capabilities to study each atmosphere with current and future ground-based and space-based telescopes.

## 3.1 Vertical Profiles

### 3.1.1 Temperature-Pressure profiles

In planetary atmospheres, where the vertical winds are generally small compared with those in the horizontal direction, we may assume that pressure and altitude are related by hydrostatic equilibrium, where the vertical pressure difference $dp$ across a slab of air at altitude $z$ of density $\rho$ and thickness $dz$, subject to a gravitational acceleration $g$ is $dp = -\rho g dz$. Assuming the air behaves as an ideal gas, the density may be determined from the temperature $T$ and pressure $p$ as $\rho = \bar{m}p/RT$, where $R$ is the molar gas constant and $\bar{m}$ is the mean molecular weight of the atmosphere. Substituting for $\rho$ and integrating (approximating $T$, molecular weight and gravity are constant) we find $p = p_0 e^{-z/H}$, where $p_0$ is the pressure at $z = 0$, and $H = R_g^* T / g$ is the *scale height*.

The vertical temperature structure in planetary atmospheres is controlled by the transfer of heat, firstly upwards from the deep interior (or solar-heated surface for terrestrial planets) and secondly downwards, from solar radiation that has been absorbed high in the atmosphere. Deep in the atmosphere, the infrared (IR) optical depth to space is high and the heat may not escape radiatively. Hence, the air rises convectively to transfer the heat upwards. As the air parcels rise, the pressure drops and so they expand. If we assume that there is no exchange of heat between a parcel and its surroundings, then the expansion may be considered to be adiabatic and reversible. Writing entropy $S$ as a function of $T$ and $p$:

$$dS = \left(\frac{\partial S}{\partial T}\right)_p dT + \left(\frac{\partial S}{\partial p}\right)_T dp = 0$$

$$\tag{1}$$

We can replace the first partial derivative in terms of heat capacity, and the second partial derivative using one of Maxwell's relations to give:

$$dS = C_p \frac{dT}{T} - \left(\frac{\partial V}{\partial T}\right)_p dp = 0$$

$$\tag{2}$$

where $C_p$ is the molar heat capacity at constant pressure. Assuming the gas is ideal, this equation can be simplified to:

$$C_p \frac{dT}{T} = \frac{R}{p} dp$$

$$\tag{3}$$

and substituting for $dp$ from the hydrostatic equation we can rearrange this equation to:

$$\frac{dT}{dz} = -\frac{\bar{m}}{C_p} g = -\frac{g}{c_p} = -\Gamma_d$$

(4)

where $c_p$ is the *specific* heat capacity at constant pressure (J kg$^{-1}$ K$^{-1}$). Hence, we can see that we expect the temperature to fall linearly with height with a rate, $\Gamma_d$, known as the *dry adiabatic lapse rate* (DALR).

**Table 1. Mean observable planetary atmospheric properties (Giant planet compositions are given at a pressure level of 10 bar). Gas abundances are listed as % mole fraction.**

|  | Venus | Earth | Mars | Jupiter | Saturn | Uranus | Neptune | Titan |
|---|---|---|---|---|---|---|---|---|
| $H_2$ | - | - | - | 85.9 | 87.5 | 81.2 | 81.2 | 0.1 |
| He | - | - | - | 13.4 | 11.7 | 14.7 | 14.7 | - |
| $H_2O$ | - | 1.0 | 0.03 | 0.4 | 0.06 | - | - | - |
| $H_2S$ | - | - | - | 0.01 | 0.03 | 0.04 | 0.007 | - |
| $NH_3$ | - | - | - | 0.05 | 0.1 | - | - | - |
| $CH_4$ | - |  |  | 0.2 | 0.5 | 4.0 | 4.0 | 5.6 |
| $CO_2$ | 96.5 | 0.3 | 95.0 | - | - | - | - | - |
| $O_2$ | - | 21 | 0.17 | - | - | - | - | - |
| $N_2$ | 3.5 | 77.0 | 2.8 | - | - | - | - | 94.2 |
| $\Gamma_d$(K/km) | 10.7 | 9.8 | 4.3 | 2.2 | 1.0 | 1.0 | 1.3 | 1.26 |
| H (km) | 15.9 | 8.5 | 11.1 | 24 | 50 | 32 | 29 | 20 |
| $R/R_{Earth}$ | 0.95 | 1 | 0.49 | 11.2 | 9.4 | 4.0 | 3.9 | 0.4 |
| $M/M_{Earth}$ | 0.82 | 1 | 0.11 | 318 | 95 | 14.5 | 17.1 | 0.022 |
| g (ms$^{-2}$) | 8.9 | 9.8 | 3.7 | 24 | 10 | 8.8 | 11.0 | 1.35 |

Note: The solar system element abundances of Grevesse et al. (2007) were here assumed for the Giant Planets, with the following elemental enrichments (mostly following Irwin, 2009): He: 0.92, 0.79, 1.06, 1.06; O: 5.0, 10.0, 100.0, 100.0; C: 4.83, 10.86, 103.0, 103.0; N: 5.5, 10.0, 1.4, 3.9; S: 5.0, 14.0, 37.0, 54.0. The S and N abundances for the Ice Giants are those of Tollefson et al. (2021) and Molter et al. (2021), respectively. Other figures are from standard literature sources.

If the air contains condensable gases, then the adiabatic cooling will cause the condensation of cloud particles and the associated release of latent heat will result in the air cooling more slowly with height. When such latent heat release is accounted for, the rate of decrease of temperature with height is known as the saturated adiabatic lapse rate (SALR), which may be shown to be (e.g., Irwin, 2009; Andrews, 2000; Atreya, 1996):

$$\Gamma_s = -\frac{g}{c_p} \frac{\left[ 1 + \frac{1}{RT} \sum L_i x_i \right]}{\left[ 1 + \frac{1}{\bar{m} c_p T^2} \sum \frac{L_i^2 x_i}{R} \right]}$$

(5)

where $L_i$ is the *molar* latent heat of vaporisation (or sublimation) of the $i$th condensing component, and $x_i$ is the saturated volume mixing ratio (or *mole fraction*) of the component. Depending on the mole fraction of condensing species $\Gamma_s$ can be significantly less than $\Gamma_d$. Li et al. (2018) note that when more than one species is condensing at the same pressure level, additional cross terms appear in Eq. 5.

In addition to single condensate clouds, two-component clouds may also form in the atmospheres of some planets. For the giant planets, the most important of these is solid ammonium hydrosulphide, $NH_4SH$, which may form by the reaction:

$$NH_3 + H_2S \leftrightarrow NH_4SH(s) \qquad (6)$$

Finally, we note that while the assumption of constant molecular weight used to derive the lapse rate in Eq. 4 is reasonable for terrestrial atmospheres, it is less applicable to the giant planet atmospheres, which can contain large fractions of condensable species in an otherwise low molecular weight $H_2$-He atmosphere. Here the lapse rate can be severely affected by the sudden drop in molecular weight going through a condensation cloud as shown by Guillot (1995) and Leconte et al. (2017), leading to superadiabatic profiles that inhibit convection. In addition, Irwin et al. (2022) noted that the decrease in molecular weight at the methane condensation level in itself greatly increases the static stability at these pressure levels, further inhibiting convection.

Higher in the atmosphere, the infrared opacity is low and we need to consider radiative heat balance. For a planet at a distance $D$ from the Sun (with $D$ measured in astronomical units, AU), the total power absorbed by the planet is:

$$P_{abs} = (1 - A_B)\pi R_p^2 F_S / D^2 \qquad (7)$$

where $F_S$ is the solar constant at Earth's distance from the Sun ($\sim$1.37 kW m$^{-2}$), $R_p$ is the planet's radius, and $A_B$ is the *Bond Albedo*, the fraction of sunlight energy reflected by the planet in all directions. Assuming no other sources of heat, this absorbed power must be balanced by the thermal radiation emitted to space, which may be calculated as (e.g., Irwin 2009):

$$P_{emm} = 4\pi R_p^2 F_{IR} = 4\pi R_p^2 \sigma T_B^4 \qquad (8)$$

where $T_B$ is the *effective radiating temperature, or bolometric temperature*. Equating the total absorbed power to the emitted power we can rearrange for $T_B$:

$$T_B = \left(\frac{(1-A_B)F_S}{4\sigma D^2}\right)^{1/4} \qquad (9)$$

At high levels in the atmosphere, the opacity of the overlying air becomes progressively smaller. If we consider a very thin layer of opacity $\varepsilon$ high in the atmosphere at local temperature $T_S$ with negligible atmospheric opacity above it, then the atmosphere below effectively emits as a black body of temperature equal to the bolometric temperature $T_B$. However, heat absorbed by the thin layer may be emitted both upwards and downwards and thus in equilibrium, we find:

$$\varepsilon \sigma T_B^4 = 2\varepsilon \sigma T_S^4 \tag{10}$$

Hence, the limiting temperature of the thin layer, known as the *"skin temperature"*, is given by (Goody and Yung, 1989):

$$T_S = \frac{T_B}{2^{1/4}} \approx 0.841 T_B \tag{11}$$

With more detailed analysis it can be shown that the temperature in the upper atmosphere should tend gradually to this stratospheric temperature via the Milne-Eddington equation (e.g., Atreya, 1986):

$$T_S^4 = \frac{T_B^4}{4}(2 + 3\tau) \tag{12}$$

where $\tau$ is the IR optical thickness between the layer and space, or 'optical depth'. Calculating $T(\tau)$ from Eq. 12 we find that the rate of increase of temperature with optical depth as we go down through the atmosphere is initially small and much less than the lapse rate. Hence, the upper atmosphere is convectively stable, since parcels displaced vertically will be colder than the surroundings and tend to return to their original altitudes. However, $\tau$ increases rapidly with depth in the atmosphere as the pressure increases exponentially, and so the radiative equilibrium vertical temperature gradient eventually exceeds the lapse rate, at which point convection in the atmosphere quickly reduces the temperature gradient to the adiabatic lapse rate. The boundary between the convective and radiative regions is known as the *radiative-convective boundary* and typically occurs at a pressure of 0.1 bar (Robinson and Catling, 2014) for all solar system planets, with the single exception of Mars, whose atmosphere is very thin.

While for some planets, the temperature above the radiative convective boundary follows a simple Milne-Eddington radiative equilibrium profile, for many planets we find a minimum of temperature near the radiative convective boundary and then temperatures rising above. The increasing stratospheric temperatures are typically caused by solar radiation through either absorption or photolysis of certain gaseous constituents, or by the absorption of aerosols. Regions of the atmosphere where the temperature increases with height are especially stable to overturning circulation, since any parcels of air that might ascend would cool adiabatically, find themselves much colder and denser than the surrounding air, and sink quickly back to their original level. Hence, the atmosphere in these regions is stably stratified and is known as the *stratosphere*. In contrast the air at deeper levels, where the temperature is governed by the adiabatic lapse rate is constantly convectively overturning and so is known as the *troposphere* (after the Greek word for 'turning'). The minimum of temperature at ~0.1 bar is thus referred to as the *tropopause*.

### 3.1.2 Chemical compositions

The planets in our Solar System form two distinct classes. In the inner solar system are found the terrestrial planets, which are small planets formed predominantly of refractive, rocky materials, further from the Sun are found the giant planets, which are much larger, much less dense and are composed predominantly of gaseous elements such as hydrogen and helium combined with heavier materials such as $H_2O$ and $NH_3$. The reason for this dichotomy in planet types is believed to stem from when the planets formed from the circumstellar disc surrounding the forming Sun, where temperatures were high nearer the Sun and much cooler further away. Following the generally accepted core-accretion model (e.g., Pollack et al., 1996), planets first form by solid particles in the disc clumping and sticking together to form planetary embryos that eventually grow large enough to gravitationally attract material. This lead to a period of runaway growth that eventually becomes so rapid that even gaseous constituents in the disc can be trapped. Near the Sun, only rocky materials were available to form embryos which did not grow quickly enough to directly gravitationally collapse the local nebula. Further from the Sun, beyond the 'ice line', frozen water, ammonia and methane was also available, enabling the rapid growth of much larger embryos that could eventually trap huge quantities of hydrogen and helium from the circumsolar disc. This period of planetary formation is believed to have been halted by the T-Tauri phase of the Sun (e.g., Drouart et al., 1999), when the solar wind became very strong and essentially 'blew way' all remaining gas. In this view Saturn and Jupiter (which are mostly composed of hydrogen and helium) can be thought of as planets where the runaway gravitational collapse of the local nebula was successful. Further away from the Sun, Uranus and Neptune, which would have formed more slowly in the lower density circumsolar disc, were not able to grow fast enough to as effectively trap hydrogen and helium before the T-Tauri phase halted any further growth. Hence, these two planets are predominantly made of heavier constituents such as water, ammonia and methane, with only the outer quarter of the planets composed mostly of hydrogen and helium. This T-tauri phase would also have stripped the inner, terrestrial planets of their primordial atmospheres.

Hence, in this model, the inner solar system planets are solid rock and their atmospheres are secondary atmospheres that have formed later from volcanic outgassing from the interior, and also impacts from comets, formed predominantly of water, methane and ammonia ice. Further from the Sun, the outer planetary atmospheres are those left over from formation and are predominantly composed of $H_2$ and He, with increasing fractions of heavier elements as we move out from the Sun from Jupiter to Neptune.

### 3.1.3 Clouds and Aerosols

As material is transported upwards through the troposphere of a planetary atmosphere, the temperature drops and for some gases the partial pressure becomes equal to the saturated vapour pressure (SVP). As the constituent condenses, cloud layers may form and the gaseous abundance of the condensing molecules will fall as the saturated vapour pressure. To a first-order approximation, the general shape of the abundance profile of a condensable species may be determined from an Equilibrium Cloud Condensation Model (ECCM) by considering a parcel of deep air that is lifted right up through the atmosphere without mixing with surrounding air. At first the mixing ratio remains fixed at its deep level, but at a certain level it becomes equal to the saturated volume mixing ratio (VMR) and thus the gas can start to condense. Moving the parcel to higher altitudes – and thus lower temperatures – more and more of the gas condenses to form aerosols and thus the mixing ratio profile follows the saturated VMR curve. At the tropopause, the temperature stops decreasing with height. If we assume that cloud particles condensed at lower altitudes are not carried with the parcel, but instead remain where they are or fall through the atmosphere, then the VMR cannot rise again above the tropopause by re-evaporation of the aerosols and instead remains fixed at the tropopause value. Hence the tropopause acts as a 'cold trap' to molecules that condense in the troposphere and limits their stratospheric abundances.

While ECCMs are useful for estimating the approximate level of the cloud bases, and thus where the volatile species start to condense, they do not model well the rate of decrease of gas abundance with height since they assume no horizontal mixing of the parcel with the surrounding air, and they do not factor in any precipitation physics. In addition, the technique gives no indication on the vertical extent or optical thickness of the cloud, which has to be inferred from other simple expectations. In general, the cloud structure will depend on two things: 1) the rate of uplift, or vertical mixing; and, 2) the rate of formation of cloud aerosols and their size, which governs how quickly they fall back down through the atmosphere towards warmer regions where they may again evaporate. It is also worth mentioning here that cloud formation is greatly facilitated by the presence of solid particles upon which gaseous constituents can condense. These particles are known as *Cloud Condensation Nuclei* (CCN). For terrestrial planets, such CCNs can be provided by surface dust, while for the Earth, we can also add sea salt and smoke particles. However, for the giant planets, where there is no solid surface, the CCNs are believed to instead derive from photochemical products, generated in the stratosphere and mixed down. Hence, we find that cloud condensation for these planets is intimately related to the photochemistry in the stratospheres. Hence, we can see that accurately estimating the vertical distribution and size distribution of the condensed particles, both of which are vital parameters for correctly interpreting remote sensing observations, are extremely difficult microphysical problems.

The simplest approach to the question of estimating the cloud density is that of Lewis (1969) and Weidenschilling and Lewis (1973), in which as the particle ascends, any gas that condenses is assumed to form cloud particles which are left behind in this level as the air parcel continues on its ascent. A more realistic approach to quantifying the cloud structures is that of Ackerman and Marley (2001). This model attempts to model the vertical dependence of both the condensed and vapour phases of a condensable gas, via a balance equation between the upward turbulent mixing of both the condensed and vapour mole fractions of a condensing gas, $q_t = q_c + q_v$, and the downward transport of condensate $q_c$ through precipitation:

$$-K \frac{\partial q_t}{\partial z} - f_r w_* q_c = 0.$$ (13)

Here, $K$ is the eddy diffusion coefficient, $w_*$ is the convective velocity scale = $K/L$, $L$ is a 'mixing length', defined as $L = H \times \max[0.1, (dT/dz)/\Gamma_{adiab}]$, where $H$ is the scale height, and $f_r$ is freely tuneable 'rain' coefficient. Below the cloud, $q_t = q_{deep}$ and $q_c = 0.0$, while in the cloud $q_c = \max[0, (q_t - q_{SVP})]$ and $q_v = q_t - q_c$. Substituting for $w_*$, this equation simplifies to

$$-L\frac{\partial q_t}{\partial z} - f_r q_c = 0 \qquad (14)$$

In this approximation, when $f_r$ is small, the mole fraction of condensed phase is uniform with height, while as $f_r$ increases, the condensed cloud deck becomes more vertically confined. If we further prescribe the particle size distribution, then we can compute the optical depths of the clouds, with a larger value of $f_r$ implying, but not forcing larger particle sizes. While the Ackerman and Marley parameterisation allows an ECCM model to be used to constrain the vertical profiles of both vapour and condensate, it is still an approximation. In particular it assumes there are abundant cloud condensation nuclei upon which condensates may form and it makes no account of photochemical effects on the condensing gases, nor on how the chemical composition of the CCNs themselves may affect the thermal properties of the condensed ices.

Above the radiative-convective boundary, convective clouds do not usually form unless they are the result of massive deep convection that 'overshoots'. However, this part of the atmosphere is not generally aerosol-free since in these low-pressure regions incident sunlight is able to penetrate and photo-dissociate molecules to yield photochemical products, which are sometimes extremely complicated, depending on the molecules available to be photo-dissociated. These photochemical products can clump to form hazes, which can then 'seed' the condensation of clouds lower in the atmosphere as we noted earlier. As we shall see, the observed aerosols in the giant planet atmospheres are dominated by hazes derived from the photochemical products of $NH_3$, $PH_3$ and $CH_4$.

### 3.2 Venus and Mars

The atmospheres of Venus and Mars are secondary atmospheres outgassed from the interior and are composed predominantly of $CO_2$. As Venus is slightly closer to the Sun (~0.7 AU) than the Earth, and Mars slightly further away (~1.5 AU), we might expect from a simple radiative equilibrium model that Venus's atmosphere will be slightly warmer than the Earths, and Mars's slightly cooler. Indeed, both planets would be classed as 'habitable' if the Solar System was detected as an exoplanetary system. However, planetary atmospheres are rarely as simple as our *ab initio* expectations as we shall see.

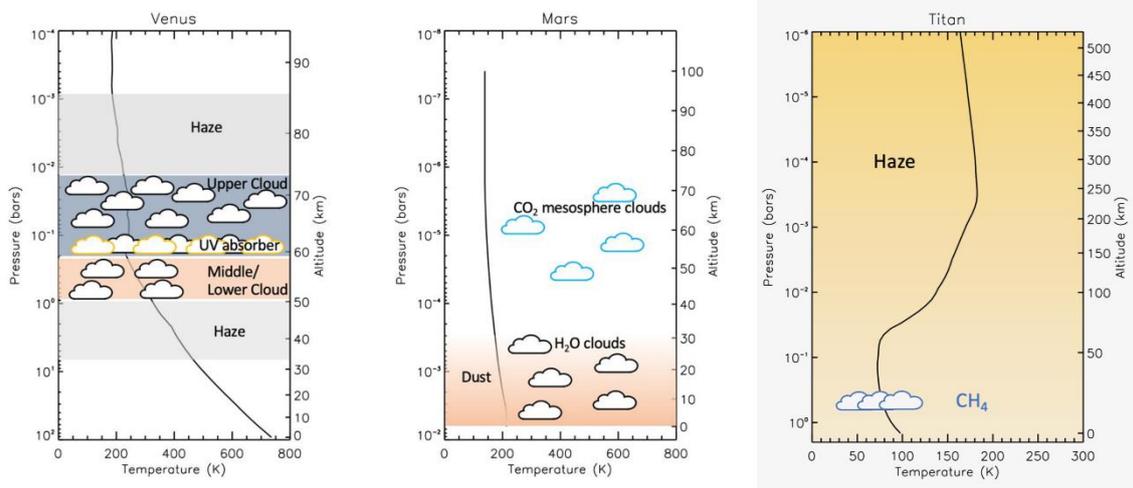

**Figure 2.** *Venus, Mars and Titan Temperature-Pressure and Cloud Profiles*

### 3.2.1 Venus

The atmosphere of Venus is the most extreme of the terrestrial planets of the solar system with a surface pressure of ~90 bar and temperature of ~740K. The atmosphere is composed almost entirely of $CO_2$, with very little $H_2O$ and is thought to have been caused by a runaway greenhouse effect (Komabayashi, 1967; Ingersoll, 1969; Kasting, 1988). Both $CO_2$ and $H_2O$ are 'greenhouse gases' in the sense that they are transparent at visible wavelengths, and thus allow sunshine to penetrate through and heat the ground, but both have strong absorption bands at infrared wavelengths. The presence of these gases in an atmosphere thus raises the altitude of the tropopause, where the infrared optical depth is approximately 1.0, since the temperature increases from the tropopause towards the ground following the adiabatic lapse rate (Eq. 4), which is roughly constant with altitude, an increased tropopause height leads to an increased surface temperatures.  For example, the greenhouse effect of the Earth's atmosphere increases the surface temperature from $T_B$=255K to ~290K, allowing abundant liquid water and organisms to flourish. However, since Venus lies closer to the Sun, its early atmosphere would have been warmer, which would have allowed more water to evaporate into it. Since $H_2O$ is a strong greenhouse gas this would have warmed the atmosphere further, allowing more water to evaporate into it leading to more greenhouse warming and a runaway situation where the oceans evaporated completely. While this was happening, $CO_2$ and more $H_2O$ would have been added continuously through volcanism. On Earth, this $CO_2$ dissolved in the oceans and formed carbonate rocks, and was thus recycled. However, since Venus's oceans are thought to have evaporated early, this trapping was not possible and so the increasing $CO_2$ abundance added to the greenhouse warming. Eventually, almost all of Venus's water was evaporated into its atmosphere, photo-dissociated and lost to space as hydrogen. As a result, the tropopause on Venus occurs at an altitude of ~60 km above the surface at a temperature and pressure of roughly 220 K and 0.1 bar respectively. Below this level the temperature increases quasi-linearly with depth following a dry adiabatic lapse rate of ~10 K/km, leading to the massive temperatures of 740K seen at the surface. Above the tropopause, aside for some small cloud-heating effects, there is no very distinct stratosphere since there is a lack of short-wavelength absorbers and $CO_2$ , which is stable to photodissociation, cools the atmosphere by thermal radiation very effectively at longer wavelengths.

With Venus's hot atmosphere composed mostly of $CO_2$, there are few gases such as $H_2O$ to condense to form clouds. Instead, the aerosols we see in Venus's atmosphere are believed to be composed of concentrated sulphuric acid droplets, with $H_2SO_4$ thought to be derived from the photodissociation products of $CO_2$, $H_2O$ and $SO_2$, the latter perhaps emitted by volcanic activity. The aerosols are spread through a series of cloud/haze layers which locally are not particularly dense, but are spread over a huge height range from 49 – 63 km (e.g., Titov, 2018), and so have very large opacity at visible wavelengths. However, although these optical depths are large, the aerosols are mostly conservatively-scattering and so sunlight can still be scattered all the way down to the surface and thus heat it, which leads to the temperature increasing linearly with depth all the way to the ground.

The vertical structure of Venus's atmosphere is reviewed in detail by Titov et al. (2018). Some *in situ* sounders (e.g., Venera, Pioneer Venus) detected a sharp increase in extinction just below 50 km, suggesting an optically-thick cloud from 47-50km, but some probes did not, suggesting variability. The general consensus is that the top of the atmosphere is dominated by an upper cloud (60 – 70 km), populated by 'mode 1' (r ~0.2 micron) and 'mode 2' (r ~1 micron) particles (Fig. 2). This cloud is believed to be where the sulphuric acid particles are produced, and this layer also contains the mysterious UV-blue absorber (0.3 – 0.5 microns), variations of which give rise to the well-known cloud-top features seen at blue and UV wavelengths (Fig. 2). The UV-Blue absorber is responsible for the absorption of roughly half the solar irradiation and this gives rise to strong local heating at the base of this cloud, which is variable with position, and which leads to a stable temperature profile, resistant to convective overturning and thus defines the tropopause of Venus's atmosphere at ~ 60km. The next cloud down is the middle cloud, which is often separated from the upper cloud by a gap of 1 – 2 km at ~56km. The cloud density increases to maximum at ~ 50km, with the middle cloud running into the lower cloud with no very clear distinction. Here there are possibly trimodal particles (1, 2 and 3 (r ~3-4 microns)), but 'mode 3' might just be tail of 'mode 2' (Toon, 1984). Here the T-profile is more adiabatic, suggesting convective overturning. The middle and lower clouds have been probed *in situ* by Vega probes and balloons.

Above and below the main clouds there is evidence for hazes, of unknown origin. The hazes below ~40 km are particularly intriguing since at these temperatures $H_2SO_4$ is not thermally stable. Very close to the surface, there is tentative evidence from some of the Venera probes of a near-surface haze at 2 – 5 km (Grieger et al., 2014), although the evidence for this is still being debated.

Observations of near-IR spectral windows from the ground and from missions such as NASA's Galileo spacecraft, show variability of this cloud, which appears 'backlit' against the thermal radiation scattered up from the hot surface and deep atmosphere. The morphology is detectable in narrow-wavelength windows of very low $CO_2$ absorption at wavelengths as low as 1 micron. There is even a suggestion that when the clouds are especially thin that the thermal radiation from the surface may be visible to ground-based astronomers observing Venus with the naked eye and hence explain the occasional 'ashen light' that is occasionally reported from Venus's dark side near to the planet's inferior conjunction with the Sun. However, the calculated intensity of this surface thermal emission at visible wavelengths is below the expected threshold of the human eye and as this phenomenon has never been imaged on film or digitally, it may be an optical illusion. Alternatively, it may possibly be some sort of 'airglow' caused by the interaction of Venus' atmosphere with the solar wind (e.g., Gray et al., 2014). At shorter wavelengths, variations in the abundance of the UV-Blue absorber give rise to the familiar 'V' and 'Y' clouds visible from spacecraft such as Pioneer Venus and more recently Akatsuki (see section 3.3.1). To the naked eye, however, Venus appears as a smooth, white, featureless disc.

**3.2.2 Mars**

While Venus, which lies closer to the Sun than Earth, appears to have suffered a runaway greenhouse, conditions at Mars, which is smaller than the Earth and further from the Sun seems to have followed the opposite course. After formation, Mars probably had a warm, dense atmosphere, with abundant $CO_2$ and water, and there is evidence in the observed rock formations that liquid water flowed on the surface of this planet in relatively recent geological history. However, Mars receives less solar heating than the Earth and its smaller size will have meant that it accumulated less heat during formation (and also accrued a small quantity of radio-isotopes to generate internal heat from radioactive decay). This means that convective overturning of the mantle would likely have stopped earlier, leading to the formation of a thick crust and the cessation of volcanism. In addition, this more rapid cooling would have led to the collapse of Mars' magnetosphere due to the early cesseation of the convective 'dynamo' in Mars' core. This would have resulted in the atmosphere being more exposed to the solar wind and thus eroding even faster.

While there is ample evidence of volcanism in the past on Mars, there is no evidence of volcanic activity now. Hence, the resupply of $CO_2$ from the interior would have switched off earlier and some of the $CO_2$ that was in the atmosphere is believed to have dissolved in the oceans to form carbonate rocks and was so locked away, reducing the greenhouse warming. The lower temperatures arising would have reduced the water vapour component, reducing the greenhouse even more. In addition, Mars' low surface gravity left it more prone to thermal (Jeans) escape, which is more efficient for lighter isotopes than heavier and led to a considerable enrichment in heavier isotopes compared to Earth. In particular, the D/H enrichment suggests that Mars has lost a huge amount of its original water. As a result of all these processes, the surface atmospheric pressure has decreased from perhaps 1 bar in the past to just the $\sim 6$ mbar we see now. Although much of Mars' water has been lost, water is still present on Mars, but is mostly locked in permafrost and the planet's permanent ice caps. Mars' mean surface temperature of $\sim 220$ K is close to the expective radiative equilibrium temperature and is seen to fall by $\sim 2.5$ K/km from the surface until a poorly defined tropopause at an altitude of $\sim 40$ km, after which it follows roughly a radiative equilibrium profile. Solar heating at these levels is very low, due to the atmosphere being thin and largely transparent at visible wavelengths and hence the temperature does not rise again to form a stratosphere. Instead, the region above the tropopause is known as the mesosphere. The current atmosphere on Mars is now so cold that a considerable fraction of the $CO_2$ atmosphere freezes out at the winter poles to form seasonal $CO_2$ ice caps on top of the permanent water ice caps and this seasonal variation causes the surface pressure to vary globally from 8 mb at the equinoxes to 4 mbar at the solstices.

Most of the aerosol in Mars' atmosphere is in the form of suspended surface dust particles, which are raised from the dry surface via strong winds and lead to a background visible opacity of about 0.4 (Fig. 2). This dust loading leads in turn to the tropospheric lapse rate of $\sim$ 2.5K/km being considerably smaller than the expected dry adiabatic lapse rate of $\sim$ 4.3 K/km due to absorption of sunlight by the dust particles heating the air. The presence of dust has such a very large effect on local atmospheric solar heating that in some circumstances a positive feedback effect can happen, with increased dust leading to increase solar heating, which can lead to increased winds, which can kick up more dust, and so on. Hence, in some regions local dust storms can be generated that, under certain conditions, can amplify to global dust storms that completely shroud the planet's surface for periods of months at a time (e.g., Conrath et al., 1973; Pollack et al., 1979; Briggs et al., 1979).

Although water vapour has a low mole fraction in Mars' atmosphere, the atmosphere is so cold that the atmosphere is never very far from saturation and so water ice clouds are actually rather common (Fig. 2). Martian seasons are defined by the solar longitude, $L_S$, which is the angle between the line from the Sun to Mars and the same line at the moment of Mars' northern spring equinox. Hence, $L_S = 0°$ is the northern spring equinox, $L_S = 90°$ is northern summer solstice, and so on. Mars atmosphere also needs to be understood in the context that its orbit is rather eccentric, with its distance from the sun varying from 1.38 AU at perihelion ($L_S = 251°$) to 1.67 AU at aphelion ($L_S = 71°$). This variation gives rise to strong variations in temperatures, and so cloud formation. As a result, the water clouds in Mars atmosphere are seen mostly in the 'Aphelion Cloud Belt' (e.g., Määttänen and Montmessin, 2021) from $L_S = 30°$ to 160° at low latitudes (10°S to 30°N). The most prominent clouds are associated with orographic uplift over large-scale topographic features such as the Tharsis ridge. The prevailing easterlies at the equator cause the formation of 'V' shaped clouds, with one of the volcanoes at the apex (rather like the wake of a boat). Tenuous clouds associated with Earth-like weather fronts have also been seen in the winter hemisphere and low-level fogs are a feature of the southern hemisphere at dawn and dusk, due to the rapid cooling of the surface at night. Some of this fog can even settle as a frost at dawn. Nearer the poles, the growth of the seasonal polar caps during Autumn takes place under a hood of $CO_2$ and water ice haze. This clears by early winter, but the air at the cap edge has a very stable temperature inversion and so air deflected by surface features can undergo vertical gravity-wave oscillations producing regular trains of clouds over hundreds of km. Finally, in addition to near the polar caps, thin $CO_2$ ice clouds are also seen in the mesosphere in regions of especially low temperature at autumnal mid-latitudes and also occasionally at equatorial latitudes.

### 3.2.3 Prospects for telescopic observations of Venus and Mars

Telescope observations of Venus are difficult due to its proximity to the Sun, but observations are possible in multiple wavelength bands. Venus presents its largest apparent disc size to the Earth when it is close to inferior conjunction, when we see part of Venus's disc reflecting sunlight, but also the nightside. Although the dayside reflectivity dominates visible and near-infrared observations, it is possible to view thermal emission from the surface and lower atmosphere in narrow wavelength intervals of very low $CO_2$ absorption from $1-3$ μm, even by amateur astronomers. At 1 micron, the topography of Venus' surface is observable since high altitude regions are cooler due to the adiabatic lapse rate. It is even argued that such observations have captured active volcanism on Venus' surface (Shalygin et al., 2015; Mueller et al., 2017). At longer wavelengths, the thermal emission observations can be used to probe atmospheric abundances. Recently, Greaves et al. (2020) announced the discovery of a phosphine absorption line in Venus' atmosphere from submillimeter observations, which, it was claimed, might originate from organic activity in the clouds. This discovery and interpretation has been contested by several subsequent studies, including Encrenaz et al. (2020), Snellen et al. (2020), Lincowski et al. (2021) and Villanueva et al. (2021), who note that the absorption line actually originates above Venus's clouds in the mesosphere, and is more likely to be caused by $SO_2$ absorption, which is known to be present and highly variable in Venus's atmosphere and has an absorption line at almost exactly the same frequency. This conclusion is also supported by Stratospheric Observatory for Infrared Astronomy (SOFIA) observations, which observes at wavelengths where there is not any ambiguity with $SO_2$ lines and which did not detect phosphine (Cordiner et al., 2022).

Telescope observations of Mars can provide a continuity of observations for monitoring seasonal change, such as dust storm activity, the development of Mars' seasonal polar ice caps and ephemeral clouds. In addition, high-resolution spectroscopy at visible and sub-millimetre wavelengths can be used to search for trace species such as $H_2O_2$ and $CH_4$ (e.g., Encrenaz, 2009) and also directly measure the wind speeds by detecting the Doppler shifts of lines near the planets' limb (e.g., Moreno et al., 2009). In general, the visible and near-infrared are most sensitive to the distribution of cloud particles, via reflection of incident sunshine, while longer wavelengths probe thermal emission from the ground and atmosphere and are sensitive to composition, but less sensitive to cloud, whose opacity drops rapidly with increasing wavelength. Of particular note is the search for gaseous methane, which was first detected by the Planetary Fourier Spectrometer (PFS) on ESA's Mars Express spacecraft (Formisano et al., 2004). Methane has since been detected in ground-based observations (e.g., Mumma et al., 2009), and also from surface landers (e.g., Webster et al., 2015; Giuranna et al., 2019), but it has not been detected by sensitive instruments on the Trace Gas Orbiter spacecraft (e.g., Korablev et al., 2019; Montmessin et al., 2021) and its detection and temporal and spatial variability is hotly debated (e.g., Zahnle et al., 2011; Fonti et al., 2015). If it is present, then potential sources hydrothermal activity, or perhaps even microbial life!

### 3.3 Jupiter and Saturn

The atmospheres of Jupiter and Saturn are both to first order of solar composition and thus composed mostly of $H_2$ and He. The abundance of heavier elements is greater than that seen in the Sun's atmosphere by a factor of ~3 for Jupiter and ~10 for Saturn (e.g., Mousis et al., 2018), which is consistent with the core accretion model and Saturn collapsing a smaller fraction of the gaseous nebula before the start of the Sun's T-tauri phase (e.g., Helled, 2023). Although the planets have similar radii of ~10 $R_\oplus$, Jupiter is much more massive with a mass of 318 $M_\oplus$, compared with 95 $M_\oplus$ for Saturn. As a result, the scale heights at the observable cloud tops at ~1bar are very different (~25 km for Jupiter, and ~50 km for Saturn), which has great implications for the visibility if the clouds as we will see. The temperature-pressure profiles of Jupiter and Saturn are consistent with our *ab initio* expectations with tropopauses at 0.1 bar, adiabatic profiles below and pronounced stratospheres, which are caused by the absorption of solar radiation directly by methane, and also by photodissociation. Both Jupiter and Saturn are seen to be radiating more energy than they absorb from the Sun, which is believed to be caused by the continuous slow release of primordial heat, gathered during formation and still radiating away (e.g., Pearl et al., 1990). As a result, their bolometric temperatures are greater than the radiative equilibrium temperatures we would expect from Eq. (9).

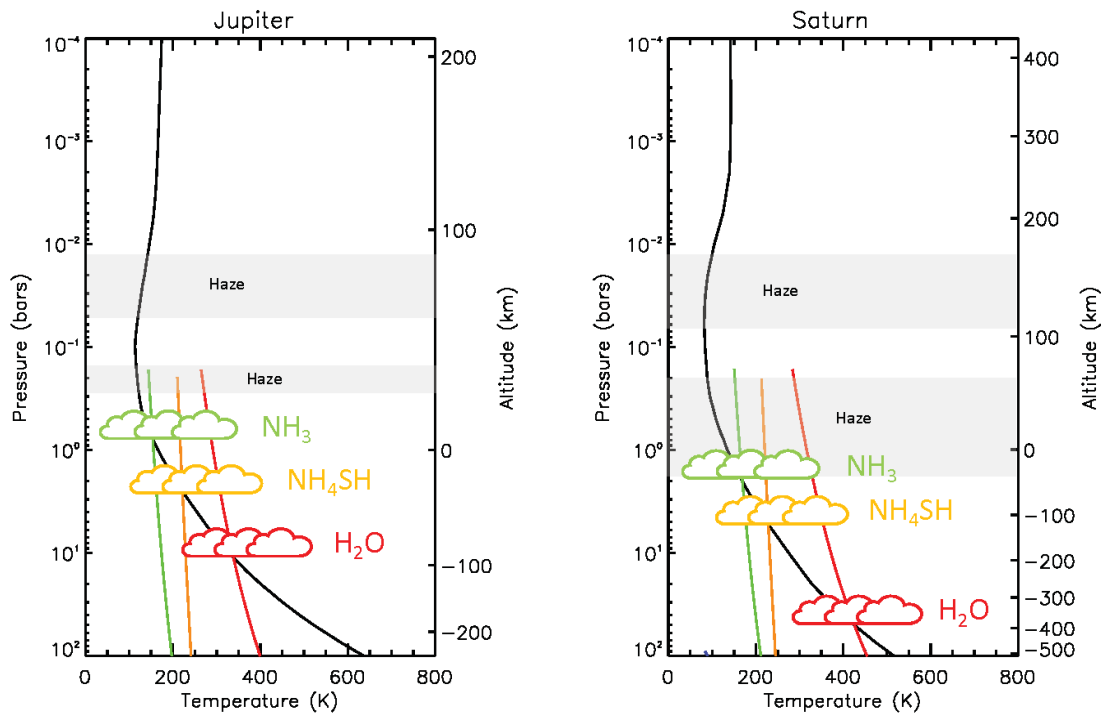

**Figure 3.** Jupiter and Saturn Temperature-Pressure and Cloud profiles

### 3.3.1 Jupiter

Applying an ECCM to the atmosphere of Jupiter leads to the expectation of the following cloud layers (e.g., Atreya, 2005a): an aqueous ammonia cloud based at 8 bar, an $H_2O$ ice cloud based near 6 bar, an $NH_4SH$ cloud based at 2-3 bar and an ammonia ice cloud based at ~0.7 bar. However, the cloud layers seen in Jupiter's atmosphere are not quite as simple as that predicted by ECCMs. Instead, numerous analyses of Jovian observations (reviewed by Dahl et al., 2021) find that the main cloud deck, presumably composed mostly of $NH_3$ ice, perhaps mixed with other ices, is based deeper at roughly $1-1.5$ bar. At these levels pure ammonia ice will sublimate and hence it seems that the clouds are not pure condensates at all, but instead heavily mixed with photochemical products that affect their sublimation temperature. In addition, the clouds arising from condensed ammonia, water, etc., are expected to be formed of bright highly-scattering particles. However, the aerosols in Jupiter's atmosphere are observed to be coloured by blue-absorbing particles of debated composition, known as chromophores (after the Greek for 'colour carrier'). These chromophores are believed to be generated by photochemical dissociation of methane, ammonia and phosphine at higher altitudes, which mix down to cloud levels to both seed cloud condensation and also either mix with the pure condensates, or coat them (known as riming) making them darker. Hence, the observed clouds range in colour from white/ochre in the bright 'zones', where the air is thought to be rising, bringing fresh condensable gas from below, to the brown colours seen in the darker 'belts', where the air is thought to be sinking back to down and concentrating the photochemical products. We also see the scarcity of pure condensates in high-resolution reflection spectra of Jupiter. If ammonia ice were present then we would expect clearly identifiable ammonia ice absorption features, but these appear to be largely absent. The exception to this is in small, bright areas of rapid convection, where Galileo/NIMS observed Spectrally Identifiable Ammonia Clouds (SIACs) (Baines et al., 2002). Higher in the atmosphere diffuse haze layers are seen in the upper troposphere, thought to arise from the photolysis of upwelling ammonia and phosphine, and thinner hazes are seen in the stratosphere, arising from photolysis of methane to generate ethane, acetylene and other hydrocarbon products

The Galileo mission included an atmospheric entry probe, which it was hoped would provide a definitive characterisation of the vertical cloud structure. The probe was targeted at the northern edge of the bright equatorial zone, which is a region where the clouds are usually thick. However, occasional breaks in the clouds are seen at this latitude, which look dark from Earth (because of the lower cloud reflectance), but bright at 5-microns, which detects light thermally emitting from deeper in the atmosphere (~10 bars) and scattered up through the clouds. Most unfortunately, the Galileo probe entered in the middle of just such a '5-micron hotspot' and returned detailed observations of an extremely anomalous sample of Jupiter's atmosphere (Orton et al., 1998)! Just how anomalous can be seen in the cloud density profile returned by the probe's nephelometer experiment, which showed low opacity, vertically thin clouds at 1.8 bar (assumed to be $H_2O$ ice), 1.4 bar (assumed to be $NH_4SH$) and 0.4 bar (assumed to be $NH_3$ ice).

The zonal cloud structure of belts and zones is mostly symmetric about the equator, although the Northern Equatorial Belt is generally darker than its counterpart in the southern hemisphere, which instead is adjacent to Jupiter's Great Red Spot which is noticeably red. However, the visibility of the belts and zones varies greatly from year to year (e.g., Rogers, 2009) as cloud activity varies in what is clearly a very active dynamical system. Jupiter's atmosphere contains numerous anticyclonic vortices, or storm systems, of which the most well-known is the Great Red Spot (GRS) at 22°S, which has existed in its current form since at least 1831. The GRS has a diameter roughly that of the Earth, and extends from the troposphere at 0.1 bar to pressures perhaps as great as 100 bar, as determined by Juno/MWR observations (Bolton et al., 2021). Another notable anticyclone vortex in Jupiter's atmosphere is Oval BA (near 33°S) that resulted from sequential mergers of three similar anticyclones that first appeared in 1939-40 (Sánchez-Lavega 1999, 2001). Oval BA turned red in 2005 (Simon-Miller et al., 2006) due probably not to major dynamical changes but to the formation of a red chromophore in a periphery ring (Pérez-Hoyos et al. 2009; Wong et al. 2011).

The Jovian atmosphere is seen to be widely coloured from near-white in the zones, to a brown colour in the belts and the red colour seen in the GRS and Oval BA. Sromovsky et al. (2017) suggest that all these colours can be explained by a single colour-carrying compound, a so-called 'universal chromophore', based on material generated in the laboratory by photolysing mixtures of $NH_3$ and $C_2H_2$ using ultraviolet radiation (Carlson et al., 2016). Baines et al. (2019) found that this chromophore was most likely located in a narrow layer in the upper troposphere just above the level of the ammonia clouds (the so-called 'Crème Brûlée model'), while Braude et al. (2020) found that to model the belts and GRS the chromophore was spread more within the cloud, with a steeper blue-absorption gradient than the Carlson et al. chromophore. Work is ongoing in this area. Anguiano-Arteaga et al. (2021, 2023) found that in the visible the GRS and surroundings can be explained by a two- layer model. At the top of this model, lies a stratospheric haze with a base near the 100 mbar level, an optical depth $\tau$ at 900 nm of the order of unity and particles with a size of 0.3 μm. Beneath this haze lies a more vertically extended tropospheric haze of micron sized particles with $\tau$ ~10 (900 nm) extending down to 500 mbar. Both haze layers show a stronger short wavelength absorption, and thus both act as chromophores. The altitude difference between cloud tops in the GRS and surrounding areas in this model was found to be ~10 km.

Finally, before the arrival of the Juno mission at Jupiter in 2016, observations of Jupiter's atmosphere revealed that the abundance of ammonia was high in the zones and low in the belts (certainly for the SEB, EZ and NEB, but less clearly at more polar latitudes), which was consistent with the view of upwelling in the zones and downwelling in the belts. It was expected that the ammonia abundance would tend to a uniform value with latitude below the cloud formation layers. However, Juno microwave (MWR) observations found that deeper than 1-2 bars the abundance of ammonia is depleted down to very deep pressures (20-30 bars) at all latitudes except the equator (Li et al. 2017, Moeckel et al. 2023), especially in the North Tropical Belt. What causes this depletion is still the source of considerable debate, but Guillot et al. (2020) suggest that at the pressures and temperatures found in Jupiter's atmosphere ammonia vapour can dissolve into water ice to form a special phase of material they name 'mushballs', which quickly grow to form hail-like particles that can strip the atmosphere of ammonia down to the very great depths seen in the Juno/MWR observations. Although this interpretation is still debated, it is clear that the physics of cloud formation in the cold, hydrogen-dominated planets of the outer solar system may be considerably more complicated than the simple expectations of ECCMs.

### 3.3.2 Saturn

With a similar composition and temperature-pressure profile, the expected vertical profile of clouds in Saturn's atmosphere is very similar to that of Jupiter, but the lower atmospheric temperatures mean that: 1) the expected cloud bases are deeper; and 2) the increased scale height of Saturn's atmosphere leads to the clouds being more vertically extended. ECCM simulations lead to the expectation of the following cloud layers (e.g., Irwin, 2008): an aqueous ammonia cloud based at 18 bar, merging into an $H_2O$ ice cloud based near 10 bar, an $NH_4SH$ cloud based at 5 bar, and an ammonia ice cloud based at ~ 1.8 bar. However, once again, there is a clear gap between the ECCM predictions and the actual clouds/hazes seen in Saturn's atmosphere. The vertical structure of the observed clouds seen in Saturn's atmosphere, determined from ground-based, HST, and Voyager 2 observations, is reviewed by Pérez-Hoyos and Sánchez-Lavega (2006). Multiple solutions are consistent with the observations, which vary from study to study, but can be summarized as being composed of three components: 1) a deep cloud based at p > 1 bar; 2) a thick upper tropospheric haze at 100 – 500 mb; and 3) a thin stratospheric haze at p < 100 mb (West et al., 2009). As for Jupiter, the upper tropospheric haze likely arises from the photodissociation of upwelling ammonia and phosphine, while the stratospheric hazes are likely photochemical productions from the dissociation of methane.

The higher abundance of upper atmospheric haze in Saturn's atmosphere, whose opacity does not vary rapidly with latitude, and the large scale-height of Saturn's atmosphere, gives rise to a much blander appearance at visible wavelengths, with a bright equatorial zone seen, but otherwise limited variation. However, Saturn's zonal winds do vary rapidly with latitude, pointing to the presence of upwelling zones and downwelling belts (see section 4). The deeper atmospheric structure only reveals itself at longer wavelengths, where the reflectivity of the small haze particles decreases, and especially at 5 microns, where we see thermal emission from the deep atmosphere leaking up through gaps in the overlying clouds revealing cloud structures as dark features. Cassini/VIMS observations show that the deeper structure varies very rapidly with latitude and is highly detailed (Choi et al., 2009). Wave-like features are also seen such as the 'ribbon-wave' seen by Voyager 2 at 41°N (planetocentric) and subsequent ground-based observations, and the North Polar Hexagon (NPH), first seen by Voyager 2 in 1986 (Godfrey and Moore, 1986) and since observed by Cassini and from the ground (Sánchez-Lavega et al., 2021b). The NPH was observed by the VIMS (Baines et al., 2009) and CIRS (Fletcher et al., 2008) instruments on Cassini, showing that the feature extends over a very wide range of pressures from ~6 bar (VIMS) up to the tropopause at 0.1 bar (CIRS) (Sayanagi et al., 2018). The haze and cloud structure within the NPH was further explored from Cassini/ISS visible/near-IR observations by Sanz-Raquena at al. (2018) who found the opacity of the tropospheric haze to be greatly depleted and suggested that this is a region of downwelling. At even higher altitudes, detached multilayer hazes were seen above the NPH by Cassini/ISS limb observations (Sanchez-Lavega et al. 2020), extending from 0.5 to 0.01 bar and believed to be composed of hydrocarbon ices, condensed from photochemical products of methane in the stratosphere. Hence, this atmospheric structure extends over a remarkable large altitude range in Saturn's atmosphere.

### 3.3.3 Prospects for continued telescopic observations of clouds of Jupiter and Saturn

Ongoing imaging observations of Jupiter with imagers such as HST/WFC3 have and will continue to be used to monitor the ongoing changes in Jupiter's zones and belts, and also the evolution in appearance and occasional interactions between vortices such as the Great Red Spot and Oval BA. Meanwhile, IFU observations with instruments such as VLT/MUSE (475-933 nm) are being used to better characterize Jupiter's vertical and latitudinal cloud structure (e.g., Braude et al., 2020). At longer wavelengths, observations at 5-$\mu$m reveal Jupiter's clouds dark features against a bright background of thermal emission from $\sim$ 5-6 bar, sometimes in astonishing detail, and this region is also sensitive to absorption from $NH_3$, $PH_3$, $GeH_4$ and $AsH_3$, whose spatial variation is indicative of the meridional, overturning circulation. At even longer wavelengths, thermal emission from the upper troposphere (1-0.1 bar) and stratosphere (0.01-0.001 bar) is detected that can be used to determine the temperature structure at these altitudes, which strongly constrains the dynamics of the atmospheres. Preliminary results from the James Webb Space Telescope, launched in 2021, which includes IFU spectroscopy covering 1 – 30 mm, are already totally overhauling our understanding of these atmospheres. Going forward, continuing telescope observations will be needed to support ongoing space missions such as Juno and also JUICE, which was launched in April 2023 and will arrive in Jupiter orbit in 2031. Such telescope support is vital in matching spacecraft observations targeted to specific small regions to the wider context of Jupiter's atmosphere.

The aims and goals of ongoing observations of Saturn mirror those of Jupiter, but are made more difficult by Saturn's smaller size, obscuration by Saturn's rings, and lower temperatures, which lowers the radiance at thermal infrared wavelengths. Continued visible and near-IR observations are needed to monitor the seasonal buildup and change of haze near Saturn's poles, and also monitor the potential development of Great White Spots. Once again, longer wavelength thermal infrared observations allow determination of gaseous composition and atmospheric temperature.

### 3.4 Titan

Titan's atmosphere is Earth-like in its basic composition, consisting mainly of nitrogen ($N_2$) and methane ($CH_4$), the latter playing the role of condensable gas to form clouds (Table 1). Photochemical reactions give rise to a rich variety of hydrocarbons and nitrogen compounds, including oxygen derivatives (Bézard et al. 2014, Dobrijevic et al., 2014). These form vertically extended layers of dense hazes up to $\sim$ 1000 km above surface (Fig. 2c) (West et al. 2014). The densest parts of these hazes concentrate in the lowest $\sim$ 600 km, being ubiquitous below $\sim$ 300 km, encompassing the troposphere and stratosphere (Hörst 2017). The spatial and temporal variability in haze optical properties, for example following the long seasonal cycle, can be monitored from the integrated brightness of the disk as a function of wavelength (from UV to near-IR), viewing geometry and solar illumination effects (Sromovsky et al. 1981; Lorenz et al., 2004). Disk-resolved spectra in the H and K bands (1.4–2.4 $\mu$m) have been used to track short-term variabilities (Nichols-Fleming et al. 2021). These hazes affect the thermal energy balance of the atmosphere and its general circulation. It is therefore necessary to determine the spatial and temporal distribution of the chemical compounds in relation to the insolation and solar activity cycles in order to interpret changes in atmospheric dynamics (Griffith et al., 2014) (see 4.2.5).

Clouds from the condensation of $CH_2$-$N_2$ occurs in the lower atmosphere (between ∼ 9 to 15 km from the Huygens probe measurements), and above 15 km and up to ∼ 45 km is methane ice condensation that leads to the formation of extensive clouds, which were first observed with ground-based telescopes before the arrival of the Cassini mission (Griffith et al. 2014, Hörst 2017; see 4.2.5). Methane follows a condensation-evaporation cycle similar in many respects to the terrestrial hydrological cycle, many aspects of which are unknown. Continuous observations are needed to determine the nature of these tropospheric methane clouds (convective thunderstorms, stratus), their spatial distribution, their cycle of condensation-precipitation-evaporation in relation to the insolation cycle, and the properties of the particles that form them.

Other condensate clouds have been observed with Cassini instruments over the polar areas (Griffith et al. 2014, Hörst 2017 and references therein). In the northern pole an extensive ethane ($C_2H_6$) ice cloud was observed in the winter hemisphere at altitudes of ∼ 30 to 60 km driven, according to models, by subsiding circulation (De Kock et al 2014; West et al. 2016). This cloud vanished progressively around the equinox in 2009 in agreement with GCM models (Rannou et al 2006). A condensate cloud of HCN was observed by Cassini first to form over the south polar region (between 2004-2008) at an altitude of 300 km tracing a compact vortex. The situation reversed and this cloud subsequently formed over the north polar area between 2012 and 2017 (Le Moullet et al 2018) in agreement with atmospheric circulation predictions and seasonal changes in hydrocarbon distributions (Vinatier et al., 2017). Monitoring of these cloud changes should be objectives of the JWST and the largest ground-based telescopes.

## 3.5 Uranus and Neptune

The observable atmospheres of Uranus and Neptune are predominantly composed of $H_2$ and He, but there is a much greater abundance of heavy elements (30 – 100 times solar): Lindal et al., 1987; Lindal, 1992; Tollefson et al. 2021, Molter et al., 2021). This is consistent with the core formation scenario, with both planets unable to form quickly enough to condense large masses of the solar nebula and so being mainly composed of heavier elements. Both planets have a similar radius of ~4 $R_\oplus$, and a mass of 14.5 $M_\oplus$ for Uranus and 17.1 $M_\oplus$ for Neptune. The similar mass and radius gives a similar scale height of ~30 km for both planets. Again, the temperature-pressure profiles of Uranus and Neptune are consistent with our *ab initio* expectations, with tropopauses at 0.1 bar, adiabatic profiles below and pronounced stratospheres, which are caused by absorption of solar radiation by methane and also photodissociation. However, while Neptune is seen to be radiating considerably more energy than it absorbs from the Sun (2.5 times as much), Uranus is seen to be in almost perfect radiative equilibrium (e.g., Pearl et al., 1990), suggesting that all its internal heat of formation has already escaped, or that internal convection is inhibited in some way (e.g., Nettelmann et al., 2016). As a result, although Uranus is considerably closer to the Sun than Neptune, its atmospheric temperature profile is very similar to that of Neptune's. An additional difference between Uranus and Neptune is that the obliquity of Uranus is a massive 98°, compared with an Earth-like 29° for Neptune. This means that Uranus's atmosphere experiences the most extreme seasonal forcing of any planetary atmosphere, with its summer pole pointing almost directly towards the Sun at solstice. Indeed, averaged over its 84-year orbit the poles receive more sunlight than the equator.

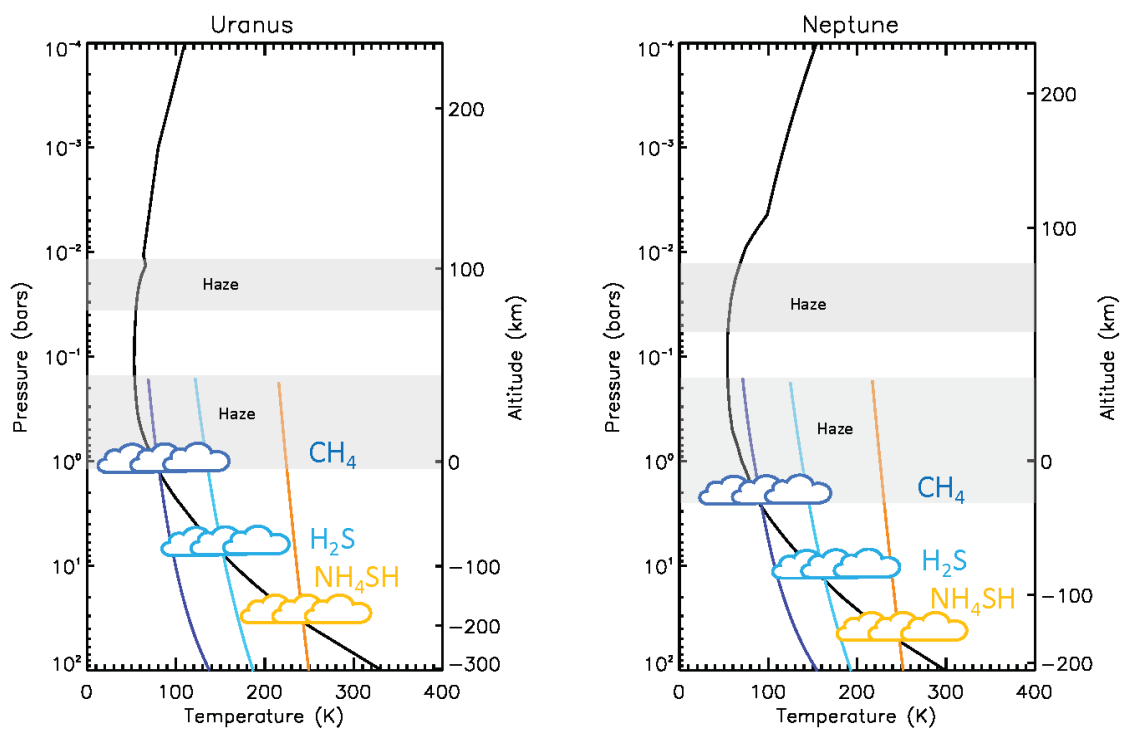

**Figure 4.** Uranus and Neptune Temperature-Pressure and Cloud Profiles

As the temperature-pressure profiles are so similar, applying an ECCM to the atmospheres of Uranus and Neptune gives similar expected structures with a predicted water cloud at ~100 bar, an $NH_4SH$ cloud at ~40 bar, either a $NH_3$ cloud or $H_2S$ cloud at ~10 bar and a cloud of $CH_4$ at 1-2 bar. For Jupiter and Saturn, the abundance of ammonia appears to be greater than that of hydrogen sulphide and so all $H_2S$ appears to combine with a $NH_3$ to form a cloud of $NH_4SH$, leaving $NH_3$ free to condense alone at lower pressures. However, for Uranus and Neptune, it appears that the opposite is true and the abundance of hydrogen sulphide exceeds that of ammonia, which is thus combined with $H_2S$ in a cloud of $NH_4SH$ at ~40 bar, leaving $H_2S$ alone to condense at 5-10 bar. The evidence for this comes first from observations at microwave wavelengths ($1 - 20$ cm), where de Pater et al. (1991) noted that the continuum absorption in observations made with the Very Large Array (VLA) suggests that the abundance of $H_2S$ must exceed that of $NH_3$ in the $20 - 40$ bar region, and later a direct detection of $H_2S$ absorption lines in the $1 - 10$-bar region from Gemini/NIFS spectra near 1.5 μm (Irwin et al., 2018, 2019). This could either mean that Uranus and Neptune accreted much more $H_2S$ at formation, which has implications for formation scenarios, or that much more $NH_3$ is combined with $H_2O$ in either: a) an aqueous ammonia cloud'; b) a deeper Water-Ammonia Ionic Ocean (Atreya et al. 2005b) in the deeper atmosphere; or c) perhaps even combined with water in a 'mushball' phase (Guillot, 2021).

Microwave observations of Uranus and Neptune find considerably lower thermal emission from equatorial latitudes than polar latitudes latitudes (e.g., Molter et al., 2021; Tollefson et al., 2019), which suggests that the abundance of $H_2S$ at the ~40 bar level is considerably enhanced at equatorial latitudes; the increased opacity forcing the microwave emission to come from lower pressure levels where the temperature is lower. Higher in the atmosphere, the abundance of methane is found to be a significant fraction of the total atmosphere (2 – 4%). For many years it was assumed that the abundance did not vary greatly with latitude, but by analysing the 800 – 850 nm region of HST/STIS observations, Karkoschka and Tomasko (2009, 2011) demonstrated that the abundance of methane also varies greatly with latitude for both planets, with abundances of ~4% at the equator reducing to ~2% at the poles. The distribution of both $H_2S$ and $CH_4$ suggests that convective overturning is greater at the equator than at the poles.

Although we have evidence that the deep $H_2S$ abundance exceeds that of $NH_3$, and so we expect a deeper cloud of $H_2S$ ice, we might also expect a very thick cloud of methane ice at 1-2 bar since methane is such a large component of the atmospheres of these planets (~4%). However, we find that once again our simple expectations from an ECCM do not survive contact with observations. Instead, the atmospheres of both planets are found to be shrouded by a medium-opacity (~ 1 – 3 at visible wavelengths) layer of aerosol (radius ~ 1 μm) at approximately the expected methane condensation level (1-3 bar), which are strongly absorbing at UV/blue and red/infrared wavelengths. This absorption indicates significant mixing or riming with photochemical materials generated in the stratosphere and upper troposphere, which have been mixed down to lower levels (Irwin et al., 2022). Deeper in the atmosphere, Irwin et al. (2022) find a layer of aerosol with a top near 5 bar, assumed to be linked with the $H_2S$ condensation level, but again composed of material with significant UV/blue and red/infrared absorption. Above both discrete layers is found an extended haze of similarly absorbing, but smaller particles (r ~ 0.1 μm) extending from the 1 – 3-bar haze layer up into the stratosphere. In addition to this basic atmospheric structure, which is found to vary little with latitude, discrete, bright, white clouds are seen in the ~100 – 600-mb region, whose scattering properties are consistent with the medium-sized (~1 μm) methane ice particles.

These appear to be caused by convective events and are found to be much more widespread and thicker in Neptune's more vigorous atmosphere than in Uranus's more 'sluggish' atmosphere. For Neptune, these upper tropospheric methane ice clouds are particularly concentrated in the latitude bands 20-40°S and 20-40°N, which suggests rapid upwelling of air to the tropopause at these latitudes. However, this circulation seems highly variable, and while these clouds were thick in Neptune's atmosphere in the early 2000's, they have been remarkably sparse in recent years. For Uranus, the upper tropospheric methane ice clouds were seen to be virtually absent during the Voyager 2 encounter, but this coincided with Uranus's southern summer solstice, when Uranus's atmosphere appears to be particularly quiescent. As Uranus moved through its orbit towards northern spring equinox in 2007, these clouds became much more commonly seen and were probably initiated by seasonal variations in solar insolation warming the lower atmosphere and generating instabilities. Now that Uranus's north pole is swinging towards the Sun (reaching northern summer solstice in 2030) the level of this convective storm activity appears to be waning again.

Higher in the atmosphere, in the stratosphere at pressures p < 0.1 bar, the photodissociation of methane is believed to generate complex hydrocarbon chains that combine together to form haze aerosols (e.g., Moses et al. 2020; Toledo et al. 2019, 2020). These are transported to lower altitudes by turbulent mixing to seed cloud condensation in the troposphere, but while in transit from the warm stratosphere to the cold tropopause at 0.1 bar (T ~55K) they also seed the condensation of gaseous photochemical products such as $C_4H_2$, $C_2H_2$ and $C_2H_6$ to form tenuous ice layers at pressures of ~0.001, 0.01 and 0.02 bar, respectively.

In the troposphere, how is it that we do not detect a thick methane ice cloud at the methane condensation level? There is a key major difference between giant planet atmospheres and the Earth's atmosphere in that condensing species ($H_2O$, $NH_3$, $CH_4$, etc.) are all *heavier* than the bulk of the hydrogen-helium air. The opposite is true for Earth's atmosphere, where water vapour is lighter than the bulk of the nitrogen-oxygen atmosphere. As a result, 'moist' air is naturally buoyant in the terrestrial atmosphere and tends to rise, whereas 'moist' air is naturally dense in giant planet atmospheres and tends to sink. Hence, while clouds may form in the Earth's atmosphere simply by moist, buoyant air rising to its condensation level, the formation of clouds in the giant planet atmospheres is more complicated. This difference is particularly acute for the main level of methane ice condensation in the atmospheres of Uranus and Neptune since methane has such a high mole fraction (~4%) in an otherwise light $H_2$/He atmosphere, that there is a significant drop in the mean molecular weight of these atmospheres at these levels. This drop is so strong that, even when we account for the heat released by latent heat of formation, the atmospheres are predicted to be statically stable and will thus resist convective overturning (Irwin et al., 2022, Hueso et al., 2020, Leconte et al., 2017, Guillot, 1995). Irwin et al. (2022) argue that haze particles generated by photochemical destruction of methane in the stratosphere and upper troposphere are transported down to the methane condensation level by turbulent mixing, and there seed $CH_4$ ice formation.

However, since the abundance of $CH_4$ is so high the formation of ice particles is very rapid, leading to the formation of large particles (r ~5mm, Carlson et al., 1988) that immediately 'snow out', falling to lower levels were the methane ice sublimates, allowing the haze particles to mix further down to lower levels where they can then seed the condensation of $H_2S$ ice at 5-10 bar. At the $CH_4$ condensation level, we are then left with a mixture of hydrocarbon haze and small methane ice particles that is confined in a region of static stability leading to the relatively featureless haze observed. Irwin et al. (2022) find this layer, which they dub 'Aerosol-2' to be twice as optically thick in Uranus's less convectively active atmosphere than in Neptune's more turbulent conditions, which serendipitously explains why Uranus appears less blue than Neptune since the absorption by red-absorbing methane gas is more obscured by the 'Aerosol-2' layer on Uranus. At the $H_2S$ condensation level, there is less constraint from observations on particle size, but again there seems to exist a mixture of blue- and red-absorbing photochemical products and a bright white ice component, which is assumed to be $H_2S$ ice.

The opacity of the main aerosol-2 layer at 1-3 bar to first order seems to vary little with latitude, which is surprising given the very large latitudinal variation seen in the abundance of methane both planets, varying ~4% at mid-latitudes to ~2% at polar latitudes. What causes this variation is the subject of some debate, but the lower methane abundance at polar latitudes does lead to the planets appearing slightly more reflective near the poles at longer, methane-absorbing wavelengths. This is most notable for Uranus, which because of its large obliquity has very extreme seasons. Uranus went through its northern summer equinox in 2007 and since then its northern pole has been swinging towards the Sun, and so also towards us here on Earth. A brightening of the polar latitudes has been seen during this period in the development of a polar 'hood' (Sromovsky at al., 2019). While part of this brightening is due to viewing the lower methane-abundance polar regions at lower zenith angle (e.g., Toledo et al., 2018), there is evidence that the 1 – 2-bar aerosol layer is also becoming more reflective (Sromovsky at al., 2019; James et al., 2023), increasing the visibility of this layer at polar latitudes.

Finally, dark spots and regions are a relatively common, but mysterious feature of Neptune's atmosphere, first detected in 1989 during the Voyager 2 flyby (Smith et al., 1989), which detected dark anticyclonic features such as the Great Dark Spot (GDS) and Dark Spot 2 (DS2). These features were found to be dark at wavelengths less than ~600 nm, but were not detectable at longer wavelengths. In addition, a dark belt was observed near 60°S, which has been dubbed the South Polar Wave (SPW) since its northern extent varies with longitude with a wavenumber-1 dependence. The SPW feature has a spectral dependence that is very similar to that of the discrete dark spots and Karkoschka (2011a) deduced that it was caused by a darkening of particles at pressures > 3 bar, which was later confirmed by Irwin et al. (2022), who found it to be caused by a spectrally-dependent darkening of the particles in the Aerosol-1 layer at 5 – 8 bar. The darkening mechanism for the SPW and discrete spots such as the GRS are very likely the same, but the identity of the darkening 'chromophore' is not known, and how it is introduced remains unclear. The SPW feature seems to be dynamically linked with the South Polar Feature (SPF) (Hammel et al., 1989), which are bright, short-lived upper tropospheric methane ice clouds that regularly appear on the southern boundary of the SPW at 70°S at the longitude of the SPW's maximum northern extent. Karkoschka (2011b) suggests that the SPW and SPF are both static with respect to a new internal rotation rate of 15.9663 hours, which is considerably shorter than the System III rotation rate of 16.108 hours determined from Voyager 2 radio data (Lecacheux et al., 1993). Deeper bright clouds are also occasionally seen (Irwin et al. 2011) at similar latitudes to the SPF, which may perhaps be linked.

Unlike Jupiter's Great Red Spot, dark spots in Neptune's atmosphere are not long-lived and are observed to drift towards the equator and dissipate after a couple of years. However, they appear every few years in both the northern and southern hemispheres, with the most recent example NDS-2018 (northern dark spot 2018) discovered in 2018 (Simon et al., 2019). Dark spots appear to be much less common in Uranus's atmosphere, but one was discovered in HST imaging in 2006 (Hammel et al., 2009).

### 3.5.1 Prospects for continued telescopic observations of clouds of Uranus and Neptune

The atmospheres of both planets are very cold (~50K at the tropopause and ~90 K at 1 bar) and so their thermal emission spectra are extremely difficult to view from Earth. However, filter-imaging observations from mid-infrared imagers such as VISIR at the Very Large Telescope (e.g., Roman et al., 2019, 2022) can reveal latitudinal variations in the atmospheric temperatures and circulations of both planets, but spectroscopic thermal-IR observations have been very limited in the past to the flyby observations of Voyager/IRIS (e.g., Hanel; et al., 1986), and space-based observatories: Infrared Space Observatory (e.g., Encrenaz et al., 1998), and Spitzer (e.g., Orton et al., 2014). These observations are shortly to be totally surpassed by observations with JWST with the NIRSpec and MIRI instruments that have very recently been recorded.

At shorter wavelengths, ground-based observations and observations with HST are crucial to explore ongoing seasonal variations in the distribution of gases and aerosols in the atmospheres of Uranus and Neptune. Since the Voyager 2 encounters with these planets in 1986 and 1989 respectively, we have observations covering only 44% and 21% of the orbital periods of Uranus (84 years) and Neptune (165 years). Uranus passed through its northern spring equinox in 2007 and is currently heading towards its northern summer solstice in 2030. Meanwhile, Neptune passed through its northern winter solstice in 2005. The characterisation of seasonal changes is of particular importance for Uranus. The Voyager 2 encounter with Uranus in 1986 was shortly after its southern summer solstice in 1985. At the time of the encounter, Voyager/ISS images showed very little variation of reflectivity with latitude and very few discrete clouds, with the atmosphere seeming almost quiescent. However, as Uranus approached it northern spring equinox in 2007 a broad polar 'hood' of brighter reflection about Uranus's south pole faded to a single zone of brightness at ~40S and bright dynamic storm clouds were seen at mid-latitudes by HST. During the northern spring equinox, the bright zone at ~40S faded while a corresponding zone started to form at ~40N. This zone has since expanded towards the north pole forming a new north polar `hood'. If we assume symmetry between the northern and southern hemispheres then in 2031 Uranus should look almost identical to its appearance to Voyager in 1986. Observations of the formation, evolution and development of Uranus's northern polar hood are thus a high priority for atmospheric scientists. However, while in 1986 we only had filter imaging, we now have Integral Field Unit (IFU) spectrometers, such as the VLT/MUSE instrument, where each pixel of the image is a medium spectral resolution spectrum form 473 – 933 nm, which allows the cloud structure and distribution of methane to be determined with high precision (e.g., Irwin et al. 2021). At longer wavelengths, observations with JWST/NIRSpec and VLT/ERIS (the successor to VLT/SINFONI) will allow IFU observations of the 1.5 $\mu$m region, which can be inverted to determine the latitudinal variation of $H_2S$ in both atmospheres (e.g., Irwin et al., 2018, 2019).

### 4. Solar system: Atmospheric dynamics

In this section we present a review of the dynamics of the atmospheres of the planets Venus, Mars, Jupiter, Saturn, Uranus, Neptune, and of the satellite Titan. We describe first the basics of atmospheric motions followed by the capabilities of imaging them with current and future telescopes. We then review our current knowledge of their general circulation (global wind regimes) as well as a selection of the most relevant dynamical phenomena in each of them. We comment on a list of studies that can be made of these phenomena using ground-based and space-based telescopes.

## 4.1 Atmospheric dynamics: Fundamentals

Motions in planetary atmospheres are described by the laws of fluid dynamics on a rotating body with spherical (spheroidal) geometry and boundary conditions imposed by the nature of the planet. In all cases, the upper atmosphere is an open boundary to outer space (the exosphere), and the lower boundary is either a rigid surface (terrestrial planets, dwarf planets and satellites) or a deep lower boundary with a poorly defined transition in the case of giant fluid and icy planets (Sánchez-Lavega, 2011). The global motions are driven by the solar radiation and in the case of the giant planets by an additional internal energy source. Here we describe the dynamics of the neutral atmosphere, a compressible fluid that obeys the perfect gas law. We consider only bodies with substantial atmospheres from the lightest of Mars to the massive and deepest of Jupiter (for exoplanets see section 5). The following equations describe the atmospheric motions (see, e.g., J. R. Holton (2004), Vallis (2006)). For a flow element with velocity $\vec{u}$ and density $\rho$, the continuity equation is written

$$\frac{d\rho}{dt} + \rho\nabla \bullet \vec{u} = 0 \tag{15}$$

and for an ideal gas

$$P = \rho R_g^* T \tag{16}$$

The momentum equation in the rotating frame of the planet is

$$\frac{d\vec{u}}{dt} = -\frac{\nabla P}{\rho} - g\hat{k} - 2\vec{\Omega}\times\vec{u} - \vec{\Omega}\times(\vec{\Omega}\times\vec{r}) + \vec{F}_f \tag{17}$$

where $\vec{F}_f$ represents the frictional force, $-2\vec{\Omega}\times\vec{u}$ is the Coriolis force and $-\vec{\Omega}\times(\vec{\Omega}\times\vec{r})$ is the centrifugal force and $\hat{k}$ is the unit vector perpendicular to the body surface. The thermodynamic energy equation (conservation of energy) is written as

$$\rho C_p \frac{dT}{dt} - \frac{dP}{dt} = \nabla(K_T\nabla T) - \nabla\vec{F}_{rad} + Q \tag{18}$$

Here, $C_p$ is the specific heat, $K_T$ the thermal conductivity, $\nabla \vec{F}_{rad}$ the divergence of heat flux by radiation, and $Q$ includes internal energy sources (diabatic, latent and frictional heating). Each term of (18) has units of power per unit volume (Wm$^{-3}$). Note that ($d/dt$) in the left hand side represents the total derivatives.

Vertical coupling in the atmosphere takes place due to energy, momentum and mass transport caused by the large-scale circulation and atmospheric wave propagation and dissipation and by disturbances and weather-related processes. In the absence of hazes, clouds and dust their signature is detected in the atmospheric composition and in the temperature, pressure and wind velocity fields.

## 4.2 Dynamical phenomena in planetary atmospheres

We present case studies of a selection of representative dynamical phenomena in planetary atmospheres traced by clouds (aerosols in general, to include dust) as observed in optical and near infrared images from ground- and space-based telescopes.

### 4.2.1 Venus

The main dynamical elements of Venus's atmospheric dynamics are the global solar-antisolar circulation in the high atmosphere, the zonal superrotation and the meridional Hadley cell at cloud level, the polar vortices and a variety of waves and eddies at mid- and equatorial latitudes. Venus is a slow rotator and in equation (17) the dominant terms in the force balance are the pressure gradient and the centrifugal force resulting in a cyclostrophic balance regime (Sánchez-Lavega, 2011)

$$\frac{u^2 \tan \varphi}{R_p} = -\frac{1}{\rho} \frac{\partial P}{\partial y} \tag{19}$$

being $R_p$ the planet radius. Combining (15-18) gives a thermal wind equation for this balance

$$2u \tan \varphi \frac{\partial u}{\partial z} = -\frac{R_g^* R_p}{H} \frac{\partial T}{\partial y} \tag{20}$$

The vertical integration of the left hand-side allows the wind velocity to be calculated from measurements of the meridional temperature gradient $\partial T/\partial y$. Venus' atmosphere in under a super-rotation state since the rotation period of the atmosphere is about 4 days compared to 243 days for the planet (Archinald et al. 2018). From $\sim 45°$N to $45°$S the zonal velocity is nearly constant at 65 km altitude with $u \sim 100$ ms$^{-1}$ and at 50-55 km altitude with $u \sim 75$ ms$^{-1}$, decreasing poleward at both heights linearly with latitude (Sánchez-Lavega et al., 2017) (Figure 5a). These are the polar areas dominated by the presence of large, complex and unstable polar vortices (Garate-Lopez et al., 2013). In the vertical, the horizontal velocities increase from 0 ms$^{-1}$ at surface to 100 ms$^{-1}$ at 65 km (Figure 5b). The scatter between the wind profiles in Fig. 5a gives an idea of the stability and temporal variability of the super-rotation but including the possible effects of thermal tides and wave motions, and small altitude differences in the cloud tracers. The long-term variability in the zonal and meridional wind profiles in the day side (upper clouds at 70 km) can also be studied by means of Doppler spectroscopy (Machado et al., 2017). Continuous and accurate measurements of wind speeds are required to constrain the wind variability and the nature of the changes, particularly close to the polar regions.

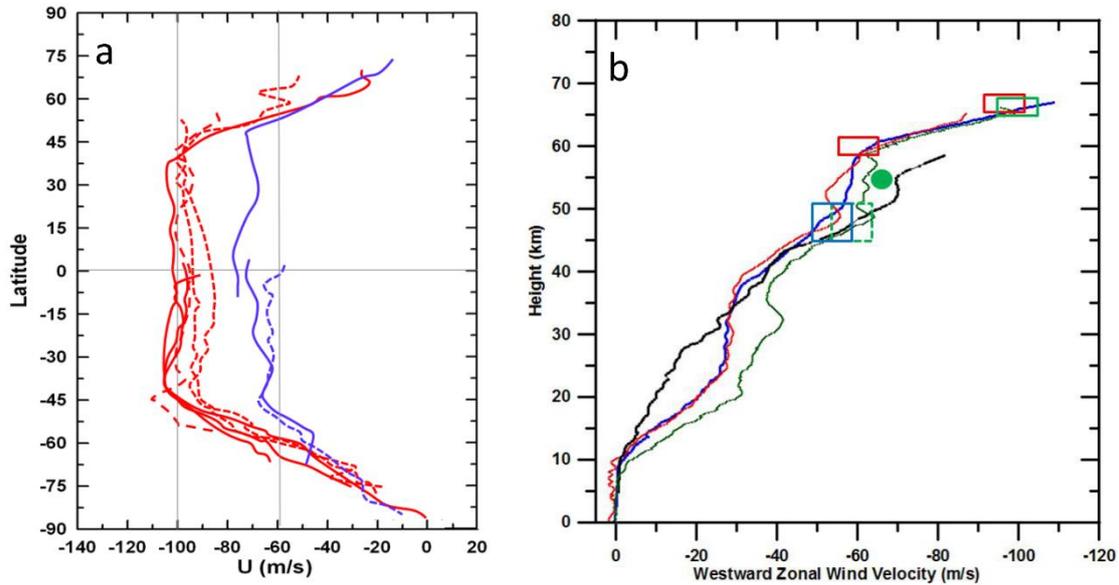

**Figure 5**. *Venus winds. (a) Meridional profiles of the zonal wind velocity measured from cloud tracking at UV wavelengths (red lines, altitude 65–70 km) and at 980 nm (blue lines, altitude ∼60 km). UV data from the following missions: Mariner 10 (1974), Pioneer-Venus (1980-1982), Galileo (1990), Venus Express VIRTIS (2006–2012), Venus Express VMC (2006–2012). Wavelength 980 nm data from: Galileo (1990), Venus Express VIRTIS and VMC (2006-2012). (b) Vertical profiles of the zonal winds measured by the Pioneer-Venus probes (9 Dec., 1978) called: Sounder (green), North (black), Night (red), Day (blue). The green dot is from the Vega 1 and 2 balloons (11–15 June, 1985) at latitudes (+7.3∘, −6.6∘). The rectangles are averaged velocities from Venus Express VIRTIS daytime (red) and night-time (blue).*

Different kinds of dynamical features develop in the cloud layers from ∼ 40 km to 70 km altitude that can be observed depending of the selected spectral wavelength range (Fig. 6) (Peralta et al. 2017). On the day-side, images obtained with small telescopes using filters for the UV-Violet (wavelength $\lambda$ = 380-420 nm) and red (> 650 nm, in particular at 900 nm), allow us to follow changes in the winds and in the planetary-wave activity taking place in the upper cloud layer (65-70 km) and in the middle cloud (55-60 km), respectively (Sánchez-Lavega et al., 2016; Peralta et al., 2015, 2023). Another objective is the study of the stability and long-term behaviour of the north-south global wave observed at 10 $\mu$m located at an altitude of 65 km and its coupling with the equatorial topography (Fukuhara et al., 2017). The nature of the UV-absorbing aerosol that is mixed with Venus's upper cloud remains a major mystery (Pérez-Hoyos et al., 2018; Lee et al., 2021). Ground-based studies of the long-term photometric variability of Venus' integrated disk brightness in the UV, coupled to the observed cloud morphology (waves, convection) and wind velocities can shed light on its origin (Lee et al., 2022).

Night-side images at infrared wavelengths ($\lambda$ = 1.74 µm, 2.25-2.5 µm) show the cloud opacity to thermal radiation and give information of dynamical phenomena (waves, winds) in the altitude range 44-48 km (Machado et al., 2022). In addition, observations at 5 µm depict slow motions of waves on the night side. At longer infrared wavelengths ($\lambda$ = 8 – 12 µm) thermal images allow the simultaneous study of the day and night sides of the upper cloud from the colder polar areas to the equatorial dynamics (Sato et al., 2014). Coordinated observations from the UV to the infrared using ground-based facilities can be used to search for correlations between features in the thermal opacity maps and in the visual images to address the dynamical coupling between layers in the 45-70 km altitude (Titov et al., 2018), and search for vertical transport associated with winds, convection and waves. Venus exhibits a variety of waves from the planetary-scale with periods between 3.9 and 5.3 days (Lee et al. 2020) combined with those stationary relative to orographic forcing (Sánchez-Lavega et al., 2017; Peralta et al., 2020). Their vertical extension and coupling to the super-rotation is another subject to be explored with ground-based telescopes.

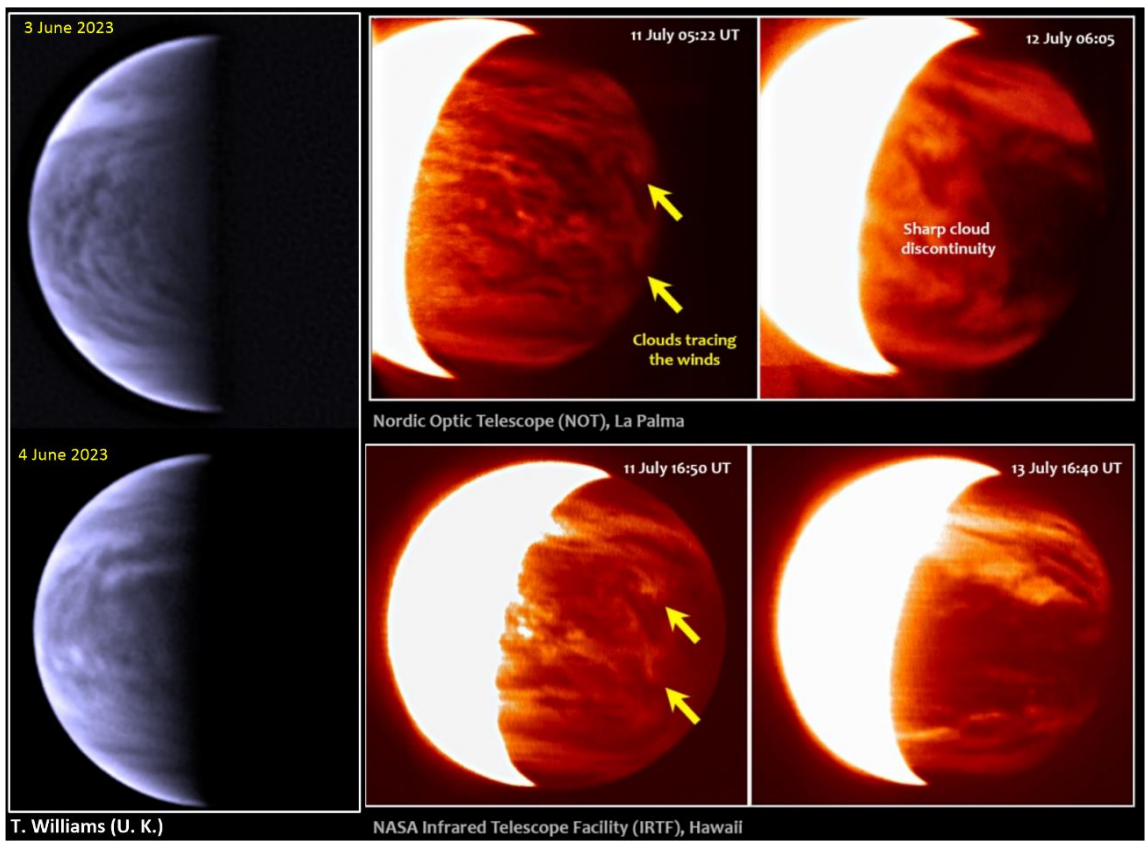

**Figure 6.** *Ground-based images of Venus showing the clouds at two heights in day and night time. Left panel: images taken by Tom Williams (Wiltshire, U.K.) in UV (360 nm) with a 40.6 cm diameter telescope using the "lucky-imaging" method on 3 and 4 June 2023. Clouds are at ∼65-70 km altitude on the day side. Right panel: Images taken at a wavelength of 2.2 µm of the night-side of Venus obtained with the NOT telescope (La Palma, Canary I.) and with the IRTF (Hawaii) on the same date, but separated by ∼10 hr.*

*They show two different views of Venus clouds at ~ 45-50 km altitude produced by opacity to thermal radiation from below. Credit: NOT/NASA, IRTF.*
https://www.not.iac.es/news/www/news/venus-daytime-notcam.html

Other observations from the ground relevant to dynamical phenomena are (Gray et al., 2022): (a) images at $\lambda$ = 1.27 $\mu$m of the oxygen airglow emission with peak intensity at ~ 95 km height (90-110 km) to show maps of winds and waves (changing patterns allow us to derive motions); (b) Survey for possible optical "flashes" with durations of ~ 1-2 s produced by bolides (never seen on Venus); (c) Survey for lightning events produced by ultrashort OI emissions at $\lambda$ = 557.7 nm (Takahashi et al., 2023).

### 4.2.2 Mars

Like Earth, Mars is a rapid rotating planet and in the momentum equation (17) the dominant terms in the force balance are the Coriolis and pressure gradient forces and the motions are to first order in geostrophic balance (Sánchez-Lavega, 2011)

$$\frac{\partial u}{\partial z} = -\frac{g}{f\,T}\frac{\partial T}{\partial y}$$
$$\frac{\partial v}{\partial z} = \frac{g}{f\,T}\frac{\partial T}{\partial x}$$

(21)

The zonal (*u*) and meridional (*v*) velocities follow the thermal wind equation

$$\frac{\partial \vec{V}_g}{\partial z} = \frac{g}{fT}\hat{k}\times\nabla_h T$$

(22)

Its vertical integration allows wind velocity to be calculated from measurements of the horizontal temperature gradients (Mitchell et al. 2019; Barnes et al. 2017). Fig. 7 shows the seasonally changing vertical and meridional structure of Martian zonal jets.

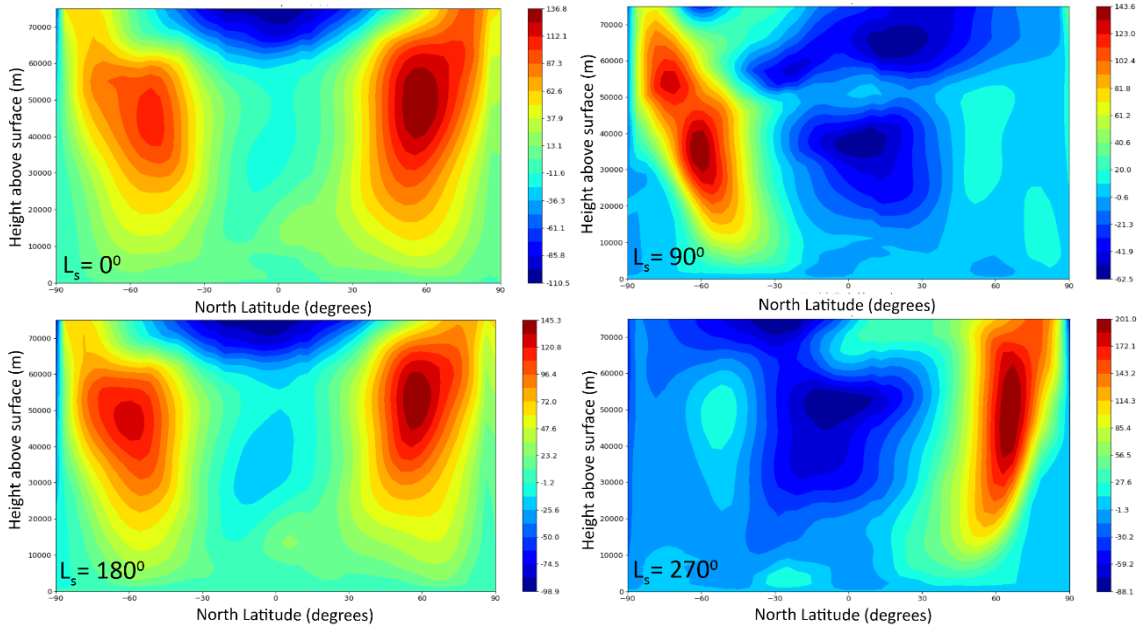

**Figure 7.** *Mars zonal winds u from the Mars Climate Database. Color diagram of the velocity in ms$^{-1}$ (note the change in the scale for each panel) in a plot of latitude (90 °N to 90 °S) and height (0 to 75 km) at local time 12 hr and areographic longitude 0 ° for the four seasons (solar longitude $L_s$ = 0 °, 90 °, 180 °, 270 °). MCD v6.1 climatology average solar scenario. Credits: LMD/OU/IAA/ESA/CNES.*

Despite the large number of space missions orbiting Mars and operating on its surface, observations with telescopes can still provide important information on different dynamic phenomena on the planet. Complementarily, these observations can provide the context and the framework to interpret the studies from orbiting spacecraft and surface missions. A very important area of research using ground-based telescopes is high-resolution infrared spectroscopy ($\lambda \sim$ 1-12 μm). Since the Martian atmosphere is very tenuous, the absorption lines from molecular species are very narrow and a high spectral resolving power is required over a large spectral interval for different types of studies (Encrenaz, 2009). Ground-based telescopes provide these qualities allowing sampling of the full observable Mars disk, mapping the compounds' abundance and dust optical depths, and their relationship to dynamics. A global picture of the distribution of trace species (e.g., $H_2O$, HDO, $H_2O_2$, CO, $CH_4$, HCl and others; Encrenaz et al., 2004) can then be correlated with the chemistry and atmospheric dynamics and their seasonal and inter-annual variability (Daerden et al., 2019). Examples of global dynamical processes affecting the chemistry, compound abundances and their distribution on the large-scale, are the meridional motions related to the seasonally changing Hadley-cell, the evolution of the Aphelion Cloud Belt (from solar longitudes $L_s \sim 0° - 180°$), the dusty period ($L_s \sim 180° - 360°$) and the occasional development of Global Dust Storms GDS (Guzewhich et al., 2020; Khare et al., 2007). The method also allows for short-term phenomena and diurnal processes to be investigated just by looking at systematic differences across the East-West direction. The first infrared spectra and images from the JWST so far released clearly show the potential that this telescope can provide in the future in these studies.

Images obtained with telescopes can complement very well those of spacecraft at high resolution as during the rapid onset, development and evolution of GDS (Sánchez-Lavega et al., 2019). Tracking the motion of the dust storm edges allows us to retrieve the wind speeds and directions. Other synoptic and planetary-scale phenomena on Mars, such as cloud bands from the Aphelion Cloud Belt, the North and South Polar Hoods and regional-scale dust storms, are additional targets in particular when Mars is close to opposition with the planet reaching its maximum size. HST images of Mars can go a step further due to the good resolution and can address the study of particular phenomena at synoptic-scale (> 1000-2000 km), for example the double-annular cyclone (Cantor et al., 2002; Barnes et al. 2017; Sánchez-Lavega et al., 2018) (Fig. 8). Large ground-based telescopes can study, at high spatial and spectral resolution, transient and smaller-scale atmospheric phenomena (> 250-500 km). Another territory for telescope survey in the optical range is the presence and characterization of high-altitude phenomena detaching at the limb and terminator (Sánchez-Lavega et al., 2015; Lilensten et al, 2022).

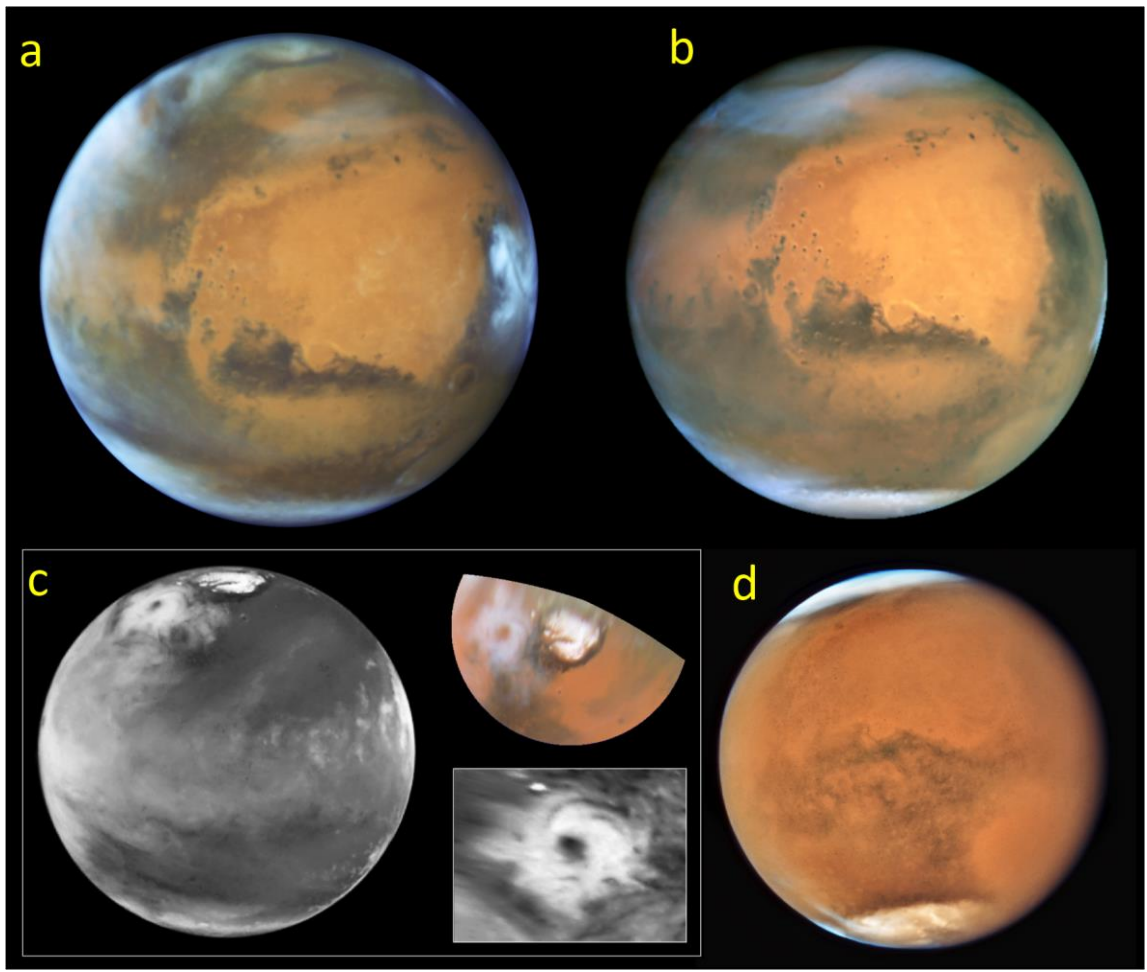

**Figure 8.** *Mars atmospheric features as observed by the Hubble Space Telescope. (a) Water ice clouds at morning and evening (12 May 2016, $L_s$ = 151°); (b) Polar water ice clouds (26 June 2001, $L_s$= 184°); (c) Annular-double cyclone with polar projection (27 April 1999, $L_s$ = 130°); (d) Great Dust Storm GDS2018 (26 July 2018, $L_s$ = 218°). Credits: NASA, ESA, and STScI.*

### 4.2.3 Jupiter

Jupiter is a rapidly rotating planet and the atmospheric motions are outside the equator in geostrophic balance (eqs. 21-22). Jupiter's wind velocity profile at the upper cloud level has been precisely measured since the Voyagers 1 and 2 flyby in 1979 showing few changes in time (Fig. 9a). The atmosphere of Jupiter and its vigorous dynamics at upper cloud levels (~1-5 bar, Fig. 3) are one of the clear research targets for telescopic observations. On the long-term one objective is the study of the intrinsic variability of the zonal wind profile u(y) at cloud level (Tollefson et al., 2017), its vorticity (du/dy) and vorticity gradient (d²u/dy²), and its relation to the number, size and properties of cyclones, anticyclones, waves and storms. This will allow to search for momentum and energy transfers between these disturbances and the jet profile (Ingersoll et al., 2004).

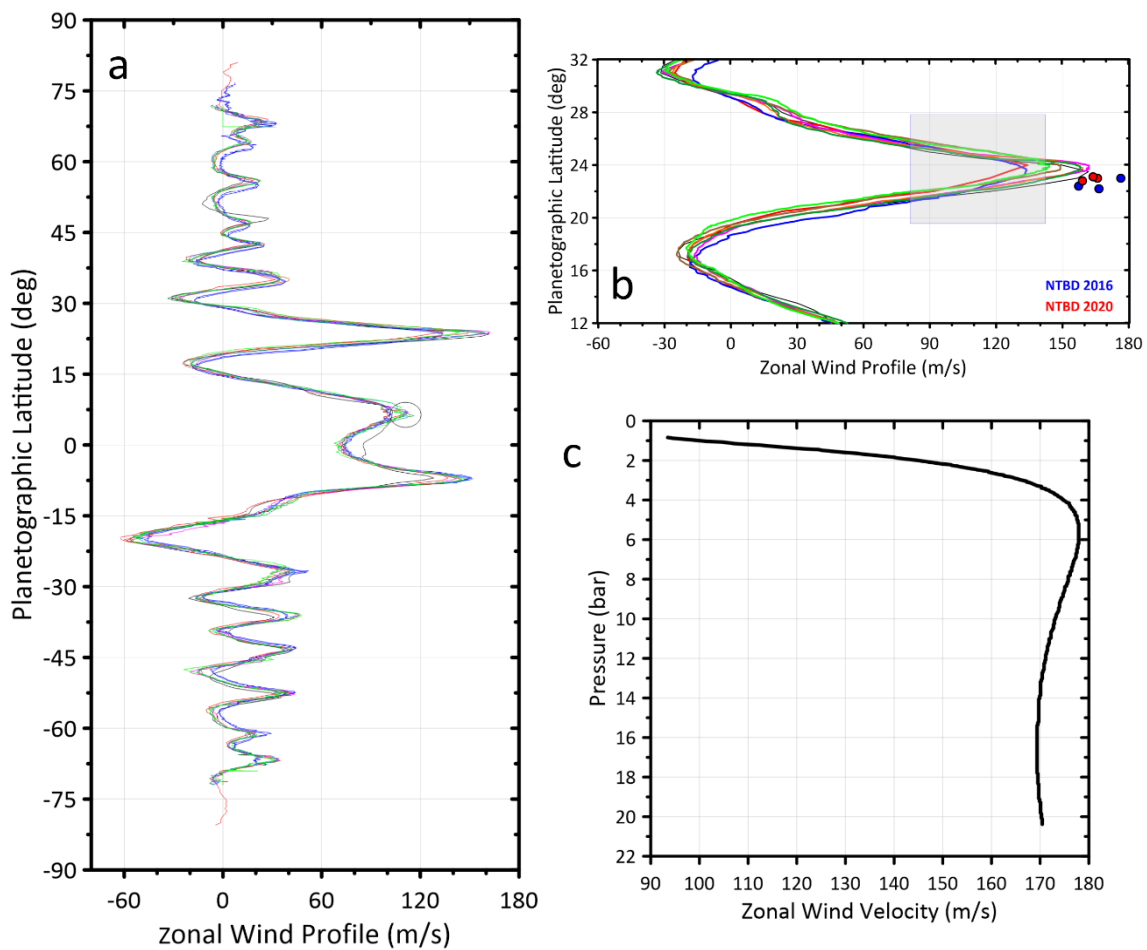

**Figure 9.** *Jupiter's winds. (a) Meridional wind profile from cloud tracking with cameras onboard different spacecraft and years: Voyager 1-2 in 1979 (black, Limaye 1986); Cassini in 2000 (red, Porco et al. 2004); HST in 1995-2000 (blue, García-Melendo and Sánchez-Lavega, 2001); HST in 2009 (magenta), 2012 (brown), 2015 (orange), 2016 (purple and green), all these from Tollefson et al. 2017. The circle marks the location of the Galileo probe in 1995. (b) Meridional wind profile in time of the NTB jet from 1979 to 2016 (same color code and references as in (a)). The blue dots are the velocity of the*

*three plumes of the NTBD in 2016 (Sánchez-Lavega et al., 2017) and the red dots are for the NTBD plumes in 2020. The shaded area is where the winds are largely perturbed by the disturbance. (c) Vertical profile of the zonal wind at $7°$ as measured by the Galileo probe (Atkinson et al. 1998).*

A second objective is to understand the physics and cyclic nature of the two major Jupiter's planetary-scale disturbances with a medium-term duration ($\sim$ month): the South Equatorial Belt Disturbance (SEBD) (Rogers, 1996; Sánchez-Lavega and Gómez, 1996; Fletcher, 2017; Fletcher et al., 2017) and the North Temperate Belt Disturbance (NTBD) (Sánchez-Lavega et al, 2011, 2017). These unpredictable disturbances lead to a global change from a "zone" (high albedo band) to a "belt" (low albedo band) (Pérez-Hoyos et al. 2012, 2020). They start with the outbreak of bright spots or storms produced by moist water convection at 16°S for the SEBD, where the zonal wind velocity is u = 0 ms$^{-1}$, and at 23°N for the NTBD in a peak of the intense eastward jet where u = 165 ms$^{-1}$ (Fig. 9b). The SEB eruptions are considered major when they are preceded by the fade of the belt that becomes for a short time ($\sim$ 0.5-1.5 years) a "white" band with a zone-like aspect (Pérez-Hoyos et al., 2012; Fletcher et al., 2011). The outbreak generates a complex pattern of waves and turbulence that propagates according to the wind profile u(y) until they encounter the initial source and mix to form a new low albedo belt (Fig. 10).

The SEBD source is initially a bright compact spot of fresh ice that grows rapidly in size by divergence at top clouds of the convective updrafts being sheared apart in few days (Hueso and Sánchez-Lavega, 2001). The NTBD has multiple sources (2-3 sequential outbreaks separated by a few days) each with a plume-like shape that last for longer periods (Fig. 10). This means that the convective sources are sustained in time, probably by flow convergence at the water cloud level at about 5 bar. Observations of the details of this phenomenology at high-resolution and at multiple wavelengths with high-spectral resolution will allow us to look for the involved chemical compounds in the storm source and in particular for the presence of ammonia-ice and water-ice particles at the top of the bright clouds.

Fig. 11 shows a calendar of the historically recorded outbreaks for both disturbances that gives an idea of their cyclic behavior (Sánchez-Lavega 1989; Sánchez-Lavega and Gómez, 1986). The temporal interval between consecutive onsets of the SEBD major episodes (i.e., those preceded by a "fade" of the belt) is 6.1 ± 3.6 years (with extremes 1.8 – 15 years) and for the NTBD is 10.7 ± 9.4 years (with extremes 3.7 – 34.6 years). This points toward a thermodynamic energy cycle that involves water vapor distribution (Li and Ingersoll, 2015). Moist water air could be transported at depth to the two latitudes where convergence and storm trigger occurs (16°S, 23°N).  There, air is forced to ascend by buoyancy, condensing, and forming the visible storm (Ingersoll et al. 2000; Hueso and Sánchez-Lavega 2001; Palotai et al., 2023).

To better understand the thermodynamic behaviour of water in Jupiter and its probable relation with the jet system, we need to complete the statistics and properties of convective storms that develop in $\sim$ hours to few days at scales of $\sim$ 500 – 2000 km. Storms of this type are observed regularly, for example in the North Equatorial Belt (latitude 12°N) (Wong et al., 2023) and in cyclonic systems at mid-latitudes (Iñurrigarro et al., 2020; Hueso et al. 2022), among other places. We need to monitor in the long-term the number and size of these storms as a function of latitude (jets) and in relation to vortex activity, as well as their recurrence time-scale. These and the SEBD and NTBD phenomena are the key to understanding moist convection processes in hydrogen-dominated atmospheres and how heat and momentum exchanges with surroundings occur. Medium-size telescopes can be used for such a monitoring task.

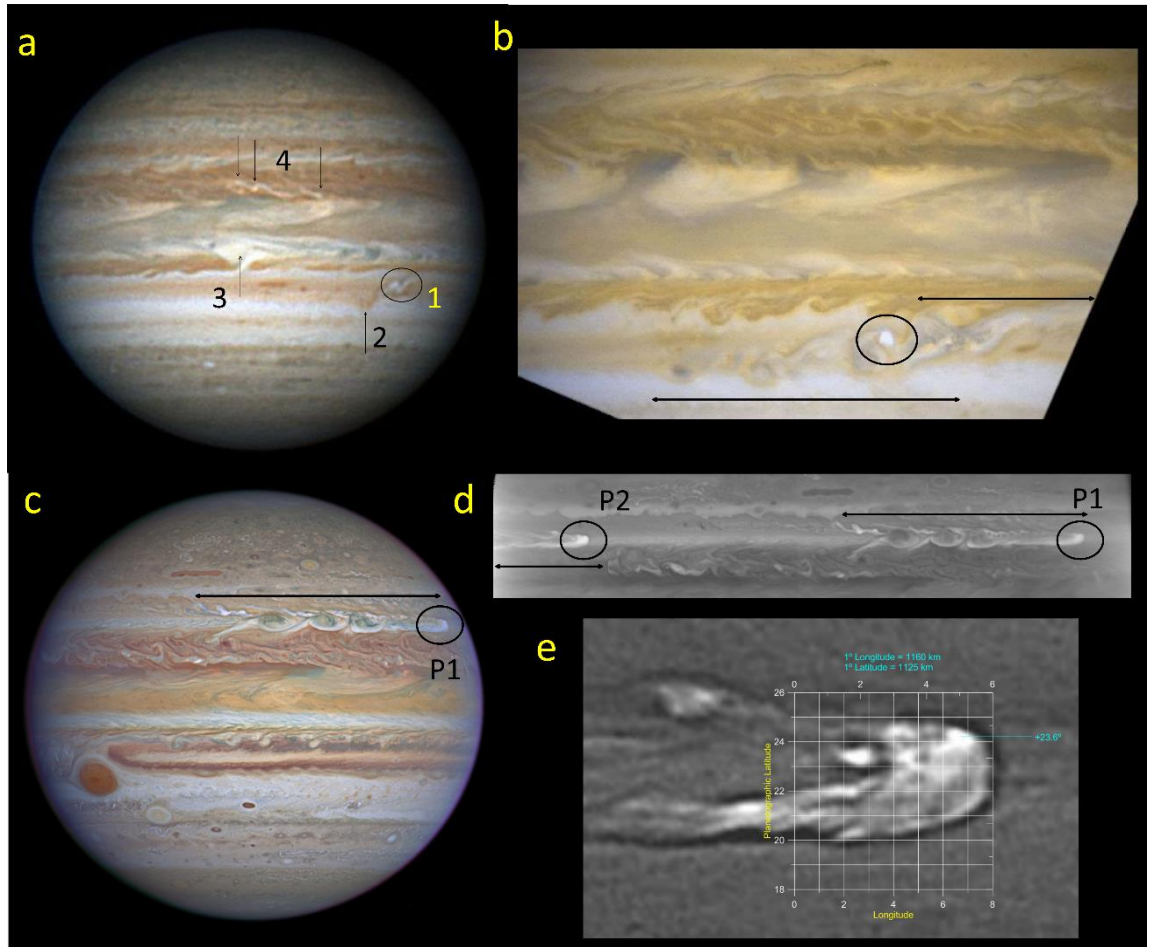

**Figure 10.** *Convective major storms in Jupiter. (a-b) Outbreak of different bright spots of a South Equatorial Belt Disturbance SEBD (1-circle, latitude -16.5°S) and subsequent generation of cyclone series and turbulent patterns along the lines (b). Other features shown in (a) are: a column from a South Tropical Disturbance (STrD) (2-arrow, latitude -22.5°S); an unusual plume in the South Equatorial Zone (3-arrow, latitude -6.9°S); and convective storms in the North Equatorial Belt (4-arrows, latitudes 9°N to 13°N). Image data: (a) 20 May 2007 (Christopher Go); (b) 5 June 2007 (A. Sánchez-Lavega and A. Simon, Hubble Space Telescope). (c-d) The North Temperate Belt Disturbance (NTBD) on 20 September 2020 from Hubble Space Telescope (OPAL program). (c) The circle marks plume P1 and the line the disturbance extent following this plume, (d) Plumes P1 and P2 and their zonal disturbances at a wavelength of 467 nm, (e) details of the cloud structure of plume P1 in a methane band filter at 889 nm.*

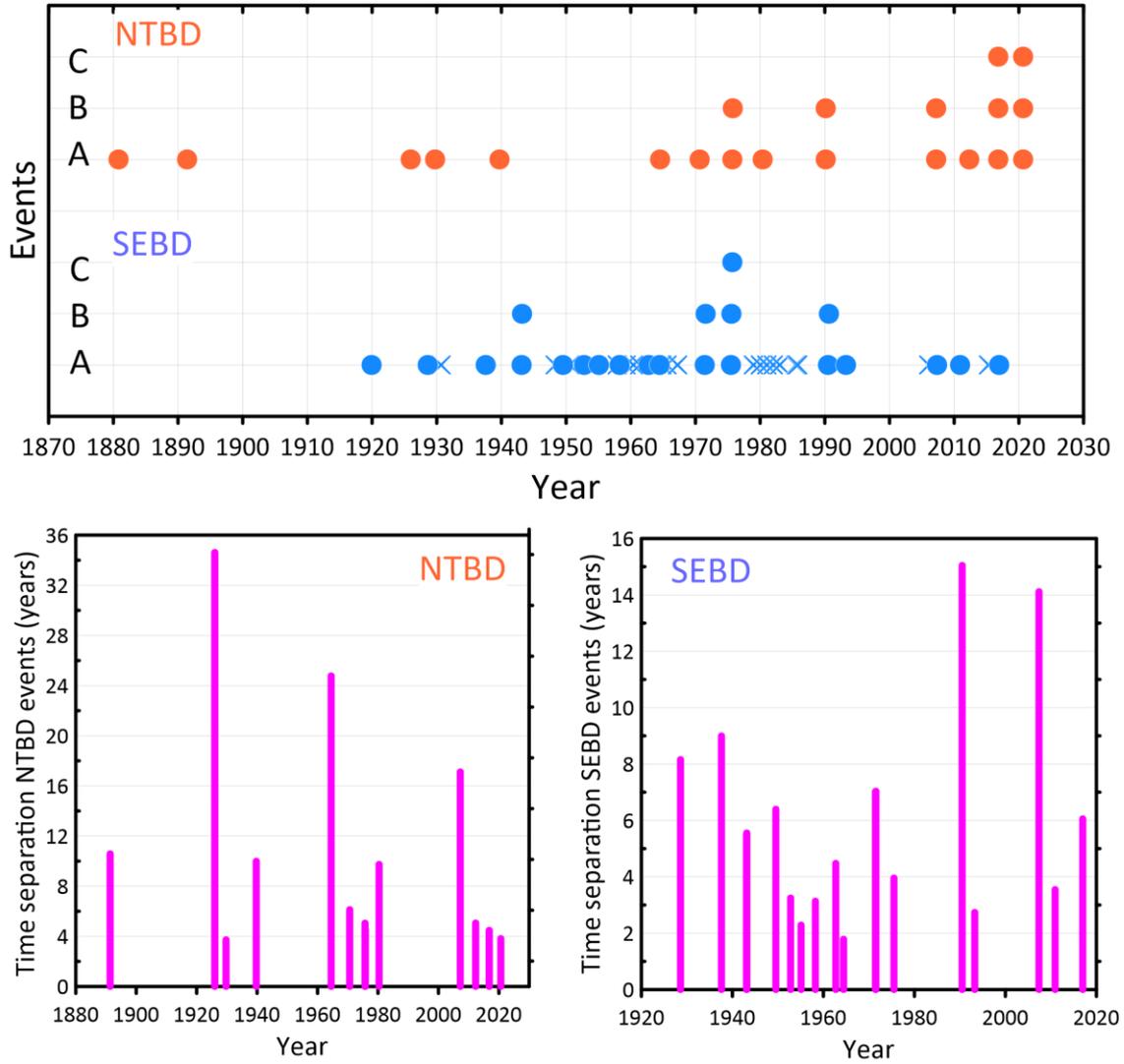

**Figure 11.** *Upper plot: temporal distribution of the historically reported North Temperate Belt (NTBD, orange) and South Equatorial Belt Disturbances (SEBD, blue). Multiple events are denoted as A, B, C. Lower plots: time separation between NTBD (left) and SEBD (right) events.*

The third objective are at the two major Jovian anticyclones, the Great Red Spot (GRS) and oval BA. The current GRS is an ellipse with a length (east-west) axis 13,900 km and width (north-south) axis 9,700 km, and including the surrounding circulation area known as the Hollow of 20,700 x 17,400 km (Sánchez-Lavega et al., 2022). Its maximum tangential velocity is ~ 120 - 130 ms$^{-1}$ and peak vorticity $<\xi> = Circulation/Area \sim$ -10$^{-4}$ s$^{-1}$ (about 7 times the ambient wind shear), is concentrated in a ring interior to the red oval (Wong et al., 2021). Its dynamical nature, temporal evolution in terms of vortex stability against interactions with other features, size evolution, the 90-day oscillation for the GRS and colour variability, and other properties are not yet understood (Hueso et al., 2009; Fletcher et al., 2010; Sánchez-Lavega et al., 2018; Simon et al., 2018; Sánchez-Lavega et al., 2022). Combined multi-wavelength simultaneous high-resolution images and spectra of the two anticyclones are necessary to explore these issues over a wide range of altitudes.

The Equatorial Zone (EZ) that extends meridionally between the latitudes of the two eastward jets at $\sim$ 7°N and 7°S exhibits haze variability and changes in the deep cloud structure whose coupling to dynamics are not understood. Recently, a narrow jet at the center of the equator and above the clouds, in the upper hazes, has been detected with JWST (Hueso et al. 2023). This means an intense vertical wind shear at the equator with winds increasing with altitude up to stratospheric levels (Cavalié et al. 2021, Benmahi et al. 2021). In addition, the dynamics of the EZ at tropospheric level is complex due to the presence of a variety of waves (Ingersoll et al., 2004) and their possible coupling to the stratospheric oscillation, also known as the Jupiter Equatorial Oscillation (JEO) (Antuñano et al. 2021, Orton et al., 2023). The hot spots and plume activity at 7°N (Ingersoll et al., 2004) and the large-scale isolated disturbances at 6°S (Sánchez-Lavega and Rodrigo, 1985), add more complexity to the equatorial dynamics. A goal of future research would be to understand and disentangle the relationship and possible coupling between all these elements of the rich equatorial dynamics of Jupiter.

Ground-based telescopes are essential to monitor possible asteroid and comet impacts and their debris in Jupiter's atmosphere (Harrington et al., 2004; Hammel et al., 1995, 2010, Sánchez-Lavega et al., 2010, 2011). Small telescopes in the range 20-50 cm aperture have been successfully used to capture bolide impacts of objects less than $\sim$ 15 m in size (detected as short light flashes $\sim$ 1 sec duration, Hueso et al., 2010, 2013 and 2018. In order to improve the statistics of fireball impacts, their nature and origin, the surveillance of Jupiter with small telescopes and the use of fast cameras operating in the "lucky imaging" mode, represent the most efficient method. In the event of a major impact, high-resolution images are required to follow the evolution of the debris cloud and its dispersion by the winds to track their velocity in altitude. High-resolution spectra at optical and infrared wavelengths are needed to determine the chemical composition and properties of the aerosol debris (Harrington et al., 2004; Fletcher et al. 2011; Orton et al. 2011).

## 4.2.4 Saturn

Saturn is a rapidly rotating planet and the atmospheric motions outside the equator are in geostrophic balance (eqs. 21-22). As commented in section 3.2, the rotation period of the planet is not precisely determined and the range of measured values is shown in Fig. 12a. As a reference system for the wind profile we use the System III as proposed by the International Astronomical Union (IAU). Fig. 12b shows the measurements of the zonal wind profile based on cloud tracking in 1980-81 during Voyager 1 and 2 flybys and from Cassini orbit in 2004-2009 (Sánchez-Lavega et al., 2000; García-Melendo et al., 2011). We also show how the wind profile changes with the different rotation periods proposed. The lowest rotation period makes the profile more symmetric.

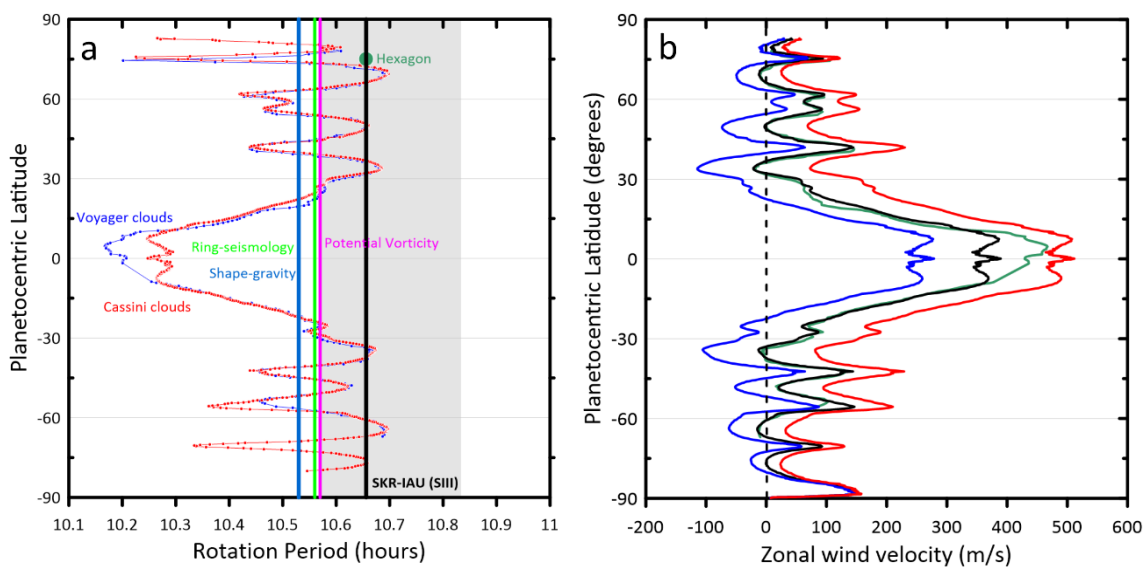

**Figure 12.** *Saturn's rotation periods and wind velocity profiles. (a) Rotation periods. Cloud tracking from Voyager 1 and 2 images (blue line; Sánchez-Lavega et al., 2000) and Cassini images (red line; García-Melendo et al., 2011) up to latitude 80°. Close to the pole, the rotation period of the polar vortex at ∼ 89° (north and south) reaches a minimum of about 6.15 hours. Saturn Kilometric Radiation (SKR) is used to define the System III reference system (black line, from radio periodicities; Desh and Kaiser 1981). The grey shadowed area is the range of temporal and hemispheric variability of the SKR periodicity from 1980 to 2016 (Carbary et al., 2018). The Hexagon wave rotation period is the mean from 1980 to 2022 (Sánchez-Lavega et al., 2021). The rotation period from the zonally averaged potential vorticity is the magenta line (Read et al., 2009). The rotation period from shape and gravity of the planet is the blue soft line (Anderson and Schubert, 2007; Helled et al., 2009). The rotation period from ring seismology is the green line (Mankovich et al., 2019; Millitzer and Hubbard, 2023). (b) Zonal wind profiles for different Saturn's rotation periods. SKR-IAU System III (10 h 39m 22.4 s) for Voyager 1 and 2 (green; Sánchez-Lavega et al., 2000) and Cassini ISS (black; García-Melendo et al., 2011). Short rotation period (10 h 32 m, blue) and long-rotation period (10 h 47 m 06 s, red) for Cassini ISS winds.*

Figure 13 shows the profiles in detail and the latitudes where mid-scale and major convective storms known as the Great White Spot (GWS) phenomenon, occur (Sánchez-Lavega et al., 2018). Particularly interesting is the broad equatorial jet between latitudes ∼ 20°S to 20°N where high temporal and spatial variability has been observed. This includes the presence of a narrow, intense central jet high in the atmosphere as traced by features detected in images with a methane band absorption filter at a wavelength of 890 nm (García-Melendo et al., 2010, 2011; Sánchez-Lavega et al., 2016; Hueso et al., 2021). The jet structure is probably related to Saturn's Stratospheric Oscillation SAO (Guerlet et al., 2011; Fletcher et al., 2017) but this needs to be confirmed. It is fundamental to disentangle the reason of the temporal differences, if due to vertical wind shear and cloud tracers detection at different altitudes, or if due to true temporal changes in the zonal flow (seasonal or not). Imaging in the UV and in the different methane absorption bands at red and infrared wavelengths can be used to look for these changes at different altitudes above the upper cloud level (West et al., 2009). At 5 μm, the lower cloud at P ~ 2-4 bar is detected from the opacity it produces to thermal radiation and the wind profile looks like similar to Voyager 1-2 (Figure 13, Studwell et al., 2018). This suggests that Voyagers' equatorial profile was deeper than the Cassini one. Higher in the stratosphere, the winds are retrieved from mm-Doppler interferometry (Benmahi et al. 2022) and the long-term imaging at infrared wavelengths will serve to track seasonal changes in the temperatures and cloud opacity. Application of the thermal wind equation (3.8) allows us to retrieve the winds and the SAO and its temporal changes (Guerlet et al., 2011; Fletcher et al., 2017).

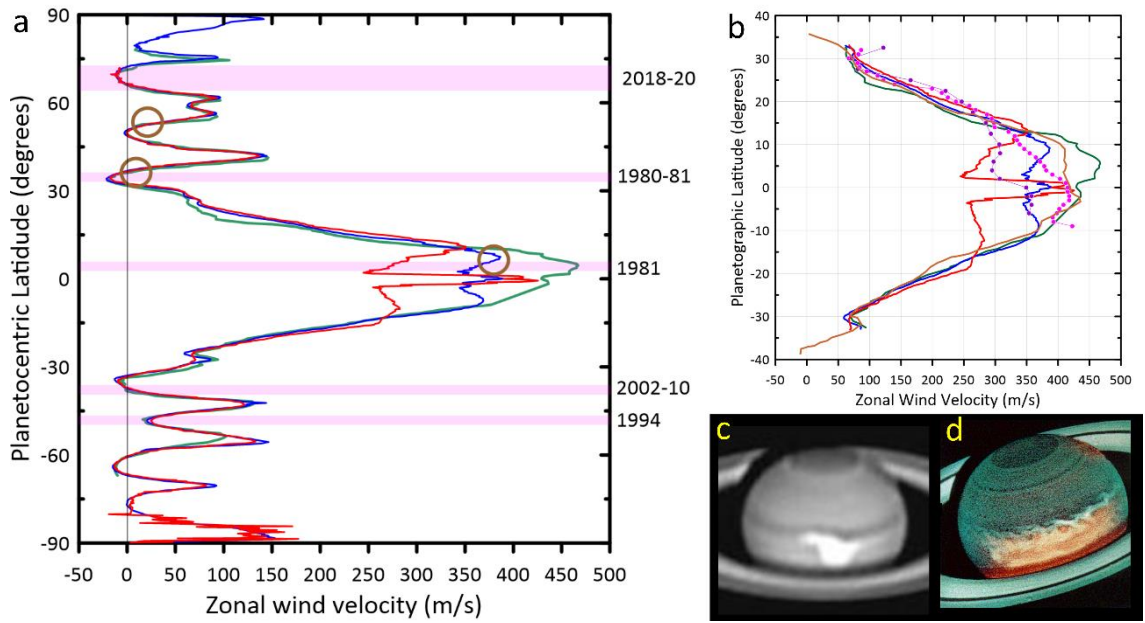

**Figure 13**. *(a) Saturn's zonal wind profile in different epochs and heights in the SKR-IAU System III (10 h 39m 22.4 s). Green from Voyager 1 and 2 (Sánchez-Lavega et al., 2000), blue is the lower cloud profile and red the upper cloud profile both for Cassini (García-Melendo et al., 2011). The magenta bands show the latitudes of mid-scale convective activity observed between 1980 and 2020. The brown disks are the location of major storms (Great White Spots, GWS). (b) The Equatorial wind profile at different epochs and heights. The color code is as in (a) and includes a deep wind profile (at the level of the NH₄SH cloud, brown code) from VIMS-Cassini profile (Studwell et al., 2018). It also shows the profile changes due to the activity of the GWS in 1990: dot-magenta profile is at the lower cloud and dot-purple at the upper cloud (Barnet et al., 1992). (c) Image of the GWS 1990 on 2 October 1990 from Pic-du-Midi Observatory (Sánchez-Lavega et al., 1991) and (d) in 20 November 1990 from HST (Barnet et al., 1992).*

The GWS is a rare, exceptional and energetic planetary-scale disturbance that has been reported in the northern hemisphere in the following years: 1876 (2°N), 1903 (36°N), 1933 (5°N), 1960 (58°N), 1990 (2°N, activity extending until 1995, Sánchez-Lavega et al., 1996), 2010 (36°N) (Fisher et al. 2011; Sánchez-Lavega et al. 2011, 2018; García-Melendo et al. 2013; Sayanagi et al. 2013). The outbreak of the storm is due to moist convection (Sánchez-Lavega and Battaner 1987; Li and Ingersoll 2015) and triggers the zonal propagation of a complex disturbance that fully encircles the planet propagating into the stratosphere in 2010 (Fletcher et al. 2011). These immense perturbations leave their imprint over time on the distribution of ammonia in the upper troposphere (1 - 20 bar), below the visible clouds, and thus play an important role in the global dynamics of the atmosphere (Li et al, 2023). It remains to be seen whether such events also occur in the southern hemisphere. The visibility of this hemisphere started in ~ 2022-2023. Given the continuing increase in the extent and coverage of amateur observations of the planet, the next few years will be a good opportunity to see if any GWS erupts. Less energetic storms have, however, been observed in the south polar region (56°S) in 1996 (Sánchez-Lavega et al., 1996) and in mid-latitudes 35°S-40°S of the southern hemisphere during the Cassini early exploration (Fig 13a) (Dyudina et al. 2007). Mid-scale storms of this kind have also been observed in the north polar region (65°N) in 2018 (Sánchez-Lavega et al., 2020). We can therefore expect their occurrence in the southern hemisphere in coming years. High-resolution images from ground-based telescopes with AO and HST and JWST images will serve to track their dynamics. Thermal infrared images and spectra will allow us to retrieve the temperature field and high-resolution spectra will allow to monitor the changes in the chemistry and the detection of possible ices transported from deeper levels (ammonia ice and water ice, Sromovsky et al. 2013).

Unlike Jupiter, Saturn does not have large and long-lived vortices (Ingersoll et al., 1984) if we exclude the polar cyclones that seem to be permanent (Sayanagi et al. 2018; Antuñano et al. 2018). Images from Cassini and HST and ground-based telescopes show that the longest-lived vortex was the one generated during GWS 2010 (Sayanagi et al., 2013; Hueso et al., 2020). This anticyclone had a size of ~ 10,000 km and changed location in latitude from 42°N to 46°N. Other cyclones and anticyclones (sizes above ~ 1000-2000 km) occur at sub-polar and temperate latitudes (~ 75°N to 65°S) (Ingersoll et al., 1984; Vasavada et al., 2006; García-Melendo et al 2007; del Río-Gaztelurrutia et al., 2018). At temperate latitudes, dark oval anticyclones have been observed to be generated by convective storms (Dyudina et al. 2007). Thus, one future objective is to study the vortices formation rate and dissipation time scales, their motions compared to the mean zonal wind profile and meridional migrations, and spectral reflectivity, among others.

While the northern hemisphere is characterized by the presence of singular and permanent planetary waves as the hexagon and ribbon waves (del Genio et al. 2009; Sayanagi et al., 2018), this does not seem to be the case in the southern hemisphere. Dynamically, the hexagon is a unique sinusoidal wave confined between latitudes 75.3°N and 76.3°N that as a whole moves slowly relative to System III but encloses an eastward jet with peak speed 105 ms[-1] (Baines et al. 2009, Sánchez-Lavega et al. 2014, Antuñano et al. 2015). Wavy patterns have been reported in some southern hemisphere belts in tropical and mid-latitudes from Voyager high-resolution images but they were of low amplitude and not observable with ground-based telescopes (Sánchez-Lavega et al. 2000). The reason for this difference between hemispheres remains to be confirmed with detailed and systematic observations of the southern hemisphere. The hemispheric comparison, in relation to the zonal wind profile, should serve to explore the nature of Saturn´s waves at cloud level.

### 4.2.5 Titan

Much of our knowledge of the atmospheric dynamics of Titan comes from the exploration performed during the flybys of the spacecraft Voyager 1 and 2 in 1981-82, and from the numerous Cassini orbiter flybys between 2004 and 2017, and from the Huygens probe descent in January 2005 (Flasar et al. 2014; Lebonnois et al. 2014; Griffith et al. 2014; Hörst et al. 2017). The vertical wind profile was retrieved during Huygens descent (at one time and one place, from surface to 150 km altitude) (Bird et al. 2005; Folkner et al. 2006). It has been complemented with retrievals of zonal velocities based on the central flash observed during stellar occultations (Sicardy et a. 1999; Hubbard et al. 2006; Sicardy et al. 2006), on Doppler spectral-shifts measurements of molecular spectral lines in the optical and thermal infrared in the stratosphere ($\sim$ 100–500 km), (Luz et al. 2006; Kostiuk et al. 2010) and on millimeter range observations in the upper mesosphere and thermosphere ($\sim$ 500–1,200 km) (Moreno et al. 2005, Lellouch et al. 2019). These measurements indicated prograde winds, reaching peak velocities of $\sim$150–200 ms$^{-1}$ above 100 km with a strong jet of $\sim$ 340 ms$^{-1}$ at 1,000 km. These winds present important latitudinal and seasonal variations. Measurements of Titan's atmospheric motions based on cloud tracking in the troposphere (heights below < 40 km) are scarce and within the range $\sim$ 0 – 40 ms$^{-1}$ (Griffith et al. 2014; Hörst et al. 2017), (Fig. 14).

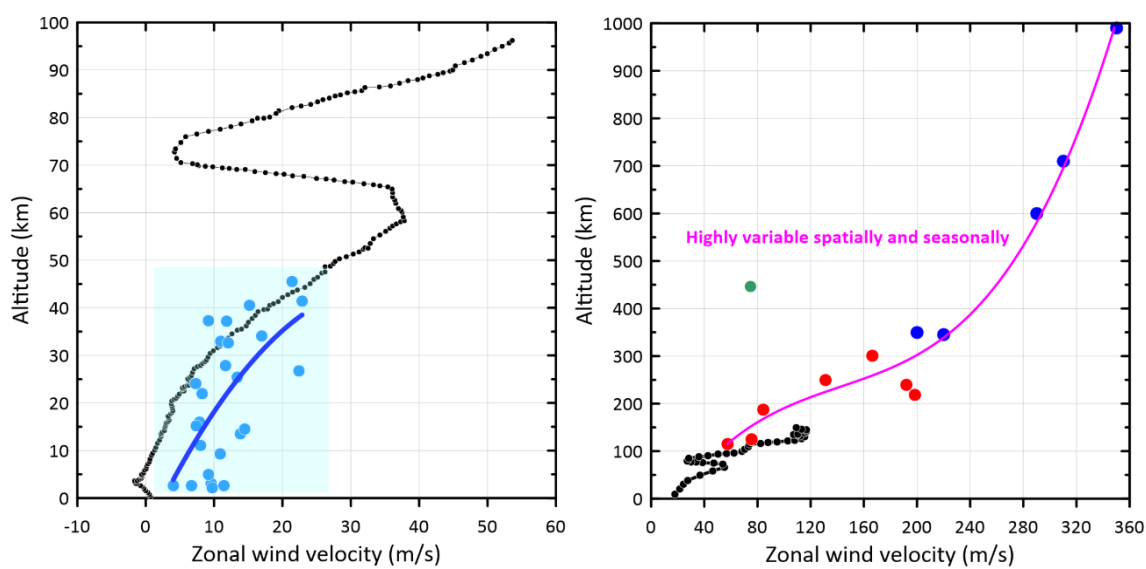

**Figure 14.** *Vertical profiles of the zonal winds in Titan. On the left (altitude 0-100 km): the black line is the profile measured during the descent of the Huygens probe (Bird et al., 2005; Folkner et al., 2006). The blue dots are wind measurements from cloud tracking (from Corlies et al., 2019). The overall uncertainty of these measurements is marked with the light blue background. The dark blue line represents a fit to the wind values. On the right (altitude 0-1,000 km, equatorial zonal winds): the black dots are the Huygens probe descent measurements. The blue dots correspond to Doppler shift measurements in emission lines (Lellouch et al., 2019) and the red and green dots include other Doppler shifts, thermal winds and stellar occultation estimates (from Hörst et al., 2017). The magenta line is a fit to the blue and red dots. Above 100 km the winds change strongly spatially and over time.*

Since Titan is a slow rotator (period 15 days and 22 h), the force balance in the momentum equation (eq. 17) reduces to the Coriolis force, centrifugal force and pressure gradient force (Sánchez-Lavega, 2011)

$$\frac{u^2 \tan \varphi}{R_p} + f u = -\frac{1}{\rho} \frac{\partial P}{\partial y} \qquad (25)$$

and the thermal wind equation is given by

$$\frac{\partial u}{\partial z}\left[\frac{2u \tan \varphi}{R_p} + f\right] = -\frac{R_g^*}{H} \frac{\partial T}{\partial y} \qquad (26)$$

Zonal winds u(z,φ) are obtained from the altitude-latitude temperature maps retrieved from inversion of thermal radiation measurements, and the vertical integration of the above equation (Lebonnois et al. 2014). In early 2000, images obtained with large telescopes of the 8-10 m class in diameter using adaptive optics (AO) at a wavelength of 2.12 μm (filters K, K', H2(1-0)), showed the ability to detect methane ice clouds in the troposphere with sizes > 500 km at heights below 40 km altitude (Griffith et al. 1998; Coustenis et al. 2001; Roe et al. 2002, 2005; Griffith et al. 2014) (Fig. 15). They have been continuously monitored since then with large-telescopes and AO (Nixon et al. 2016) and recently with the JWST (NASA blogs, 2023). These clouds also leave their imprint in the reflectivity of absolutely calibrated near infrared spectra (Griffith et al. 2014). Combining this kind of observations will allow us to make a statistical study of their spatial distribution according to the long seasonal cycle. Deriving their physical properties (sizes, temporal scales of evolution, longitude and latitude location) could allow us to explore the mechanisms behind their formation and their relation to insolation and predictions by General Circulation Models (GCM; Lebonnois et al. 2014). Among the mechanisms proposed to form clouds are moist convection (Hueso and Sánchez-Lavega 2006, Schaller et al. 2009, Schneider et al. 2012), wave phenomena, and stratification (Griffith et al. 2014). Tracking of their displacements can be used to retrieve the wind velocity vectors in Titan's lower atmosphere.

Another important aspect will be to determine trends in the hazes optical depth (described in section 3.6) and their possible coupling to the lower methane cloud formation and the ethane and HCN polar clouds and to atmospheric dynamics, as predicted by GCMs (Rannou et al. 2006) as discussed in section 3.6.

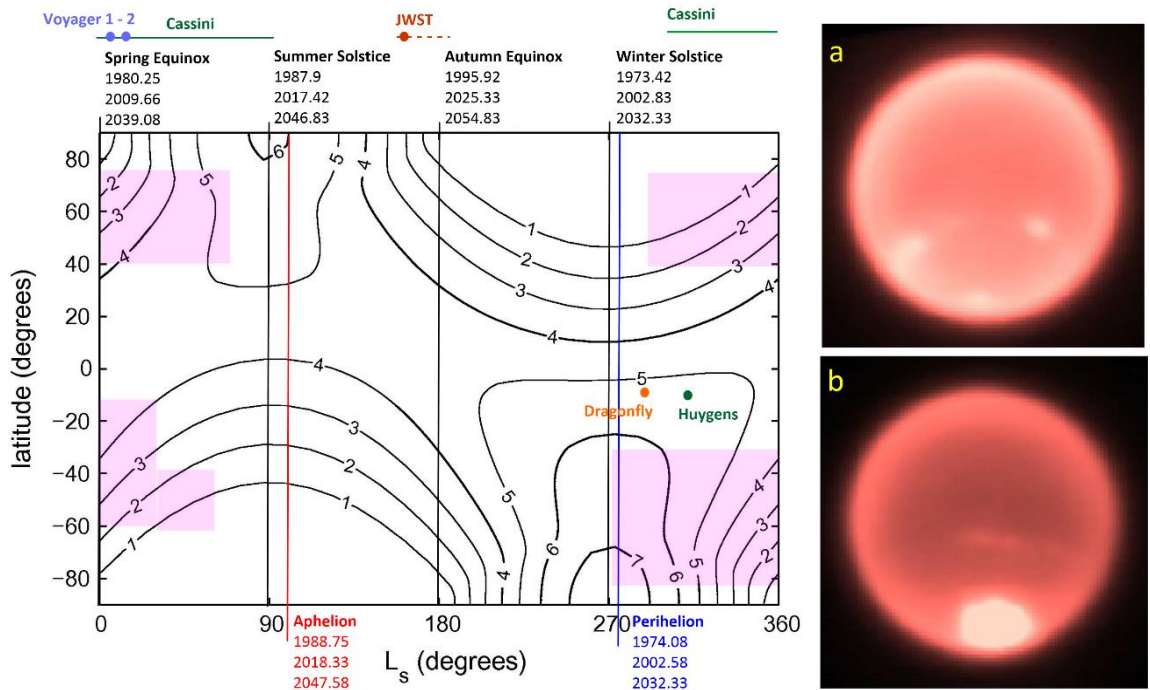

**Figure 15.** *The plot at left shows contours of the daily insolation in W/m² at the top of Titan's atmosphere along Saturn's orbital period of 29.5 years (Lora et al., 2011). The seasons and the aphelion and perihelion dates are indicated for three Saturn's years. The magenta areas show where most clouds were observed in the period 2003-2014 (Nixon et al., 2016). Also indicated are the epochs of Voyager 1 and 2 flybys (September 1980, June 1981), the coverage by the Cassini mission (2004-2017), and the start of JWST in September 2022. The Huygens probe descent (December 2005) and Dragonfly future mission (2034) are also indicated. At right, example of images obtained with the W. M. Keck Observatory with Adaptive Optics system with the NB2.11 (2.11-2.15 μm) filter on 2 September and 3 October 2004 showing scattered clouds at the southern hemisphere and the outbreak of bright clouds (Roe et al., 2005). From the Keck Observatory Titan Monitoring Project. Images available at: https://www2.keck.hawaii.edu/science/titan/*

### 4.2.6 Uranus

Uranus has a retrograde sidereal rotation period as determined by the rotation of the magnetic field of -17 hr 14 min 24s (Archinal et al. 2018). At the upper cloud level the motion of the atmospheric features is zonal and dominated by a retrograde equatorial jet and two high latitude prograde (eastward jets) (Karkoschka 2015; Sromovsky et al. 2015; Hueso and Sánchez-Lavega 2019; Sánchez-Lavega et al. 2019) (Fig. 16). Since the features used to retrieve this profile were observed at different wavelengths, there is uncertainty in their altitude location and vertical wind shears could lead to the observed dispersion. An important objective is therefore to derive the height location of the wind tracers and their accurate tracking over long time periods in order to reduce errors in the velocity measurements. Because of the planet's rapid rotation, outside the equator, these motions are in geostrophic balance (eqs. 21-22).

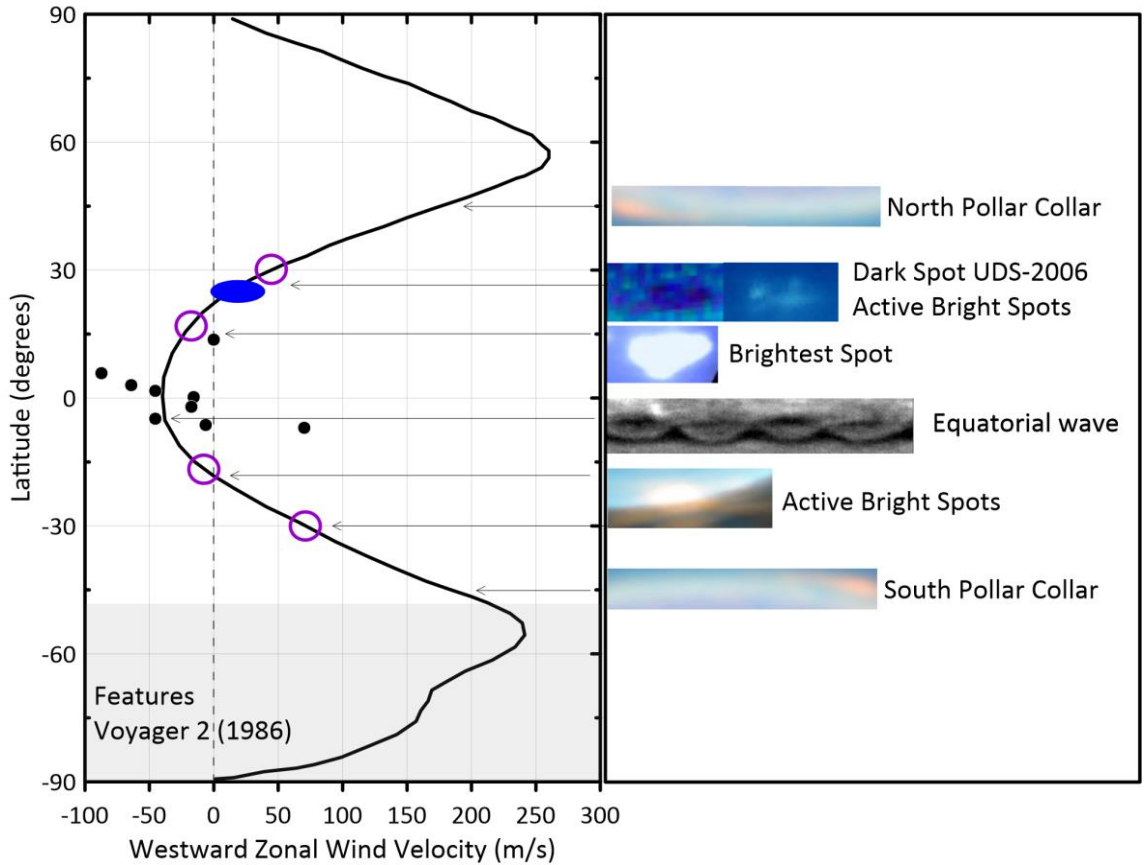

**Figure 16.** *Uranus averaged zonal wind profile as derived by L. Sromovsky (Sánchez-Lavega et al., 2019). The dots show the dispersion in the equatorial wind measurements. The shadowed region are the winds measured by Karkoschka (2015) from Voyager 2 images. Different remarkable features are show at right (see Fig. 17): a dark spot (blue oval), bright spot activity (magenta circles) and the location of the bright polar collars.*

Discrete bright spot activity with a size > 1,500 km are frequent in high resolution images taken in filters isolating methane absorption bands, in particular those centered at 0.89 μm and 2.2 μm (Fig. 17) (Karkoshka 1998; Hammel at al. 2009; Sromovsky et al. 2015). Some of them, close to the equator, are very bright and could be due to moist convection in the upper clouds (Fig. 17e-f) (de Pater et al. 2015; Irwin et al. 2016; Hueso et al. 2020). A fundamental objective with large telescopes and AO is therefore their characterization and how they relate to the wind profile and to the slow evolution of the seasonal insolation cycle with season durations of ~ 21 years. The size and shape, evolution (brightness and area), lifetime, and whether they are isolated or form a series of spots (Fig. 17a), are properties to be determined. We also need their spectral reflectivity from ultraviolet to near infrared to infer aerosol properties in these weather systems. A cluster of smaller discrete bright spots suggestive of moist convection was also observed in the north polar region (latitudes > 60°) and it remains to be seen if this is a permanent feature (Sromovsky et al. 2015). The detection of a north polar cyclonic vortex confined at latitudes > 80° using the VLA at radio wavelengths 0.7-5 cm (Akins et al., 2023) requires complementary AO-imaging and thermal infrared observations to confirm its depth to about 10 bar and infer other properties as the circulation speed and long-term variability.

There is only one dark spot reported in Uranus that showed high contrast at short wavelengths (~ 400 nm) (Hammel et al. 2009b) (Fig. 17b-c). With a size L ~ 2,000 km, this dark oval is probably an anticyclone vortex with accompanying bright high clouds. Their rarity could be due to a bias because of their low contrast with the surrounding clouds and the difficulty of imaging at short wavelengths. Search for these features should be a priority to understand vortex behavior in ice giant atmospheres. An equatorial wave has been observed in AO images (Fig. 17d) (Sromovsky et al. 2015). It was tracked during at least 2 years, and it is therefore of interest to know whether it is a permanent or transitory phenomenon, and what is its nature. This is therefore another objective of observation to be pursued. Complementarily, we also need to search for other wave phenomena at other latitudes (Sromovsky et al. 2015).

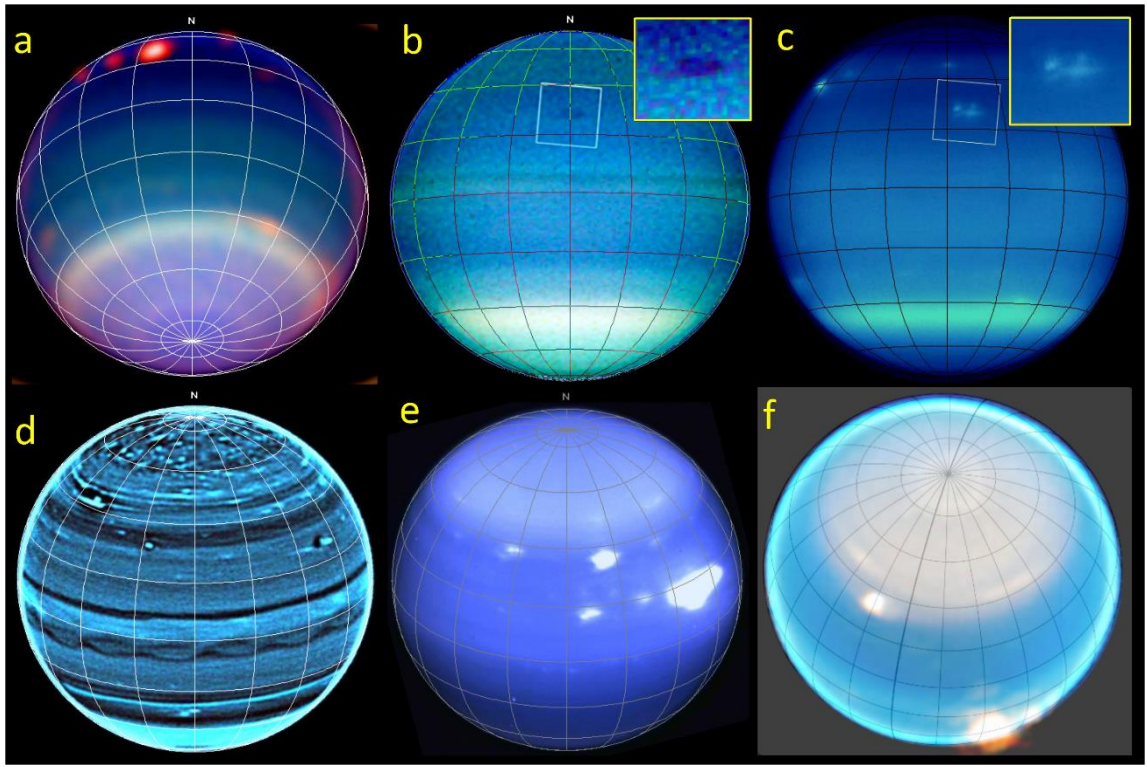

**Figure 17.** *Uranus meteorological features from telescope images in a changing view of the planet. (a) August 8, 1998 HST-NICMOS. The blue, green and red components of this image correspond to exposures taken at near-infrared wavelengths of 0.9, 1.1, and 1.7 μm. Credits: Erich Karkoschka (University of Arizona) and NASA. (b) Dark spot anticyclone UDS-2006 imaged with HST-ACS on August 23, 2006. Wavelengths: 0.550 μm (assigned to blue), 0.658 μm (assigned to green) and 0.775 μm (assigned to red). (c) Bright companion to Dark spot. False-color image taken with the Keck's NIRC2 camera on 30 July 2006 at a wavelength of 1.6 μm. The insets shows a magnified view of the Dark spot and the complex of bright features associated with it. Credit: NASA/ESA/L. Sromovsky, from Hammel et al. (2009). (d) High resolution image from Keck II telescope on 25 July 2012 at a wavelength 1.48-1.78 μm, from Sromovsky et al. (2015). (e) The Bright Spot and other spots on 6 August 2014 at 1.6 μm from de Pater et al. (2014). Credits: Keck Observatory/UC Berkeley/I. de Pater. (f) JWST Near-Infrared Camera (NIRCam) Feb. 6, 2023, colour press image made by combining data from two filters at 1.4 and 3.0 μm*

*(shown here as blue and orange, respectively). Credits NASA/ESA/CSA/STSc/Joseph DePasquale.*

The long-term tracking of Uranus' integrated brightness (1972-2015) at a wavelength of 472-551 nm shows changes in the upper haze layer, apparently following the solar 11-year activity cycle (Aplin and Harrison 2017). Whether this affects the circulation in the stratosphere and the dynamical activity of features in the troposphere is unknown and needs to be followed up in the coming years.

### 4.2.7 Neptune

As for Uranus, at upper cloud level the motions are purely zonal dominated by an intense and broad retrograde equatorial jet of -400 ms$^{-1}$ and two high latitude prograde (eastward jets) of +250 ms$^{-1}$, as retrieved from cloud tracking (Sánchez-Lavega et al. 2019; Hueso and Sánchez-Lavega 2019) (Fig. 18). The rotation period of the planet is 16 hr 06 min 36 s (Archinal et al. 2018) as determined from the periodicity of the magnetic field. Outside the equator, the atmospheric flow is in geostrophic balance (eqs. 21-22).

Differences in velocity of atmospheric features can reach 350-500 ms$^{-1}$ in the equatorial troposphere, which has not been observed on any other gaseous planet. This scatter in the velocity data at equatorial latitudes strongly suggests that the cloud tracers are at different heights in the presence of strong vertical wind shear (Tollefson et al. 2018; Chavez et al., 2023). If some features are at the $CH_4$ cloud formation level ($P_1 \sim 1$ bar) and others at the $H_2S$ cloud level ($P_2 \sim 7$ bar), their vertical separation is $\sim 50$ km (2 scale heights H) and the corresponding wind shear is $du/dz \sim 0.007–0.01$ s$^{-1}$. These are high values when compared to the Coriolis frequency at 10° latitude, $f = 2\Omega \sin\varphi \sim 3.8\times10^{-5}$ s$^{-1}$ and to the meridional wind shear $du/dy \sim 9.2\times10^{-6}$ s$^{-1}$ derived from the mean wind profile (fig. 18). The existence of intense vertical wind shear was also observed in the stratosphere and at higher altitudes as measured from Doppler-shift of molecular lines (Carrión-González et al. 2023) and derived from temperature data and the thermal wind equation (Fletcher et al., 2014). The features in the equatorial jet could also form part of a wave with its own phase speed moving relative to the zonal flow. Disentangling the nature of the velocity differences of these atmospheric features, which implies constraining their height and dynamical nature, represents a major objective of future studies.

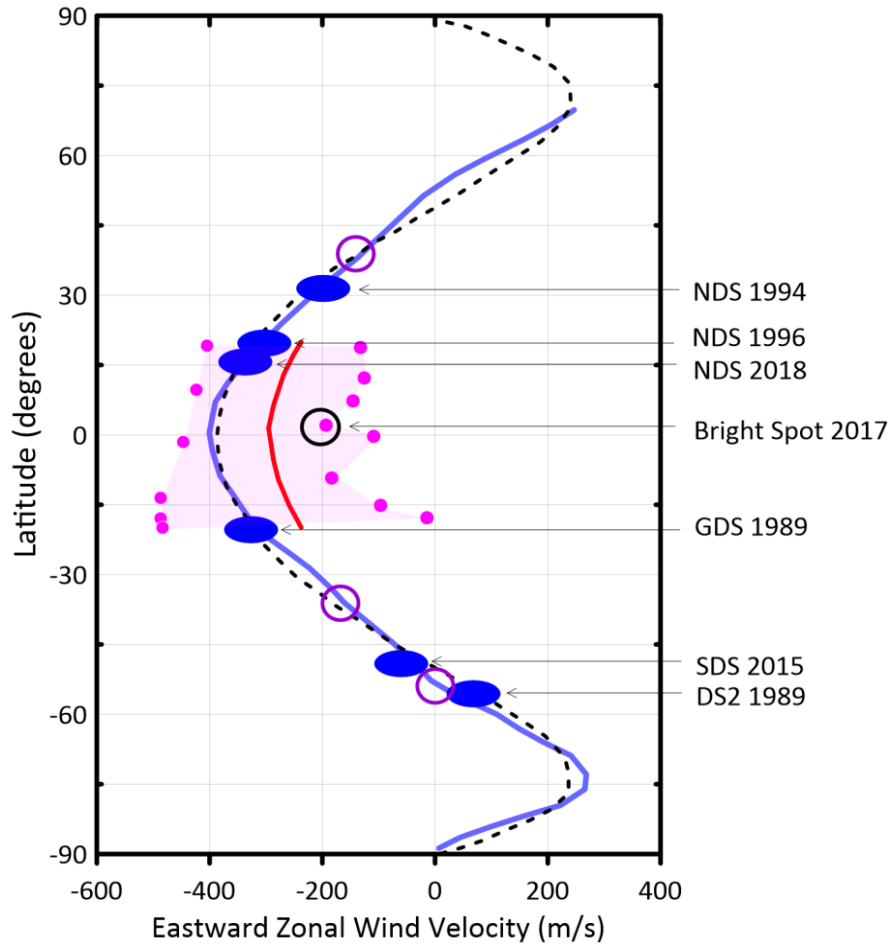

**Figure 18.** *Neptune zonal winds and location in latitude of major features. Includes observations in an ample range of wavelengths ($\sim 0.4 - 2.1$ μm) sounding features at different altitudes in the atmosphere. The blue line is the average wind profile derived by L. Soromovsky from a variety of sources (Sánchez-Lavega et al., 2019). The black dashed line is a fit to the data, extrapolated to unobserved northern latitudes (poleward 70 °N), by Soyuer et al. (2023). In the equator, the red line is from a fit to the motions of high altitude features by Tollefson et al. (2018) and the magenta points and magenta shadowed region corresponds to a variety of features at different altitudes from Molter et al (2019). The blue ovals are the dark anticyclone vortices so far observed and are identified by their acronym and discovery year. The purple disks corresponds to latitudes of preferred bright cloud activity and the black disk corresponds to the largest bright spot so far observed (Molter et al., 2019).*

The atmosphere of Neptune is very active in showing discrete bright spots at all latitudes when observed in the $CH_4$ absorption bands at 0.89 μm, 1.1 μm and 2.2-2.3 μm where they have high contrast with the background cloud deck (Hueso and Sánchez-Lavega 2019) (Fig. 19). These spots have typical sizes of $1,000 - 4,000$ km and sometimes extend over an entire circle of latitude forming bright bands. Voyager 2 imaged these spots at optical wavelengths and were found sometimes accompanying the large dark ovals but located above them (Smith et al. 1989; Ingersoll et al. 1995). These bright clouds could form in the upwelling air forced by the dark oval vortex circulation. Besides these cases, other bright spots could be formed by moist convection and by waves of different types (Hueso et al. 2020). Seasonal forced changes, if they exist, should be in the long term since season duration ~ 41 year.

Unlike Uranus, the atmosphere of Neptune shows regularly the presence of large dark oval spots (DS) with sizes ~ 4,000 - 15,000 km in length. From 1989 to date, six DS have been observed and studied (Wong et al. 2018, 2022; Simon et al. 2019). These oval spots are dark and well contrasted at blue wavelengths, and have been observed over a wide range of latitudes between 20° and 55° north and south. Their lifetime is ~ 1 and 5 years, limited at least in some cases by their migration in latitude due to beta effect (the gradient of the Coriolis force term, $\beta = df/dy$), and destruction when approaching equatorial latitudes. The DS are anticyclones according to the ambient meridional wind shear, show changes in shape and exhibit oscillations in their latitude-longitude position (Sromovsky et al. 1993; Ingersoll et al. 1995). A recent study using MUSE/VLT spectra (475-933 nm) showed that the darkening at short wavelengths has its origin in the $H_2S$ cloud at ~ 5 bar where other type of bright spots were also observed (Irwin et al. 2023).

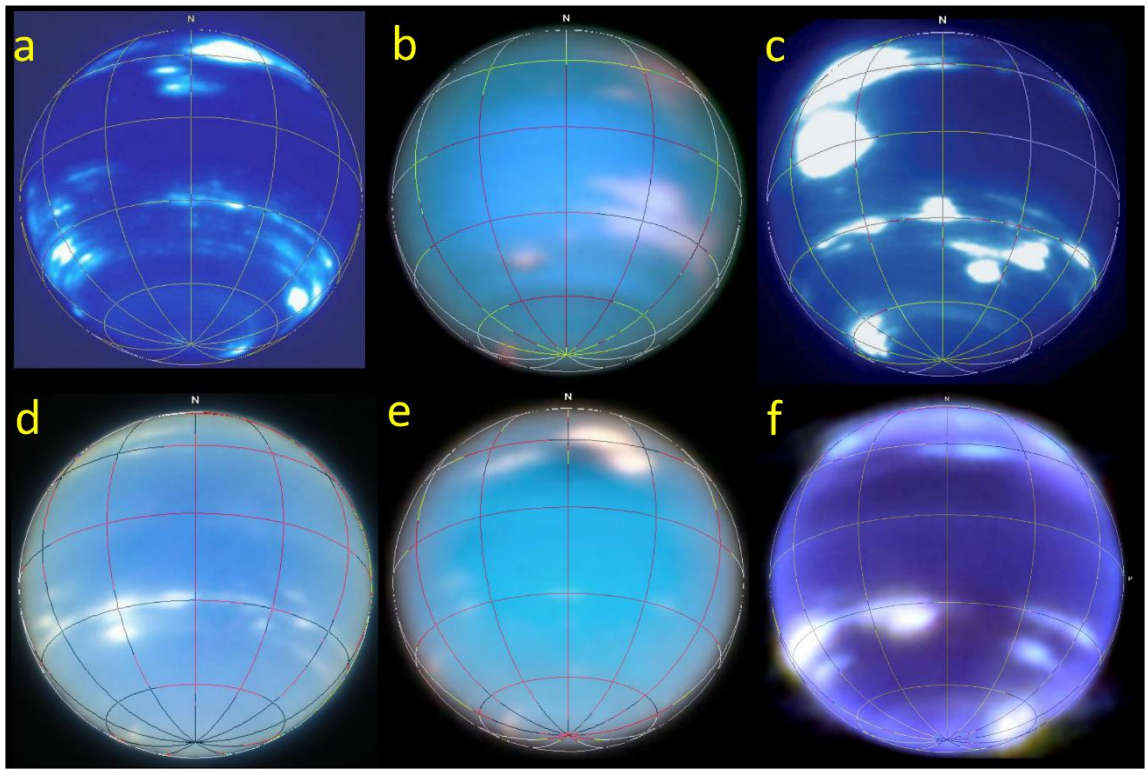

**Figure 19.** *Images of Neptune from different telescope facilities and wavelengths. (a) Keck Observatory, 26 July 2007, filter H (1.63 $\mu$m). Credit: UC Berkeley. (b) HST WC3/UVIS, 16 May 2016, blue and visible wavelengths. Credit: NASA/ESa/UC Berkeley. (c) Keck Observatory, 26 June 2017, filter H (1.63 $\mu$m). Credit: UC Berkeley. From Molter et al. (2019). (d) ESO VLT GALACSI/MUSE, image released 18 July 2018, wavelength range 591-920 nm. Credit ESO/VLT MUSE. (e) HST WFC3/UVIS, 5 November 2018, wavelengths 467, 547, and 845 nm. From Simon et al. (2019). (f) JWST NIRCam press image, 12 July 2022, 1.4, 2.1, 3.0, 4.6 $\mu$m. Credits: NASA/ESA/CSA/STScI.*

A polar hot spot related probably to a cyclonic vortex, is present on Neptune's south pole, as observed in the near- and mid-IR as well as at radio wavelengths (Orton et al. 2007; de Pater et al., 2014; Fletcher et al., 2014). Long-term thermal infrared observations using complementary AO-imaging are necessary to infer its properties including circulation and stability or variability. These studies show that ground-based and space-based telescopes are set to be in the coming years the key technique to study the atmospheric dynamics of the icy planets by means of the high and medium resolution spectroscopy and high resolution imaging.

## 5. Atmospheres of exoplanets

This section was conceived as an overview of findings and ideas in the observation and theory of exoplanet atmospheres. The choice of themes is somewhat subjective, but should hopefully point towards areas where significant activity is occurring. Priority has been given to transiting exoplanets, partly because they have been investigated more systematically than the non-transiting ones (mainly the young Jupiter-sized planets that are targeted by direct imaging). There are recent reviews that present the state of the art of directly-imaged substellar objects (Biller, 2017; Showman et al., 2020; Zhang, 2020). The present review is organized as follows. The techniques used for observation and first interpretation are discussed in section §5.1. Sections §5.2, 5.3, 5.4 and 5.5 comment on the chemical composition, aerosols, temperature and dynamics of the atmospheres, respectively. Lastly, §5.6 presents some possible future developments in the field.

### 5.1 Observational and retrieval techniques

A variety of techniques and skills are needed to extract the tiny signatures that reveal an exoplanet atmosphere. It requires a good knowledge of the instrumental systematics and of stellar activity, and of strategies to mitigate their possibly confounding effects. Clearly, an atmosphere potentially suitable for life, such as some of the TRAPPIST-1 planets (Gillon et al., 2017), is very different to the atmosphere of KELT-9 b, which has a dayside temperature of 4,600 K (Gaudi et al., 2017).

A transmission spectrum is obtained during the transit of the planet in front of its host star. It conveys the effective size of the planet at the specified wavelength. This size is dictated by the opacity of the atmosphere, provided one exists, for stellar light rays crossing from the day- to the nightside, and thus by the densities and extinction cross sections of the gases and aerosols in it. An occultation spectrum is obtained while the planet passes behind its host star, and contains in principle a combination of radiation thermally emitted by the planet and stellar photons incident on it and backscattered towards the observer. When the planet is not in transit or being occulted, a phase curve displays how the planet brightness evolves as it moves on its orbit. A phase curve contains simultaneous information on the planet day- and nightsides, and is more informative than an occultation alone. Directly imaged exoplanets do not typically transit, and an observation of them is effectively a snapshot of their hypothetical phase curves. Fig. 20 shows a few examples of what is currently possible for transmission spectroscopy with JWST and what is required to characterize the twin of an Earth-Sun system in transit.

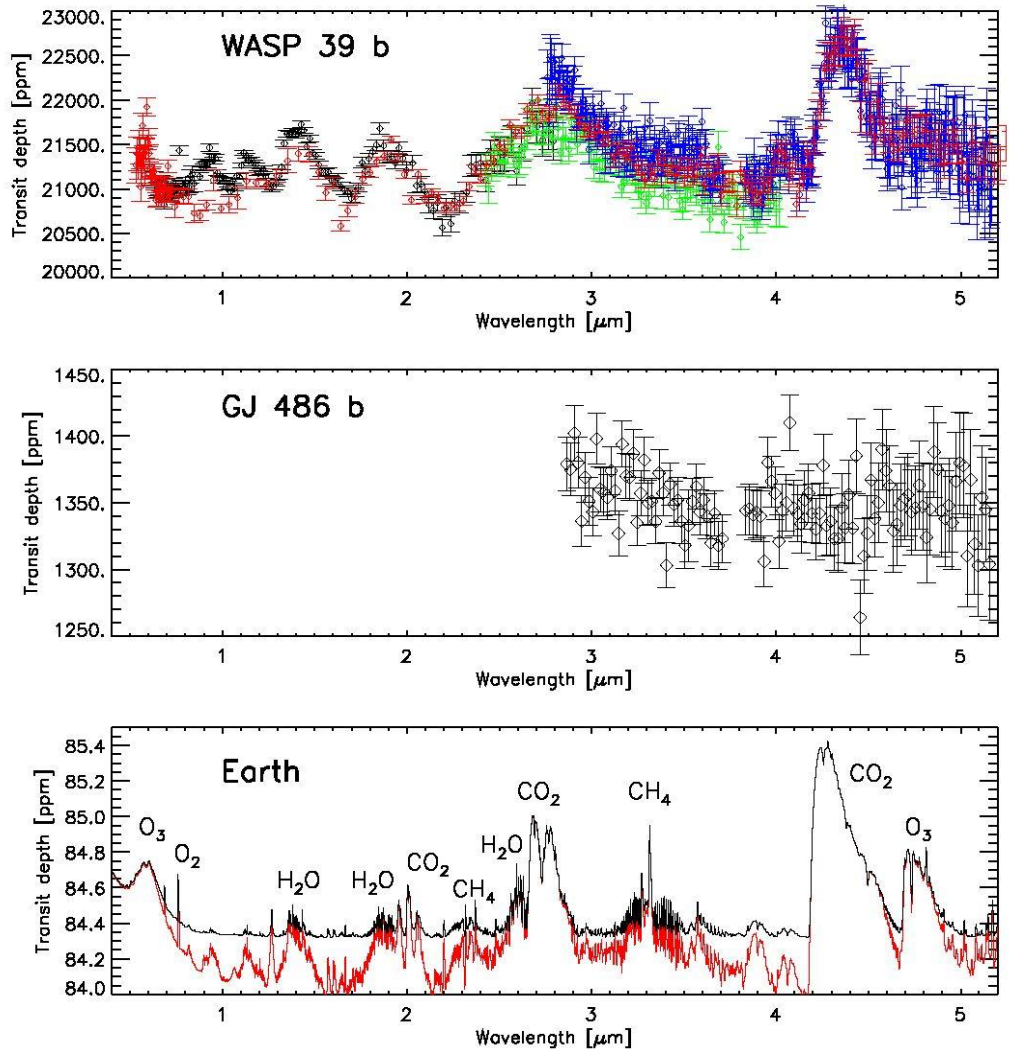

**Figure 20.** *The present and future of transmission spectroscopy. Top. Spectrum for WASP-39 b obtained as part of the JWST ERS Program for transiting exoplanets. It merges the data from 3 instruments in 4 observing modes (Ahrer et al., 2023; Alderson et al., 2023; Feinstein et al., 2023; Rustamkulov et al., 2023), and contains evidence for Na (0.59 µm), K (0.77 µm), $SO_2$ (4.05 µm) and CO (4.6 µm). A few strong $H_2O$ and $CO_2$ bands are identifiable by comparison with the Earth's transmission spectrum (bottom panel). Middle. Spectrum obtained from two transits for the super-Earth GJ 486 b as part of JWST Cycle 1 (Eureka! pipeline; Moran et al., 2023). The planet-to-star contrast is strongly enhanced by the small size of its M3.5 V-type host. The tentative mild slope shortwards of 3.7 µm might be caused by water absorption in the atmosphere of the planet or by starspot contamination. Further observations, in particular at shorter wavelengths, are needed to confirm that the slope is real and distinguish between the two proposed origins. Bottom. Detecting and characterizing the exo-twin of the Earth-Sun system is a top priority. The small scale height of the atmosphere and comparatively large stellar size result in atmospheric features ≲1 ppm (black curve). Refraction of the stellar light*

*on its passage through the atmosphere of the planet causes a major reduction in the trough-to-peak contrast (red curve: synthetic model omitting refraction, García Muñoz et al. (2012)). This reduced contrast plus the low frequency of transits translates into the prospective investment of decades of time for the characterization of a genuine Earth twin (López-Morales et al., 2019; Hardegree-Ullman et al., 2023).*

Remote sensing theory aims to connect the different types of measurements mentioned above to the physical processes of absorption, scattering and emission in the planet atmosphere. This information is our initial window into their composition, energy budget and dynamics. The atmospheric signals for exoplanets are typically tiny with respect to the stellar brightness, on the order of hundreds or tens of parts-per-million for the systems that have been investigated, and they refer to extended regions on the planet. These considerations call for sophisticated methods to extract the relevant information from the measurements and interpret them.

The techniques for spectroscopic measurement may be classified based on resolution $R$. In exoplanet work, $R \sim 100\text{-}1{,}000$, as delivered by HST and JWST, is considered low-middle resolution. With low-resolution spectroscopy, when multiple gases contribute within a narrow region of the spectrum, disentangling their contributions becomes difficult. Photometric measurements, such as those in the Spitzer or Kepler phase curves, represent limiting cases of low-resolution measurements. The term high-resolution spectroscopy (HRS) is often used for measurements made from the ground with $R \gtrsim 30{,}000$ (Snellen, 2014). The HRS technique works with individual lines in the exoplanet atmosphere, as long as they are strong. If the lines are weak, the performance of the technique is boosted by stacking many lines into a superline, or equivalently through application of cross-correlation methods to the original data. When possible, it is advantageous to combine low- and high-resolution measurements to mitigate the limitations of each separate technique (Brogi and Birkby, 2021).

Retrieval models are tools that use the measured spectra at one or multiple phases of the planet on its orbit as input, and infer from them plausible atmospheric configurations and associated uncertainties of the physical parameters (Madhusudhan, 2018; Barstow et al., 2020; Fortney et al., 2021). They represent the first step towards the physical interpretation of the observation. Retrieval models can work on transmission spectra and produce a list of chemicals in the atmosphere and some constraints on their abundances. When applied to occultation or phase curve data dominated by thermal emission, the retrievals often set tight constraints on the temperature over a range of altitudes and, possibly, longitudes, over the planet.

Retrieval models were originally designed to infer information that depended only on altitude and represented average conditions at the terminator or the dayside. These are usually referred to as 1D models. In recent years, there is a push to build 2D/3D retrieval models that can also extract information specific to the latitude/longitude of the planet (Stevenson et al., 2014; García Muñoz and Isaak, 2015; Feng et al., 2020; Irwin et al., 2020; Chubb and Min, 2022; Yang et al., 2023). There is much diversity in the way these models treat the gas chemical composition. In some of them, the abundances are kept free without any prior constraints, and it is the model that quantifies them on a purely mathematical basis. Other retrieval models incorporate internal constraints based on thermochemical equilibrium or photochemistry. The motivation for the latter strategies is that in principle they make the inferred properties more physically plausible. The risk is that an imperfect prescription of the physics in the constraints may drive the retrieval outcome away from the true state of the atmosphere.

## 5.2 Chemical composition

### 5.2.1 Gas giants

The atmospheres of gas giant planets (defined here by $R_p \gtrsim 4R_\oplus$) are H/He-dominated (Zeng et al., 2019), and contain trace amounts of heavy elements. Their chemical modelling has received much attention (Madhusudhan, 2012; Venot et al., 2012; Moses, 2014) because it is chemistry that ultimately determines whether the different elements appear in atomic or molecular form and the wavelengths at which they are detectable. Confronting the model predictions with observations and identifying possible deviations is a major force driving spectroscopic surveys (Sing et al., 2016; Edwards et al., 2022). In general, the mass and chemical composition of gas giants do not evolve much over time (Owen, 2019), unless the planets are exposed to exceptional irradiation conditions (García Muñoz and Schneider, 2019). Their spectra thus represent windows into the mechanisms through which the planets grew in the protoplanetary disk and the gases and solids that were accreted (Madhusudhan, 2019).

The atmospheres of close-in gas giants are relatively easy to probe. In transmission spectroscopy, for example, low mean molecular weights and high temperatures result in extended atmospheres and correspondingly large in-transit signatures. Unsurprisingly, hot and ultrahot Jupiters (usually defined by their equilibrium temperatures $T_{eq}$ as having 1,000 K$\lesssim T_{eq} \lesssim$2,200 K and $T_{eq} \gtrsim$2,200 K, respectively) have been preferential targets of spectroscopic surveys. $H_2$, H and He are expected to be the dominant neutrals in the atmospheres of gas giants. The evidence for $H_2$ often comes from models, which show that $H_2$ dominates over H as the main hydrogen form at the usual pressures (~1 mbar) probed with optical-infrared (IR) observations unless the temperatures are very high, as for some ultrahot Jupiters (Lothringer et al., 2018). The H atom has been detected in absorption from its ground state in the Lyman-$\alpha$ line (Vidal-Madjar et al., 2003; Ben-Jaffel et al., 2022), and in absorption from the first and second excited states (Yan and Henning, 2018; Sánchez-López et al., 2022). The He atom has been detected in absorption from a metastable state at 10,830Å (Spake et al., 2018; Czesla et al., 2022). Both atoms might be detectable in spontaneous airglow emission, but an attempt to detect He airglow has only resulted in an upper limit (Zhang et al., 2020). The H and He absorption measurements probe the upper layers of the atmosphere. Generally, for each element there is a level, often between 10 and 0.001 mbar, depending on the temperature and irradiation conditions, that separates the altitudes where the element appears mainly in molecular or atomic form (García Muñoz, 2007; Moses et al., 2011; Lothringer et al., 2018).

The fractional abundance of a heavy element X is often expressed in the form of a ratio of number densities X/H. The O/H, C/O and K/O ratios are particularly valuable for diagnostics. O/H is used as a proxy for atmospheric metallicity, as it can be constrained through near-IR (NIR) measurements of $H_2O$, and O is the most abundant element after H and He in a solar-composition gas. Given that C and O condense at different distances from the host star, the C/O ratio provides insight into the planet birthplace in the protoplanetary disk (Madhusudhan, 2019). K/O is linked to the refractory-to-volatile accretion during formation and, possibly, to ongoing accretion of interplanetary dust (Arras et al., 2022).

Retrieval models are used to constrain the atmospheric metallicities. Assessing whether the inferred metallicities correlate with the planet masses in a way similar to the solar system planets has been used as an argument to test different planet formation scenarios (Thorngren et al., 2016; Wakeford and Dalba, 2020). As work continues to build a more reliable sample of metallicity determinations, it appears that atmospheres with super-solar metallicities and sub-solar C/O ratios are quite common (Madhusudhan, 2019).

Thermochemical equilibrium describes the atmosphere at high pressure and temperature (Venot et al., 2012; Moses, 2014). Modelling the gas composition in such conditions is often done for a first comparison with the observations even though thermochemical equilibrium may not hold at the pressures that are probed. Indeed, if vertical mixing proceeds faster than chemical equilibration, it likely results that the composition probed is more representative of thermochemical equilibrium in deeper layers than of the equilibrium conditions at the altitudes probed. Similarly, rapid horizontal mixing may cause the conditions probed at the terminators to be interpreted as the equilibrium conditions at other longitudes. This effect is potentially severe when there exist strong day-to-night gradients.

Disequilibrium chemistry, driven by cosmic-rays and photofragmentation (dissociation, ionization) at wavelengths from X-rays to the Near-Ultraviolet (NUV), also cause the atmosphere to depart from thermochemical equilibrium (Locci et al., 2022; Rodgers-Lee et al., 2023). The way that disequilibrium chemistry modifies the partitioning of C atoms into $CH_4$ and CO in the atmosphere of the Neptune-sized GJ 436 b has attracted much attention, as the interpretation of the available Spitzer data for this planet has been problematic (Morley et al., 2017). Similar disequilibrium effects are expected for other elements, and it is clear that the topic will become increasingly relevant as the quality and spectral coverage of the data improve (Tsai et al., 2023).

The $H_2O$ and CO molecules have been identified on numerous gas giants, in particular with HST and from the ground with high-resolution spectroscopy (Snellen et al., 2010; Sing et al., 2016; Madhusudhan, 2019). Indeed, models predict that these molecules take much of the O and C atoms in H/He-dominated atmospheres over a broad range of temperatures. Going beyond detection, quantifying the abundances of these and other molecules and determining the altitudes at which they are probed is key to gaining further insight into atmospheric chemistry.

Recent advances have enabled the exploration of new aspects of the molecular atmospheres of gas giants. The OH radical, a byproduct of $H_2O$ dissociation, has been detected on the daysides of ultrahot Jupiters WASP-76 b and WASP-33 b (Landman et al., 2021; Nugroho et al., 2021). A key point about OH is that it participates as an intermediate chemical in the cycles that ensure the stability of other molecules. The $^{13}CO/^{12}CO$ ratio has been identified at the young gas giant TYC 8998-760-1 b, opening up a new avenue for investigating planet formation through the accreted isotopes of an element (Zhang et al., 2021). A HRS search for $^{13}CO$ in the transmission spectrum of hot Jupiter WASP-39 b has produced results that are not robust enough to claim a detection, but they nevertheless suggest that the existent facilities may have the required sensitivity for future detections (Esparza-Borges et al., 2023).

At high altitudes, the atmospheres of close-in gas giants eventually become atomic. The transition occurs because molecular reformation becomes inefficient at low pressures, whereas photofragmentation by the stellar photons and thermal decomposition at high temperatures are increasingly rapid. The atomic atmosphere is of interest in itself, as it provides a number of observables. The extent of the atomic atmosphere before ionization sets also the effective size of the planet to the stellar XUV photons that drive atmospheric escape (García Muñoz, 2007; Owen, 2019).

The list of neutrals and ions detected in the atomic atmosphere to date includes 15-20 elements, some quite exotic, which are also useful to probe the temperature and winds of the atmosphere. A non-comprehensive list of atoms/atomic lines used in such studies include: H-$\alpha$ and other lines of the H I Balmer series; the $\beta$ (3→5) line of the H I Paschen series, Ba II, Ca I, Ca II, Cr I, Cr II, Fe I, Fe II, K I, Li I, Mg I, Mg II, Mn I, Na I, Ni I, Rb I, Sc II, Si I, Sm I, Sr II, Ti I, Ti II, and V I. Most detections have been reported at ultrahot Jupiters. This partly reflects that at very high temperatures the elements are less prone to locking into condensates while their spectra become richer and probably somewhat easier to detect with HRS. The massive outflow that some of these planets may be experiencing also facilitates the lifting of the elements to the high altitudes where they are occasionally found. Their detection should yield a complementary view of atmospheric metallicity, provided that the connection between the measurements and the elemental abundances is well understood.

There are two difficulties to establishing that connection. One is that inferring quantitative information from high-resolution spectroscopy is complicated by how the data are usually self-normalized (Brogi and Birkby, 2021). Another difficulty is that at low pressures, the atoms are probably in non-local thermodynamic equilibrium (NLTE). In NLTE, determining the populations in the atomic states requires solving a problem in which each state is connected to all others through radiation and collisions with thermal and non-thermal particles. We are beginning to address such problems (García Muñoz and Schneider, 2019; Fossati et al., 2020; García Muñoz, 2023a,b; Huang et al., 2023; Gillet et al. 2023), but are far from appreciating their full implications. There are additional ideas to search for NLTE effects in molecules such as OH (Wright et al., 2023).

### 5.2.2 Super-Earths and sub-Neptunes

Looking ahead, major advances are likely to occur soon in the understanding of planets with sizes in the super-Earth ($1 \lesssim R_p/R_\oplus \lesssim 2$) and sub-Neptune ($2 \lesssim R_p/R_\oplus \lesssim 4$) ranges. They will be driven by observations with large-aperture telescopes such as the JWST and, later, with the Extreme Large Telescopes (ELTs) currently under construction. These small planets are intriguing because they do not occur in our solar system but are very common in the general exoplanet population. Based on $M_p$ and $R_p$ alone, it is often impossible to discern if they are made of a rocky core plus an H/He envelope that is small by mass but large by volume, or if they grew by accumulating enormous amounts of $H_2O$ (or other astrophysical ices) which might have produced massive steam atmospheres (Valencia et al., 2007). The spectra of their atmospheres will give us additional insight to decide on their composition.

The $M_p$-$R_p$ diagram of Fig. 21 demonstrates visually that although very-low mass planets are typically consistent with Earth-like densities, and presumably with Earthlike compositions, the composition of super-Earths and sub-Neptunes is more difficult to constrain. The difficulty arises from both the measurement uncertainties and the intrinsic degeneracies in the models, which cannot be broken with $M_p$ and $R_p$ alone.

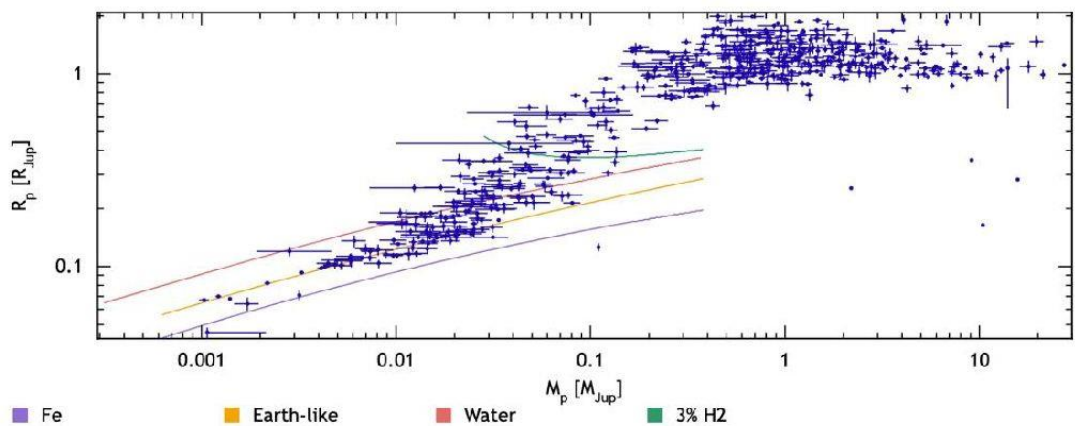

**Figure 21** *Measured radii and masses of exoplanets (Otegi et al., 2020) (update from 2023 June, https://dace.unige.ch/exoplanets/). Solid continuous lines trace the theoretical predictions for planets with the stated bulk/atmospheric compositions. This type of diagram is very valuable for a first insight into the existence or absence of an atmosphere, and its plausible compositions.*

Progress in their understanding will likely require connecting multiple aspects of their structure and history, including their long-term evolution. Investigating thoroughly a few select planets, while considering their population-wise attributes, might also help. If life is prevalent beyond Earth, the atmospheres of these planets, especially those transiting low-mass stars, may represent our best opportunities to identify its signature in the near future.

A question that often arises in the observation of super-Earths and smaller exoplanets is whether they have an atmosphere, as they are vulnerable to rapid mass loss when orbiting close to their host stars. Measuring their thermal phase curves provides insight into this question, as demonstrated with Spitzer observations. A hot-spot offset with respect to the direct view of the substellar point in the phase curve has been interpreted as caused by a thick super-rotating atmosphere, as proposed for 55 Cnc e (Demory et al., 2016), whereas the absence of an offset may indicate that the planet lacks a thick envelope, as at LHS 3844 b and K2-141 b (Kreidberg et al., 2019; Zieba et al., 2022). After losing its primary atmosphere, an ultrahot rocky planet may develop an envelope of refractories (e.g. Mg, Ca, Fe, Si, SiO) sustained by a magma ocean (Ito et al., 2015). The spectroscopic characterization of such atmospheres is setting new paths for the geophysical investigation of exoplanets.

Many of the arguments outlined for gas giants apply also to the super-Earths and sub-Neptunes for which a thick envelope may be presumed. As their atmospheres probably contain abundant O, C and other heavy elements, the possibilities are more complex, especially when combined with the diversity of host stars (from ultracool to early-type, i.e. with effective temperatures from 2,500 to 10,000 K) around which they are found (Gillon et al., 2017; Giacalone et al., 2022). Thermochemical modelling for a range of O/H and C/H ratios and temperatures shows that atmospheres dominated by $H_2O$, $O_2$, CO or $CO_2$ naturally emerge from this complexity (Moses et al., 2013; Hu and Seager, 2014).

Assessing these possibilities, in particular the occurrence of thick steam envelopes in super-critical state (Mousis et al., 2020; Madhusudhan et al., 2021), on small planets such as GJ 486 b or the TRAPPIST-1 planets calls for spectroscopic observations (Caballero et al., 2022; Ridden-Harper et al., 2023; Greene et al., 2023). Additional arguments must be invoked to break the degeneracies in the interpretation of their compositions that spectroscopy alone cannot break. The time that the atmosphere can survive before it is completely lost to space under strong stellar irradiation is one of these powerful arguments. It has been used, for example, to argue that $\pi$ Men c and Kepler-138 d have atmospheres of high mean molecular weight (García Muñoz et al., 2020, 2021; Piaulet et al., 2023), because had their atmospheres been H/He dominated, they would have been lost long ago. As the argument depends on model predictions for the mass loss rate, it is important to continue assessing the completeness of such models, especially for high-metallicity atmospheres (Ito and Ikoma, 2021; García Muñoz, 2023b). Elucidating the history of a planet is better done if the history of its host star, and in particular of its radiative output during quiescent and flaring periods, is reasonably well constrained.

There are proposals to turn the challenges in the interpretation of sub-Neptune spectra into ways of probing the surfaces below their atmospheres (Yu et al., 2021). The idea is that if the atmosphere is not too thick, incomplete recycling of the atmospheric constituents at depth may result in chemical disequilibrium potentially identifiable with spectroscopy. It is unclear though if the evidence for chemical disequilibrium in the atmosphere can be uniquely related to a shallow surface.

## 5.3 Aerosols

Clouds, hazes and dust are different aerosol types, and participate in virtually every aspect of exoplanet atmospheres (Gao et al., 2021). Clouds are typically made of liquid or solid particles of super-micron size and form via condensation from the gas phase. Hazes often refer to somewhat smaller solid particles that result from complex sequences of chemical aggregations. Clouds and hazes are formed from material in the atmosphere, whereas the atmospheric dust originates from interplanetary dust particles that disintegrate upon entering the planet (Arras et al., 2022).

Aerosols are often perceived as nuisances in transmission spectroscopy because they tend to mute any gas spectral feature, especially when the aerosol layer lies at high altitude. The sub-Neptune GJ 1214 b is well known for showing a flat transmission spectrum from the NIR to the mid-IR (MIR) (Kreidberg et al., 2014; Kempton et al., 2023), which has been interpreted as evidence for an atmospheric composition of high mean molecular weight and high-altitude aerosols (Lavvas et al., 2019; Gao et al., 2023). Planets with (nearly) aerosol-free atmospheres such as the hot Saturn-sized WASP-96 b, at which the full shape of the Na resonance line in the optical is resolvable, are the exception rather than the rule (Nikolov et al., 2018). Determining the gas abundances by means of atmospheric retrievals when the atmosphere is rich in aerosols becomes a particularly challenging problem, and the inferred abundances may even depend on the way the aerosol properties are parameterized in the retrievals (Barstow, 2020).

The aerosol extinction cross sections $\sigma_a$ depend on wavelength in ways that are connected to their composition and particle size. Transmission spectroscopy is the tool to capture those dependencies and, doing the inverse exercise, figure out the dominant aerosol in the atmosphere. At short wavelengths, say in the UV-optical for the usual atmospheric applications, $\sigma_a \sim 1/\lambda^\alpha$, with $\alpha \sim 4$ for small particles (Rayleigh limit) and $\alpha \ll 1$ for large particles (grey-absorber limit). In this approximation, the measured planet size normalized by the atmospheric scale height, a magnitude that can be determined from the transmission spectrum, varies with wavelength as $dR_p(\lambda)/H = -\alpha d\lambda/\lambda$ (Lecavelier Des Etangs et al., 2008). Constraining the scattering slope of the measured transmission spectrum, and therefore $\alpha$ by using the above equation, yields insight into the aerosol sizes if $\alpha < 4$, or shows the prevalence of Rayleigh scattering by high altitude gases/haze if $\alpha \sim 4$. The idea has been extended to consider vertical gradients in the aerosol properties that may cause the super-slopes ($\alpha \gg 4$ in the scattering slope) tentatively identified in some atmospheres (Ohno and Kawashima, 2020). The chemical groups in the aerosol particles can vibrate and produce narrow features in $\sigma_a$ at NIR-MIR wavelengths that are potentially identifiable in the transmission spectrum (Wakeford and Sing, 2015). Such observations may soon help constrain the aerosol composition at some exoplanets through the bond-specific vibrational wavelengths. The idea might also help constrain the dust composition in the comet-like tails that enshroud some possibly disintegrating planets (van Lieshout et al., 2014), thus offering one more avenue for geophysical investigations.

For the hot Jupiters that are often targeted by spectroscopic surveys, silicate clouds and hydrocarbon-based hazes are often noted as plausible composition options, partly on the basis of models that solve jointly the gas and aerosol physics in the atmosphere (Helling et al., 2008; Lavvas and Koskinen, 2017; Powell et al. 2018; Gao et al., 2020). Looking at the problem in a different way, the detection of, for example, Ca, Mg, Si or Fe in the upper atmospheres of ultrahot Jupiters supports the idea that at high temperatures and low gravities, these elements are easily lifted rather than being locked into clouds deep down (Sing et al., 2019; Ehrenreich et al., 2020; Mikal-Evans et al., 2022).

The possibilities for aerosol compositions in the atmospheres of small, temperate planets are much more diverse (Lavvas et al., 2019; Gao et al., 2021). It may become challenging to characterize these planets with transmission spectroscopy if it is finally concluded that high-altitude clouds are prevalent as for GJ 1214 b (Gao et al., 2023). Cases like that of the relatively aerosol-free sub-Neptune K2-18 b, however, may be telling us that efficient aerosol formation does not occur for all star-planet combinations (Benneke et al., 2019; Tsiaras et al., 2019). Naturally, there is interest in building tools that forecast the occurrence of aerosols. To date, such efforts have focused on the modelling of the aerosols from basic principles (Helling, 2021), or on looking for trends in planet populations (Stevenson, 2016; Estrela et al., 2022). Both approaches may be used to prioritize observation targets, but neither of them is fail proof.

In reflected-starlight measurements, the signature of the aerosols is mainly dictated by their single scattering albedo $\varpi_a$, and the vertical thickness and horizontal extent of the particle distributions. For example, at the effective wavelength of the Kepler passband (~0.65 $\mu$m), silicates are highly reflective whereas iron condensates are dark. It is thus expected that planets with extensive aerosol layers dominated by these particles will be highly and lowly reflective, respectively. Interestingly, the optical phase-curve investigation of Kepler planets has shown that typical hot Jupiters are dark (Esteves et al., 2015), which suggests that aerosols in them are either made of particles with moderate-to-low $\varpi_a$ or that otherwise the aerosol layer lies too deep and the incident optical-wavelength photons become absorbed by gases (alkalis in particular). Exceptions exist; for example, Kepler-7b is quite reflective in the Kepler passband (Demory et al., 2013), a fact that may be aided by low gravity keeping the cloud particles aloft. Furthermore, the retrieval analysis of its optical phase curve indicates that the clouds are offset towards the morning terminator, and that the particles have $\varpi_a$~1 (suggestive of silicates, perovskite or silica) and are sub-micron in size (García Muñoz and Isaak, 2015). Figure 22 elaborates on the specifics of the retrieval analysis. The effect of aerosols will be critical in future direct imaging of exoplanets in reflected starlight. Most such targets will not transit their host stars and their radii will be unknown. As both the planet cross section $\propto R_p^2$ and the planet reflectivity contribute to the planet-to-star contrast, the measured brightness will only constrain their product. Observations at multiple phases might mitigate this degeneracy to some extent, but it is likely that it will not be entirely eliminated (Nayak et al., 2017; Carrión-González et al., 2021).

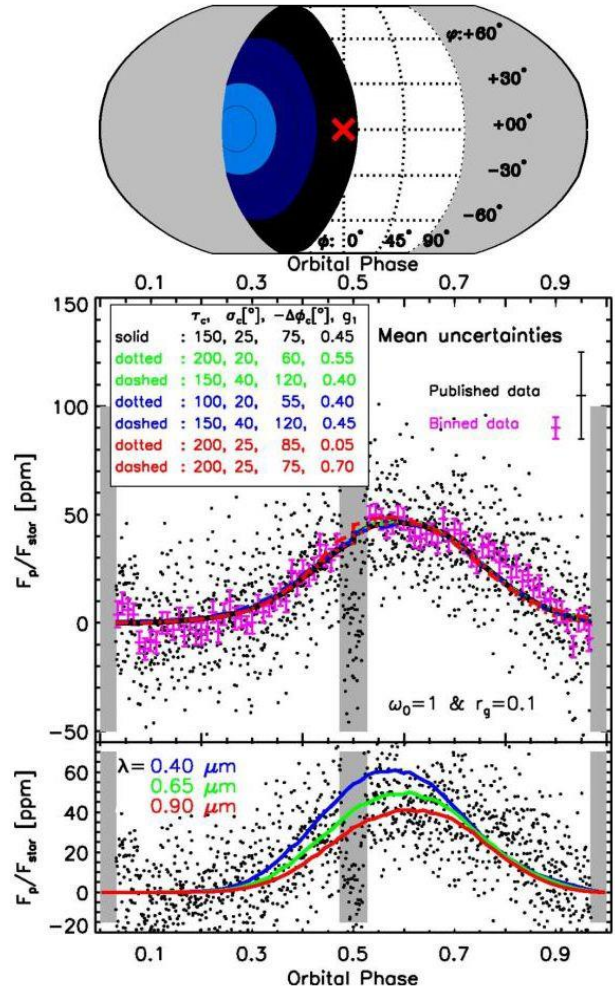

**Figure 22** *Aerosols and phase curves. Aerosols are prevalent in exoplanet atmospheres. Optical phase curves, when dominated by reflected starlight rather than by planet thermal emission, are very sensitive to the existence of clouds and their optical properties. These ideas have been tested for close-in exoplanets, and will be used in the future for characterizing directly-imaged planets in reflected starlight (Nayak et al., 2017; Carrión-González et al., 2021). Top. Cloud map inferred for the hot Jupiter Kepler-7 b based on a retrieval analysis (García Muñoz and Isaak, 2015). Middle. Kepler photometry of Kepler-7 b (black dots: original data; magenta dots: binned data, with uncertainty bars) (Demory et al., 2013), and best-fitting synthetic phase curves (dashed lines) based on the retrieval analysis. Althoug degeneracies exist in the interpretation of the measurements, the plausible solutions are consistent with extensive clouds towards the morning terminator made of highly reflective condensates. Bottom. Predicted phase curves at various wavelengths based on the optical properties inferred from the Kepler data. The scattering problem introduces a dependence on wavelength that could be tested if multi-wavelength observations become available.*

In models predicting the thermal emission from a planet, a pressure-temperature *p–T* profile may result in very different outputs if the model is allowed to develop an aerosol layer or not. This idea has been explored to show that the nightside temperatures measured for numerous hot Jupiters are probably affected by the formation of clouds and that their upper layer may lie at similarly low temperatures (Gao and Powell, 2021; Parmentier et al., 2021). As a consequence, nightside clouds may enhance the day-to-night contrast that is measured by IR phase curves.

The properties and distribution of the aerosols across the atmosphere are sensitive to condensation, and thus to the local $p$–$T$ profiles. In turn, aerosols are key at setting the $p$–$T$ profiles by controlling the amount of stellar and intrinsic energy that escapes to space and from where within the atmosphere that occurs. The aerosols also affect the atmospheric dynamics through radiative feedbacks, as demonstrated by 3D models (Roman and Rauscher, 2017, 2019; Steinrueck et al., 2023).

Low-mass planets with bulk densities as low as $0.1 \, \text{g cm}^{-3}$ have been called superpuffs. What keeps these planets bloated remains an open question, but it has been proposed that haze lifted to pressures of a few $\mu$bar by a massive outflow might be a cause (Lammer et al., 2016; Wang and Dai, 2019; Gao & Zhang 2020). Intriguingly, the transmission spectra available for these planets are featureless, which might indeed be a sign for the existence of high-altitude aerosols made of relatively large particles (Chachan et al., 2020). This idea could be tested by measuring the planet phase curve at UV-optical wavelengths near ingress/egress (García Muñoz and Cabrera, 2018), and assessing if it exhibits a Titan-like forward-scattering brightening (García Muñoz et al., 2017).

## 5.4 Temperature

Exoplanet atmospheres are heated from above by stellar radiation and from below by the internal heat released as the gravitational energy accumulated during accretion makes its way out. Stellar heating generally dominates over internal heating in the upper layers of the atmosphere for exoplanets that are close-in or sufficiently mature. Internal heating dominates for the far-out, young exoplanets that are directly imaged.

Predicting the temperature (or its generalization to multiple translational, rotational and vibrational temperatures in NLTE) requires solving the balance between heating and cooling at each location in the atmosphere. The problem is complex and in principle non-local, but it is often possible to divide it into smaller problems that are local and more easily tractable. The division is also relevant for the interpretation of observables, as discussed below.

### 5.4.1 Convective region. Radius (re)inflation

Convection is usually the main mechanism transporting energy in the deep atmosphere, and its dominance over other transport mechanisms defines the convective region. In addition to the release of gravitational energy, other processes that contribute to the internal heat flux through the deep atmosphere $F_{\text{int}}$ include Ohmic dissipation for planets with a magnetic field (Pu and Valencia, 2017), and dissipation of tides excited within planets on eccentric orbits (Jackson et al., 2008) or when the thermal inertia of the atmosphere causes a lag between the regions of maximum heating and temperature (Arras and Socrates, 2010). In models concerned with the energy balance in this region, $F_{int}$ is usually prescribed as a boundary condition (Hansen, 2008; Guillot, 2010).

Although the convective region has potentially observable implications, the advances in its understanding have been hampered by the difficulty of probing it directly. It is clear though that this understanding remains incomplete. Following the discovery of the first known transiting exoplanet, HD 209458 b, it is well known that hot and ultrahot Jupiters appear inflated with respect to the far-out planets with similar masses and ages. It is believed that the inflation originates in the convective region but the physical mechanism that causes it remains undetermined.

Figure 23 demonstrates the so-called radius inflation problem. A current view, built upon a large sample of hot Jupiters and the premise that $F_{int}$ is proportional to the incident stellar flux by a proportionality factor $\epsilon$, shows that $\epsilon\sim$0.03 for $T_{eq}\sim$1,500-1,800 K and that $\epsilon$ drops for lower and higher irradiations (Thorngren et al., 2019; Sarkis et al., 2021). In terms of the intrinsic temperature, defined by $F_{int}=\sigma_{SF}T_{int}^4$ ($\sigma_{SF}$ is the Stefan-Boltzmann constant), $T_{int}$ values as high as 700-800 K occur for some planets and inject a non-negligible flux of energy into the convection region. The same analyses show that Ohmic and thermal tide dissipation as well as potential temperature advection remain plausible mechanisms to cause the internal heating that keeps the planets inflated.

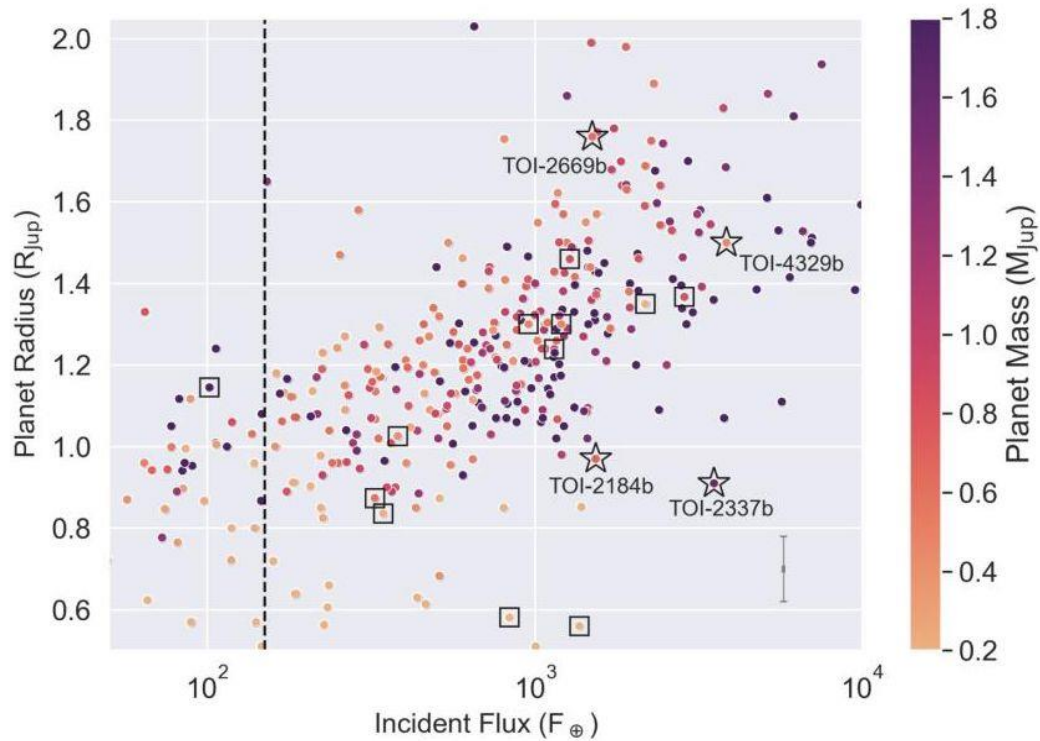

**Figure 23** *Exoplanet (re-)inflation. Strongly irradiated planets often appear larger than their less irradiated counterparts with otherwise equivalent properties. The trend is observed for planets orbiting Main Sequence stars, but there is also evidence that planets become reinflated as their host stars enter the Red Giant Branch and become increasingly luminous. This figure (borrowed from Grunblatt et al., 2022, their Fig. 11), shows the*

*increasing planet size at irradiations larger than the empirical threshold for inflation of ~150 Earth-level irradiations (dashed vertical line). Planets represented with symbols (squares or stars) orbit evolved stars. Interestingly, there are exceptions to this generally-valid trend, and for example TOI-2184b and TOI-2337b, both orbiting evolved stars, appear under-inflated with respect to what might be expected for the irradiation that they receive.*

Interestingly, exoplanets orbiting evolved stars are providing complementary insight into the radius inflation problem. Indeed, enhanced luminosity of the host star as it exits the Main Sequence and enters the Red Giant Branch should (re-)inflate a planet above its Main Sequence size provided the mechanism for radius inflation depends on the stellar irradiation. The idea has been partly confirmed by the discovery of a few inflated planets of masses comparable to and larger than that of Neptune orbiting evolved stars (Grunblatt et al., 2017, 2023). Ideally, these discoveries should motivate new explorations of the problem addressing in particular how much energy may be transferred to the convective region and how deeply that occurs (Komacek et al., 2020; Sainsbury-Martínez et al., 2023; Zhang et al., 2023).

The above ideas for the convective region in general and the radius inflation in particular, so far mostly investigated for gas giants, may have implications on smaller planets as well. Any unaccounted for heating mechanism that expands the atmosphere of sub-Neptunes will introduce additional degeneracies in the interpretation of their compositions (Pu and Valencia, 2017). This possibility and their implications have been explored by models, which show that for $\pi$ Men c ($M_p/M_\oplus$~4.5, $R_p/R_\oplus$~2.1), internal heat fluxes with $T_{int} \geq 100$ K puff up its atmosphere thus making it vulnerable to rapid mass loss (Nettelmann et al., 2011; García Muñoz et al., 2021).

**5.4.2 Radiative region**

On the dayside, radiation takes over from convection as the dominant mechanism for energy transport at a depth of ~1-1,000 bar that defines the Radiative-Convective Boundary. Radiative equilibrium and LTE define the atmospheric conditions within a few scale heights above that level. The planet IR emission originates mostly from this region, and measuring that radiation allows retrieval models to infer the *p–T* profiles.

Thermal inversions occur when the temperature increases in the direction of lower pressures. They typically develop when an efficient absorber exists whose opacity spectrum overlaps with the peak of stellar emission. For solar-type stars, this entails the absorber exhibiting strong bands in the optical. The TiO and VO molecules have been classical candidates to cause thermal inversions in the atmospheres of hot and ultrahot Jupiters (Hubeny et al., 2003), but their observational confirmation has remained problematic for some time. Recent searches using HRS may have finally detected TiO and VO at the ultrahot Jupiters WASP-189 b and -76 b, respectively (Prinoth et al., 2022; Pelletier et al., 2023).

The difficulty of detecting these molecules partly arises from incomplete spectroscopic linelists (de Regt et al., 2022), but newer cross section calculations are helping overcome that problem (McKemmish et al., 2019). Alternatively, gases such as the anion $H^-$, electronically excited H(2) or the metals Fe, Cr or Mg, which have rich absorption spectra and have been detected at a number of planets, or aerosols, might cause thermal inversions (Lothringer et al., 2018; García Muñoz and Schneider, 2019; Fossati et al., 2021).

The occurrence of thermal inversions has been firmly established in the last years by means of low- and high-resolution spectroscopy (Evans et al., 2017; Nugroho et al., 2021; Sheppard et al., 2017; van Sluijs et al., 2023; Yan et al., 2022, 2023; Cont et al., 2022). The evidence comes mainly from the detection at occultation of emission-like spectra of gases such as CO, $H_2O$, OH, Fe, Ti, Si and V. For some planets, the confirmation has been supported by observation of various gases with multiple telescopes. For the ultrahot Jupiter WASP-18b, these include CO, $H_2O$ and OH, which were detected with low and high-resolution spectroscopy obtained from space and the ground (Brogi et al., 2023; Coulombe et al., 2023; Yan et al., 2023). However, not every planet seems to develop a thermal inversion, and indeed others exhibit temperature profiles that decay with altitude or that are nearly constant over the region probed by the observations (Mikal-Evans et al., 2022). The reason is not entirely understood, but elucidating the role played in the heating and cooling of the atmosphere by the many neutral and ion atoms that have been detected at some ultrahot Jupiters might offer valuable clues.

If spectroscopy is the tool to probe in altitude the dayside temperature by inversion of the occultation thermal emission spectrum, phase curves are the tool to probe the temperatures horizontally, and combined spectroscopy-phase curves are the tool for 2D/3D-mapping. Until recently, HST has been the only telescope with the capability to produce spectrally-resolved IR phase curves of exoplanets (Stevenson et al., 2014; Mikal-Evans et al., 2022). The interpretation of these thermal emission phase curves follows the same principles as for an occultation spectrum. The key difference between the two approaches is that a phase curve reveals the combined radiation emanating from parts of the day- and nightsides that evolve as the planet moves on its orbit. With the aid of reconstruction techniques that operate on the brightness time-series, it has become possible to infer approximate 3D temperature maps from such measurements. The Spitzer telescope, as well as Kepler, CHEOPS and TESS that observe over a single optical-NIR photometric passband, have also provided valuable insight into the multidimensional temperature field of hot exoplanets (Mansfield et al., 2020; von Essen et al., 2021). Figure 24 illustrates some of these points for the ultrahot Jupiter WASP-121 b, which orbits a bright host star and has been a recurrent target of observations with multiple techniques and telescopes.

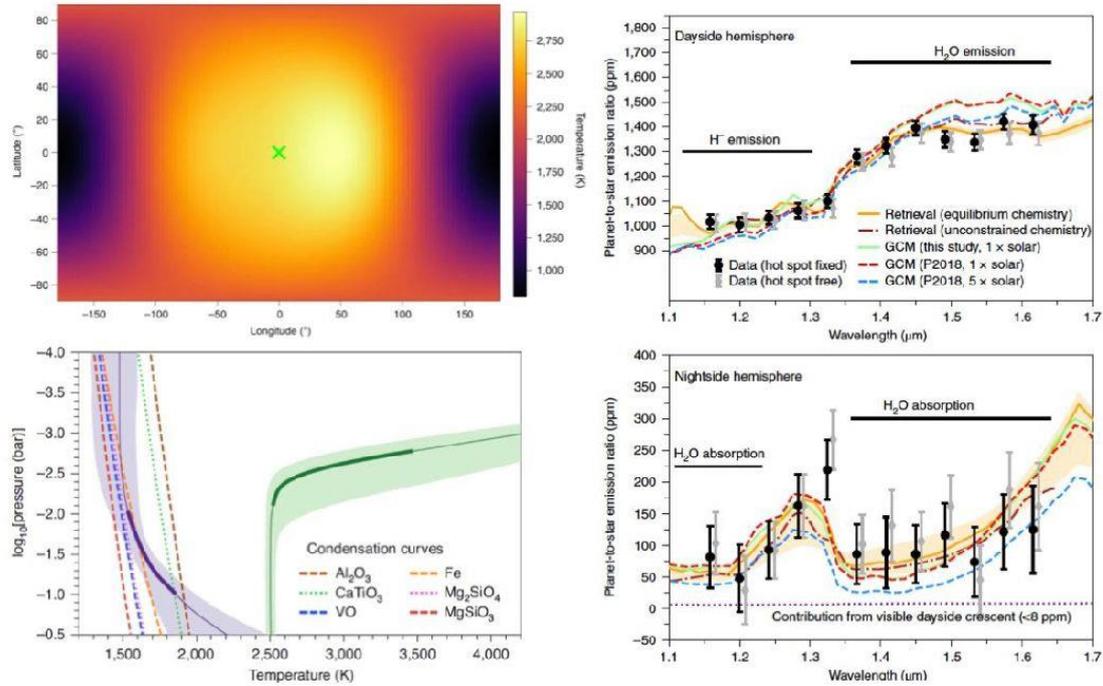

**Figure 24** *Thermal emission and 2D/3D mapping of atmospheres. The ultrahot Jupiter WASP-121 b has been targeted with multiple observing techniques to constrain its chemical composition, p–T profiles and high-altitude gas velocities. The panels here show the analysis of the spectroscopically resolved thermal phase curve measured with HST at wavelengths 1.1-1.7 µm (Mikal-Evans et al., 2022). The spectra show prominent bands of $H_2O$ (both day- and nightside) and the $H^-$ continuum (dayside) (right panels), that are used to retrieved hemispheric averages of the p–T profiles (bottom left). A variety of condensates can form on the planet nightside and they are likely to affect the brightness temperature on the nightside by shifting the emission level to the cloud top. The planet exhibits a hot-spot offset towards the evening terminator, which suggests strong super-rotation of the atmosphere.*

In steady state, the radiation emitted by the atmosphere must balance the stellar energy that is deposited plus the internal heat flux, as summarized in the equation $T_{eff}^4 = T_{eq}^4 + T_{int}^4$. $T_{eff}$ quantifies the planet thermal emission, which is in principle measurable if the IR spectrum can be determined over a broad enough range of wavelengths, as has been recently done for GJ 1214 b (Kempton et al., 2023). $T_{eq}$ is the equilibrium temperature, which depends on the Bond albedo $A_B$. This albedo quantifies the reflectance averaged over wavelength and the disk of the planet, and can be estimated if the distribution of aerosols and gases and their optical properties have been constrained. Estimates of $A_B$ can be found in the literature for a number of planets, but because they are based on a narrow range of wavelengths and phases or on indirect inferences from thermal emission measurements, they are probably tentative (Schwartz and Cowan, 2015). Both $T_{eff}$ and $T_{eq}$ can be estimated for an exoplanet if their optical-IR phase curves are measured, ideally with the help of models to fill in the gaps in the measurements. Estimating $T_{int}$ from first principles is challenging, but the trends inferred from statistical analyses might reasonably constrain it (Thorngren et al., 2019; Sarkis et al., 2021). Inversely, the mismatch between $T_{eff}$ and $T_{eq}$ might eventually reveal that the understanding of the processes contributing to the internal heat flux is incomplete.

Planets in the size range from terrestrial to sub-Neptunes with water-rich atmospheres (or in general with atmospheres rich in IR-active molecules), are susceptible to developing greenhouse conditions (Mousis et al., 2020; Aguichine et al., 2021; Pierrehumbert, 2023). The increased temperatures will significantly puff up their atmospheres, creating new challenges for the proper interpretation of the planet mass and radius measurements in terms of atmospheric composition.

### 5.4.3 Transition to the middle-upper atmosphere

The transition between the lower atmosphere (radiative region and below) and the middle-upper atmospheres often displays a variety of absorption features of, for example, Na, metals such as Fe and Mg, H I H-$\alpha$ (plus other lines of the Balmer series) and He whose wings and cores can occasionally be resolved spectroscopically. As the states participating in the lines and in the deposition and emission of radiation may be in NLTE, their populations and line opacities have to be calculated with collisional-radiative models (García Muñoz and Schneider, 2019; Fossati et al., 2021). This elaborate treatment is necessary to accurately predict the local $p$–$T$ profiles.

Heating by X-ray ($\lambda \leq 100$ Å) photons or cosmic rays has received less attention, and this has usually focused on H/He-dominated atmospheres (Cecchi-Pestellini et al., 2009; Shematovich et al., 2014; Locci et al., 2022; García Muñoz, 2023a). Especially for active stars, X-rays contribute significantly to the stellar XUV output, although this contribution remains orders of magnitude smaller than the visible-infrared output. Because the X-ray cross sections of gases are small, however, these photons penetrate well into the middle-upper atmosphere thus altering the ion-neutral balance. It would be interesting to assess the local heating by X-ray photons and cosmic rays for a range of atmospheric compositions.

### 5.5 Dynamics

An early prediction on atmospheric dynamics, confirmed subsequently by 3D General Circulation Models (GCMs) is that, subjected to strong irradiation, tidally-locked gas giants develop an eastward equatorial jet that reaches velocities of a few km/s in the direction of the planet-body rotation (Showman and Guillot, 2002; Showman et al., 2020). The specifics of the flow depend on various dimensionless numbers, in particular on the ratio $\tau_{rad}/\tau_{adv}$ between the radiative and advective times at the ~1-100 bar depth where the jet probably forms (Zhang, 2020). If this ratio >>1, the jet will efficiently redistribute the stellar energy between the day- and nightsides, shifting the hot spot towards the evening terminator. Inversely, energy redistribution by advection will become inefficient if the ratio <<1, in which case local irradiation largely dictates the thermal structure of the atmosphere. Radiation becomes efficient at high temperatures, and the expectation is that for very strongly irradiated exoplanets $\tau_{rad}/\tau_{adv}$<<1.

Two main approaches have been taken to test observationally these and other ideas on atmospheric dynamics. One utilizes the IR phase curves of the planets, connecting their shapes to the redistribution of energy by winds. The other is based on measuring one or more absorption lines in the transmission spectrum, and interpreting the shifts and widths of the lines in terms of physically-motivated wind prescriptions.

The same IR phase curves obtained with HST, Spitzer or TESS and used for mapping the planet temperatures have proven valuable for investigating atmospheric dynamics (Kreidberg et al., 2018; Mansfield et al., 2020; Wong et al., 2020; von Essen et al., 2021; Mikal-Evans et al., 2022). Important factors in the theory-observation comparison are the contrast between the maximum and minimum brightness and the hot-spot offset. Both quantities may depend on wavelength, which partly determines the depth in the atmosphere where the IR radiation originates. From the physical standpoint, the interpretation of the IR phase curves involves processes such as 3D winds capable of transporting heat and mass, condensation of gases on the nightside and the reverse process of vaporization on the dayside, and the release/expense of chemical energy in the process $H+H+M \leftrightarrow H_2+M$ (Bell and Cowan, 2018).

The theory-observation comparison of IR phase curves for gas giants have not been entirely satisfactory. For example it is unclear whether the phase curve contrast or the hot-spot offset correlate with the equilibrium temperatures (Parmentier et al., 2018). A systematic investigation of Spitzer phase curves suggests some trends in that the hotspot offset depends on the orbital period and gravity of the planets (May et al., 2022). These ideas could be further validated with measurements for additional planets. The anomalous case of CoRoT-2b (Dang et al., 2018) shows that some planets develop hot spots with westward offsets, and thus that there may be multiple competing processes in the physics behind the thermal phase curves. The investigation of a sample of IR phase curves, treated uniformly to minimize biases from the application of different data analysis techniques and from possibly different observatories, should help elucidate fundamental trends and suggest new avenues for exploration of the physics behind the measurements.

The deficiencies in past interpretation of IR phase curves have motivated the modelling of processes connected to clouds, transport of chemical energy, magnetic forces, asynchronous rotation, or the way the flow in the layers deeper than ~100 bar alters the jet strength and direction, whose implications may have been previously underappreciated (Rauscher and Kempton, 2014; Bell and Cowan, 2018; Hindle et al., 2019; Carone et al., 2020; Parmentier et al., 2021; Roman et al., 2021). Planets such as WASP-121 b, WASP-43 b or GJ 1214 b, for which high-quality phase curves exist or will be available soon are receiving particular attention (Christie et al., 2022).

Similar ideas have been explored for the ultrahot super-Earth 55 Cnc e. The Spitzer phase curve at 4.5 $\mu$m suggests an eastward hot-spot offset (Demory et al., 2016). Using a retrieval model that implements some basic physical constraints, it has been argued that these measurements are consistent with a thick envelope and $\tau_{rad}/\tau_{adv} \sim 1$ (Angelo and Hu, 2017). The high velocities required to meet this condition suggest that the envelope is made of a gas of high mean molecular weight rather than of magma. A new reduction of the Spitzer data shows however no hot-spot offset (Mercier et al., 2022), a finding that brings 55 Cnc e more in line with what has been reported for other ultrahot rocky exoplanets, and that may indicate that in reality the planet lacks an atmosphere. This example serves as a reminder of how much our understanding of exoplanet atmospheres depends on the basic processing of the telescope data.

Only weak constraints on the winds can be set when the phase curve is dominated by reflected starlight. For example, the optical phase curve of Kepler-7b has been accurately determined (see Fig. 3, and the discussion in section 1.3), and Spitzer occultations at 3.6 and 4.5 $\mu$m have confirmed that thermal emission contributes negligibly to the Kepler measurements (Demory et al., 2013). It has been proposed that the interpretation of the optical data invoking an eastward jet also requires the existence of thick clouds at the morning terminator, a finding supported by a retrieval analysis (García Muñoz and Isaak, 2015). In that interpretation, $\tau_{rad}/\tau_{adv}$>>1, which simply states that radiative cooling is relatively inefficient but does not say much about the winds (Hu et al., 2015).

The atmospheric dynamics at pressures of ~0.1-0.001 mbar has been less explored with 3D models (Showman et al., 2008), and these often omit key physical processes classically ascribed to thermospheres (Koskinen et al., 2007). At these altitudes the effect of stellar irradiation (including UV wavelengths) is particularly important, and the atmosphere may develop an axisymmetric circulation such that the gas rises at the substellar point, travels from the day- to the nightside and finally sinks at the antistellar point. Horizontal velocities ~1-10 km/s at the terminator have been predicted for such a day-to-night circulation (Showman et al., 2008; Hammond and Lewis, 2021).

At even lower pressures, the gas may flow outwards, ultimately escaping the planet. Massive outflows have been identified at a few close-in exoplanets by the extended signature that they imprint in atomic lines, for example in H I Lyman-$\alpha$ and H-$\alpha$, He I 10,830Å or C II 1,335Å (Vidal-Madjar et al., 2003; Spake et al., 2018; Yan and Henning, 2018; García Muñoz et al., 2021; Ben-Jaffel et al., 2022). We omit an in-detail discussion of outflows (Owen, 2019; Dos Santos, 2022).

The transitions between these three regimes (super-rotating flow in the lower atmosphere, day-to-night circulation at higher altitudes, outflow at very high altitudes) are surely complex, and the real flow likely exhibits all three components simultaneously. This complexity must be kept in mind for the interpretation of measurements.

In the last decade, HRS has become a powerful method to probe the atmospheric dynamics of exoplanets. The method is complementary to that of IR phase curves, and arguably more directly connected to the actual wind patterns. Early work revealed the signature of the CO molecule in the transmission spectrum of hot Jupiter HD 209458 b (Snellen et al., 2010). The signal was found with a blue shift of ~2 km/s with respect to the planet-body motion, which suggested day-to-night winds at the terminators of about that magnitude. CO is a suitable dynamical tracer because, according to the photochemical models (García Muñoz, 2007; Venot et al., 2019), it is stable and abundant in hot, solar-metallicity atmospheres. Ultrastable, high-resolution spectrographs of large spectral coverage are now available at many large- and intermediate-size telescopes, and similar measurements have been conducted for other exoplanets, targeting additional atoms and molecules.

Ultrahot Jupiters such as HAT-P-70b, KELT-9b, MASCARA-4b, WASP-33b, 76b, -189b, and -121b, with equilibrium temperatures from 2,000 K to 4,500 K, have been recurrent targets for wind characterization with HRS (Bello-Arufe et al., 2022; Borsa et al., 2021a,b; Cauley et al., 2021; Ehrenreich et al., 2020; Prinoth et al., 2022; Sánchez-López et al., 2022; Seidel et al., 2023; Zhang et al., 2022). The reasons are the same as those that motivate the search for atoms and molecules in their atmospheres in the first place. The list of targeted atoms for velocity measurements is very similar to the list presented in section 1.2. In addition to CO, the list includes the OH, TiO and VO molecules.

At high resolution, a transmission spectrum is affected by: *1)* Thermal and pressure broadening, introduced in the motion of the gases as dictated by the kinetic theory of gases and by collisions with other particles; *2)* Planetary rotation, which generally results in line broadening by a predictable amount if the planet is assumed to be tidally locked; *3)* Super-rotation of the atmosphere, which generally results in line broadening by an amount on the order of the jet velocity; *4)* Day-to-night circulation, which shifts the observed line position towards shorter wavelengths (blue shifts); *5)* An outflow, which broadens the observed line if the gas is accelerated in similar ways on the day- and nightsides; *6)* Thermal, chemical and dynamical gradients near the terminator, which may break any symmetry in *3)-5)*, especially when the planet is observed near ingress or egress. These ideas have been previously described (Cauley et al., 2021; Keles, 2021; Dos Santos, 2022; Prinoth et al., 2022; Seidel et al., 2023).

Extracting wind information from a high-resolution transmission spectrum represents an inversion exercise that uses the measured line shifts and widths (or more generally, the line shapes) as input. New retrieval models are being developed to approach the problem in a systematic way, deciding on the best-fitting wind patterns amongst a set of options, and quantifying the wind velocities and their uncertainties (Seidel et al., 2021; Dos Santos, 2022). As usual, degeneracies exist and multiple wind patterns may in principle lead to the same effect on the spectrum. Ideally, the retrieval models should incorporate the findings from GCMs or hydrodynamical escape models to set physically-motivated priors that help decide on the preferred interpretation. The retrieval models may work on the transmission spectrum averaged over the entire transit or alternatively on the time-resolved spectra. The latter strategy is advantageous, provided the signal-to-noise ratio (SNR) is high, because some of the predictions in *1)-6)* evolve as the planet transits its host star, and the time-resolved spectra may help break some of the degeneracies (Gandhi et al., 2023; Seidel et al., 2023).

The cases of the ultrahot Jupiters WASP-121 b and KELT-9 b are particularly interesting. The mid-transit transmission spectrum of WASP-121 b shows the Na I doublet near 5,890Å with no preferred blue (or red) shift (Borsa et al., 2021a). As the planet moves towards egress, however, the absorption signal splits up and the spectrum reveals a second doublet blue-shifted by ~30-40 km/s. It has been proposed that an outflow and velocities at the evening terminator of ~30 km/s cause the additional doublet (Seidel et al., 2023), which is intriguing as the velocities predicted by models are usually smaller. Further measurements over the full transit, including ingress, will hopefully shed new light on the origin of the additional Na I doublet. KELT-9b is a truly extreme ultrahot Jupiter and an almost ideal target for HRS. The absorption lines in the H I Balmer series that connect the principal quantum numbers 2→≥3 have been convincingly measured, and show no discernible blue (or red) shift. The measurements are in principle consistent with the outflow predicted by models (García Muñoz and Schneider, 2019). Intriguingly, the detection of the H I Paschen $\beta$ (3→5) line in its atmosphere is blue-shifted by ~15 km/s (Sánchez-López et al., 2022). The modelling of the H I atom suggests that the Paschen lines form deeper in the atmosphere than the Balmer lines, hinting at a change in atmospheric dynamics between the layers probed by the Balmer and Paschen series. Further measurements of this and other Paschen lines must be encouraged to consolidate these ideas.

The He I line at 10,830Å has also been used as a tracer of winds in the middle-upper atmosphere. The He I line detections have been preferentially reported at planets that are not too hot, so it provides an approach complementary to the measurements using the H I Balmer lines or other metals in ultrahot atmospheres. Some of the He I line detections exhibit a blue shift of a few km/s that suggests a combination of an outflow and a day-to-night circulation (Dos Santos, 2022). The models predict that the depth and shape of the absorption He I lines are affected by the interaction of the atmosphere with the stellar wind and by the perturbing effect of planetary magnetic field lines acting on the flow. According to these models, blue shifts are preferentially found when a strong stellar wind pushes towards the planet nightside the atmospheric gas, and when the magnetic field at the planet is sufficiently weak that the gas dynamics is not constrained by the field lines (MacLeod and Oklopčić, 2022; Schreyer et al., 2023). It is unclear how deeply in the atmosphere the day-to-night motion is initiated, but the evidence provided by the He I line represents a valuable starting point for model simulations that connect with the lower atmosphere.

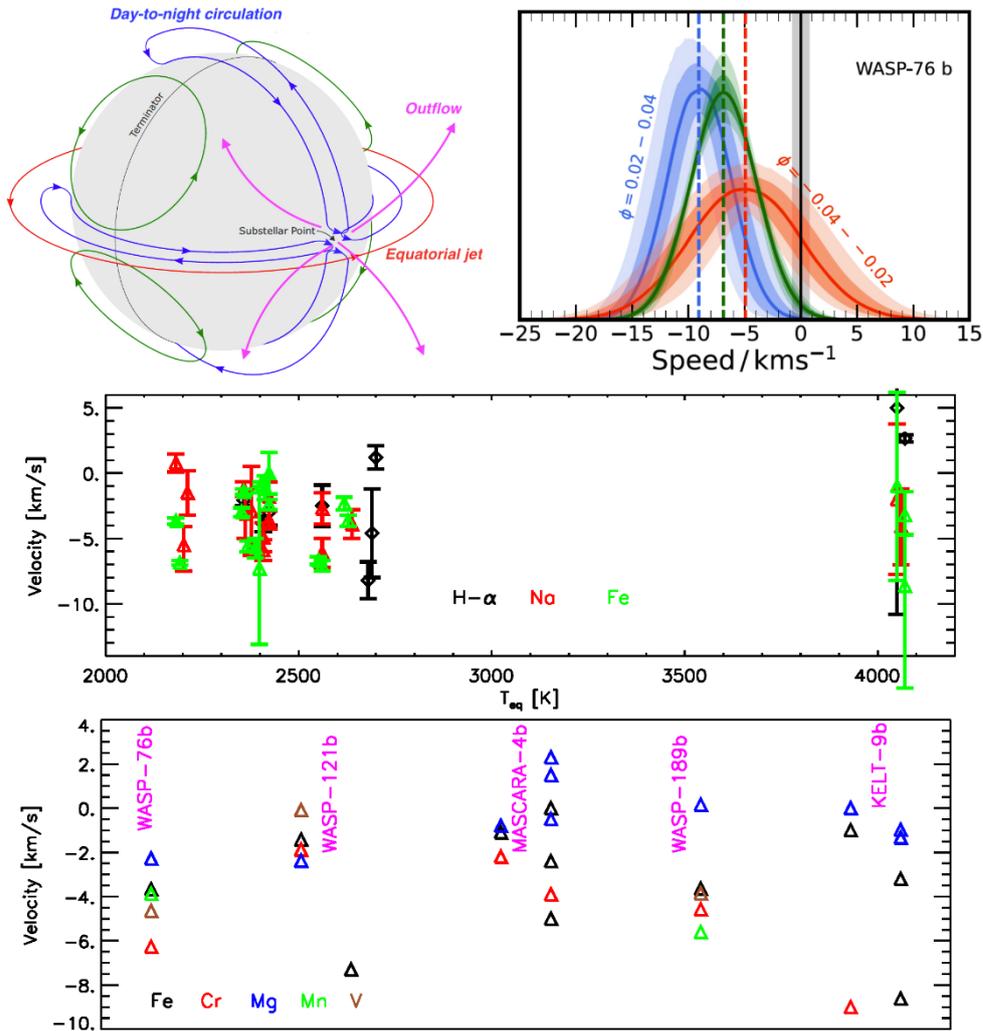

**Figure 25** <u>*Top. Left.*</u> *The atmospheric dynamics of exoplanets is often described in terms of three main motions, namely: an equatorial jet, a day-to-night circulation and an outflow. (Figure adapted from Hammond and Lewis (2021).) Right. The absorption line of a transmission spectrum encodes information on the atmospheric winds. In particular, a blue shift in the line position with respect to its wavelength at rest is interpretable as the result of a day-to-night circulation. Correspondingly, the broadening of the absorption line is interpretable as resulting from an equatorial jet or an outflow from both the day- and nightsides. The position and width of the absorption line evolve during the transit (tagged here by the orbital phase φ), which provides additional insight into type of atmospheric motion. (Figure from Gandhi et al. (2023).) <u>Middle.</u> Velocity shifts in absorption lines of H-α, Na and Fe reported for a number of ultrahot Jupiters (WASP-76b, MASCARA-2b/KELT-20b, WASP-121b, MASCARA-4b, TOI-1518b, HAT-P-70b, WASP-189b, WASP-33b, KELT-9b). Under the usual sign convention, negative velocities occur for blue-shifted lines, when the gas moves away from the star and towards the observer. Simple inspection of the data on display does not suggest a simple dependence of the velocities with the planet equilibrium temperature. <u>Bottom.</u> Velocity shifts measured for different heavy elements for a few planets with high-quality data. Uncertainty bars are omitted for clarity. Four out of the five planets exhibit blueshifts ordered as Mg>Fe>Cr, i.e. coincidentally or not, they follow the same order as their condensation temperatures.*

As part of this review, we have looked at the day-to-night velocities measured for a few atomic lines at all the ultrahot Jupiters for which we have found such measurements. Figure 25 (middle) summarizes the data for H-$\alpha$, Na and Fe. H-$\alpha$ is often interpreted as a tracer of the upper layers of the atmosphere of ultrahot Jupiters (García Muñoz and Schneider, 2019; García Muñoz, 2023a; Czesla et al., 2022). The resonance line of Na at optical wavelengths has been detected in a variety of hot and ultrahot Jupiter atmospheres. The Fe atom is a conspicuous feature in the transmission and emission spectra of ultrahot Jupiters. The data reveal that all these lines are typically blue-shifted, but there is no obvious trend between the blue shifts and the equilibrium temperature of the planets. The error bars are large, however, and there are obvious inconsistencies between the blue shifts inferred with different lines of the same element (e.g. Fe), and between the velocities inferred for the same planet by different analyses or groups. These discrepancies must be overcome to determine whether a trend exists and to possibly guide future theoretical efforts. Figure 24 (bottom) extends the analysis to other heavy elements with the aim of assessing if the blue shifts inferred for them follow an established order that is consistent between exoplanets. Interestingly, for 4 out of the 5 planets in the sample, the velocities $v$(Mg)≥$v$(Fe) ≥$v$(Cr), which is consistent with the condensation temperatures $T_C$(Mg) ≥$T_C$(Fe)≥$T_C$(Cr) (Ladders, 2003). If the trend is real, it would suggest that the measured velocities are representative of different temperatures across the day-to-night terminator. Again, robust measurements with reduced uncertainties are needed to fully exploit the information encoded in the measurements.

Vertical motions in the atmosphere can carry aerosols and heavy elements to high altitudes. There are currently no direct constraints on these motions, but the models show that they can have a significant impact on various atmospheric features. For example, strong turbulent and large-scale motions tend to keep aerosols aloft, thus enhancing their muting effect on any gas spectral feature (Wang and Dai, 2019; Chachan et al., 2020; Gao & Zhang, 2020). These motions also keep aloft any heavy elements that would otherwise settle down under the effect of gravity, thus making them accessible to the usual observational techniques. The fractionation of gases of different atomic or molecular weight occurs above the gas-dependent critical level defined by the homopause. A homopause may not exist when the planet is undergoing a massive outflow, and the fractionation by mass will be weak (García Muñoz et al., 2021). Fractionation may however occur, and the relative abundances of heavy elements in the atmosphere will vary over time, when the mass loss rate is weak or moderate and a homopause forms. There is a need to investigate the implications of such a condition on the evolving composition of small planets.

There are currently no observational constraints on the atmospheric dynamics of small planets with sizes ≤2$R_\oplus$. There is though a substantial amount of theoretical work that addresses questions concerned with the transport of energy and mass from the day-to the nightsides of these planets, and that might have implications on their habitability, for example (Wordsworth and Kreidberg, 2022).

## 5.6 The present and future

The start of scientific operations of JWST in mid-2022 is setting new standards in the investigation of exoplanet atmospheres. This statement is based on the 6.5-m diameter of its primary mirror and sensitivity to 0.6-28-$\mu$m wavelengths, the availability of imaging, spectroscopy and coronography, and a privileged L2 vantage point. It thus seems appropriate to focus on JWST for an overview of the present and immediate future of the field.

Two observing programs dedicated to exoplanets, one focused on transiting planets and one for high-contrast imaging (Bean et al., 2018; Hinkley et al., 2022), were selected as part of the Early Release Science (ERS) program conceived for the community to assess as quickly as possible the performance of the telescope and its instruments. Both programs are now completed and some results have been reported. A key conclusion from the ERS programs, and from other programs executed as part of Cycle 1, is that the telescope and instruments are performing as expected or better, with minimal effects of systematic noise and overall SNRs that are photon-limited.

The ERS program for transiting exoplanets has targeted three hot or ultrahot Jupiters. They were selected to verify the consistency between the data obtained with various instruments (WASP-39 b) and explore the instrumental responses to a bright host star (WASP-18 b) and a long-duration observation (WASP-43 b). WASP-39b ($M_p/M_J$=0.28; $R_p/R_J$=1.27; $T_{eq}$=1,100K) was observed with transmission spectroscopy and three different instruments spanning jointly 0.5-5.5 $\mu$m in wavelength (JWST Transiting Exoplanet Community ERS Team et al., 2023; Ahrer et al., 2023; Alderson et al., 2023; Feinstein et al., 2023; Grant et al., 2023; Rustamkulov et al., 2023). The spectrum (Fig. 1 top panel) shows unambiguous bands of $CO_2$, CO, $H_2O$, $SO_2$, Na and K. The occurrence of $SO_2$, rather than $H_2S$, is likely the result of photochemistry at pressures ≤1 mbar that involves various S-bearing gases and the OH radical (Tsai et al., 2023). The advantage of such a rich spectrum is evident for constraining the C/O and K/O ratios. The first analyses suggest a supersolar metallicity, a slightly subsolar C/O ratio, and a solar-to-supersolar K/O ratio. The occultation data (0.85-2.85 $\mu$m) for WASP-18 b show evidence of $H_2O$, and are consistent overall with a thermal inversion at the 0.01-1 bar level (Coulombe et al., 2023). Because the WASP-18b observation extends well into the pre- and post-occultation phases, a temperature map of the planet dayside could be inferred suggesting that the temperatures decay symmetrically with respect to the substellar point. It appears that the metallicity of WASP-18 b is approximately solar, and because there are no strong features of C-bearing molecules in the spectral region, only the soft constraint C/O<1 is derived. At the time of writing, a scientific analysis of the thermal phase curve (5-14 $\mu$m) of WASP-43 b has not been published yet. A preliminary analysis of the data confirms the long-term (~24 hr) stability of the telescope and reports on recommendations for future observations (Bell et al., 2023).

The ERS program for the high-contrast imaging of exoplanets was granted 55 hours of observing time to test all the instruments aboard the telescope in configurations relevant to the direct imaging of exoplanets, substellar companions and circumstellar disks. The coronagraphic data collected for the super-Jupiter HIP 65426 b with two different instruments (covering 2-5 and 11-16 $\mu$m each) reveal that the sensitivity achieved is superior to the expected performance (Carter et al., 2022). In the future, this should enable a variety of investigations on the atmospheres of exoplanets and substellar objects and their variability (Biller, 2017) in the way that has been demonstrated for the young, planetary-mass companion object VHS 1256 b (Miles et al., 2023). Spectroscopy of this object between 1 and 20 $\mu$m shows the presence in its atmosphere of $H_2O$, $CH_4$, CO, $CO_2$, Na and K, and significant temporal variability possibly connected to disequilibrium chemistry, turbulent mixing and clouds.

The JWST Cycle 1 observations are now well under way and breaking new ground into many aspects of exoplanet atmospheres. To keep the review concise, only a few examples specific to small exoplanets are noted. Given the pace at which new findings are being reported, it is certain that many more examples will soon appear. The examples mentioned here benefit from the enhanced planet-to-star contrasts that occur when the host star is a cool M dwarf star, and that enable the characterization of planets down to the Earth size. They also demonstrate the challenge of determining whether the planet has retained an atmosphere over its history, and of identifying any conclusive spectral features in it, especially under the effects of stellar contamination.

The transmission spectrum of GJ 486 b ($M_p/M_\oplus$~3.0; $R_p/R_\oplus$~1.3; $T_{eq}$~700 K) shows no clear spectroscopic feature at 2.8–5.2 $\mu$m (Moran et al., 2023), but it hints at a mild slope shortwards of 3.7 $\mu$m possibly attributable to $H_2O$ (Fig. 20 middle panel). The JWST data, combined with previous constraints using HRS (Ridden-Harper et al., 2023), suggest that the atmosphere of GJ 486 b might be water-rich or, alternatively, that the tentative slope at short wavelengths is due to water in cool unocculted starspots and thus not in the planet atmosphere. Observations at optical-NIR wavelengths might distinguish between these two proposed scenarios.

LHS 475 b (unconstrained mass; $R_p/R_\oplus$~0.99; $T_{eq}$~590 K) has also been observed with transmission spectroscopy at 2.8–5.2 $\mu$m (Lustig-Yaeger et al., 2023). Its transmission spectrum is essentially flat and can be used only to rule out a few scenarios such as H/He-dominated or cloudless $CH_4$ atmospheres. The data remain consistent to various extents with high-metallicity atmospheres that have high-altitude aerosols, and also with a planet that lacks a substantial atmosphere.

The spectroscopic thermal phase curve and transmission spectrum of GJ 1214 b ($M_p/M_\oplus$~8.2; $R_p/R_\oplus$~2.6; $T_{eq}$~600 K) have been measured at 5–12 $\mu$m (Kempton et al., 2023). The MIR data confirm past reports of flat transmission spectra in the NIR (Kreidberg et al., 2014). The phase curve shows distinct modulation due to different dayside and nightside temperatures (~555 and 440 K, respectively), and some absorption likely due to $H_2O$ in the corresponding spectra. Because the MIR brackets most of the thermal emission radiated by the planet, the phase curve data constrain the Bond albedo of the planet, and it is estimated that ~50% of the incident stellar radiation is re-radiated away. The view of GJ 1214 b that emerges is that of a H/He-dominated atmosphere with very high metallicity or even a water-dominated atmosphere. These scenarios seem also to require some muting of the gas spectral features by high-altitude, high-reflectivity aerosols (Gao et al., 2023).

The photometric occultation of TRAPPIST-1 b ($M_p/M_\oplus$~1.4; $R_p/R_\oplus$~1.2; $T_{eq}$~400 K) has been measured at an effective wavelength ~14.8 $\mu m$ (Greene et al., 2023). This is the innermost of the 7 planets in the TRAPPIST-1 system (Gillon et al., 2017), and probably the most vulnerable to losing its atmosphere. The occultation data indicate a dayside temperature of ~500 K that matches the theoretical prediction for a body with very low reflectance and inefficient energy redistribution between the day- and nightsides. Taken at face value, the data are consistent with a planet that has a thin atmosphere or that lost it in the past. Future, higher SNR observations should be able to test this scenario and, if proven correct, constrain spectroscopically the surface composition.

Future developments in the characterization of exoplanet atmospheres will rely on new space missions that are currently being conceived. In parallel, observations from the ground with a new generation of telescopes and instruments will create new opportunities for advancing the field.

ARIEL, the Atmospheric Remote-sensing Infrared Exoplanet Large-survey, is scheduled for launch in 2029 as ESA's 4th M-class mission (Tinetti et al., 2018, 2021). It will perform photometry in 3 bands (0.5-0.6, 0.6-0.8, 0.9-1.1 $\mu m$), and low-to-moderate-resolution IR spectroscopy from 1 to 8 $\mu m$. This spectral region is rich in rovibrational bands of molecules expected to be abundant in warm and hot atmospheres. As the first space mission dedicated to the investigation of exoplanet atmospheres, ARIEL aims to investigate the composition of ~1,000 exoplanets and, in turn, shed new light on the processes through which they form and evolve. The observations will focus on transiting exoplanets, and there are plans to produce transmission, occultation and phase curve spectra. The ARIEL observations will follow a 4-tier system, with different tiers establishing different requirements in SNR and observing time. Tiers 3-4 are reserved for planets with properties that justify a major investment of observing time.

There are three ground-based telescopes with mirror diameters larger than 20 m currently planned for or already under construction, namely the European Extremely Large Telescope (E-ELT), the Giant Magellan Telescope and the Thirty Meter Telescope. On the European side, the E-ELT (https://elt.eso.org/) is expected to see first light towards the end of this decade. Amongst the first and second generation of instruments that have been defined, there is a clear emphasis on the characterization of exoplanet atmospheres. Some of these instruments will allow for high-contrast spectroscopy of gas giants and sub-Neptunes and ultra-stable spectroscopy of exoplanet atmospheres down to the super-Earth and Earth sizes.

A main driver for exoplanet science is the search for atmospheric signatures possibly indicative of life at a few Habitable Zone (HZ) exoplanets. The HZ is loosely defined as the circumstellar region where the conditions are such that water may remain in the liquid phase at the planet surface. The probability for a planet to transit its host star is about the ratio between the stellar radius and the planet orbital distance. Its value is very small in any practical case, and entails that the vast majority of existing HZ planets will never be seen in transit. Even for those HZ planets that do transit their stars, the phenomenon of atmospheric refraction may prevent the sounding of the densest atmospheric layers. For an Earth-Sun twin system, atmospheric refraction prevents the access to the altitudes below ~12 km and thus to the troposphere (García Muñoz et al., 2012), regardless of the cloud conditions at the planet. For these and other reasons, it has been estimated that the detection with transmission spectroscopy of Earth-like levels of $O_2$, a plausible indicator of life as we know it on our own Earth, will take a few decades worth of observations even with the largest ground-based telescopes that are currently planned (López-Morales et al., 2019; Hardegree-Ullman et al., 2023).

A systematic search for the signatures of life in exoplanet atmospheres may require a new generation of telescopes with direct imaging capabilities, possibly in space. With regards to their prospective spectral coverage, two main options have been considered. They either focus on short wavelengths at which the planet signal is dominated by reflected starlight or on long wavelengths that target the thermal emission from the planet. There are plans in the international community to develop both technologies.

NASA's Nancy Grace Roman Space Telescope is planned for launch within the current decade and carries a Coronagraph Instrument that will observe in the optical (Spergel et al., 2015). The Roman telescope will serve as a Technology Demonstrator, but that may nevertheless result in the atmospheric characterization of a few gas giants (Carrión-González et al., 2021). Towards the 2040s, the Habitable Worlds Observatory (HWO) should image a variety of HZ exoplanets, including terrestrial ones, in our vicinity, for which it will have to achieve a planet-to-star contrast of ~$10^{-10}$. The HWO is under consideration by NASA as its next flagship mission after being recommended by the US Astro2020 Decadal Survey. The HWO specifics remain undecided, but it will likely be equipped with a 6-m-diameter primary mirror and will perform spectroscopic measurements from the UV to the NIR, with possibly the capacity for polarimetry. Besides its potential for characterizing stellar magnetic fields, polarimetry is a powerful technique to constrain the aerosol properties of planetary atmospheres (García Muñoz, 2018; West et al., 2022). The HWO access to UV wavelengths, which has been demonstrated to be critical for a variety of exoplanet questions, will ensure the continuity of UV spectroscopy beyond the HST times.

ESA's Voyage 2050 initiative set as a priority the development of the technology that should eventually lead to the characterization of a sample of low-mass, temperate exoplanets by measuring their thermal emission. A concept for such a technology has been proposed (https://life-space-mission.com/), which consists in performing formation-flying nulling interferometry from space in the MIR (Quanz et al., 2022). Observing at those wavelengths will yield access to a variety of molecules, some of which, such as $H_2O$, $CO_2$ and $O_3$, are often considered in biosignature studies, and should also enable the determination of the $p$–$T$ profiles of the targeted planets, thus establishing valuable constraints on their energy budgets.

Ideally, in a not-too-distant future, both concepts for high-contrast observations based on reflected starlight and planet thermal emission will become a reality. Their complementarity will pave the way for probing the atmospheres of Earth-sized planets in ways that cannot be achieved by each approach separately, while offering unprecedented detail into a variety of aspects concerning the formation and evolution of planetary atmospheres.

## 6. Conclusions

The investigation of the massive atmospheres of the solar system bodies with ground-based telescopes and with HST and JWST will be the only ones available, in the absence of space missions in the near future, in the case of Saturn, Uranus and Neptune. In the case of Venus, Mars, Jupiter and Titan they will complement those carried out in situ and from the orbit around the body according to ongoing or planned space missions. Visiting the other planets in our Solar System remains essential to expand our knowledge of their atmospheres. This statement is obvious when the technique involves taking a sample of the atmosphere, as in mass spectrometry measurements. Even when the atmosphere is sensed remotely, making the measurements from a near-orbit vantage point gives access to temporal and spatial resolutions and viewing geometries that may not be accessible with Earh-bound telescopes. We have described some of the targets on the basis of high spatial resolution imaging (at selected wavelengths) and high spectral resolution spectroscopy with spatial resolution in the planet's disc, covering a wide wavelength range from the ultraviolet to the thermal infrared. We summarise in what follows some of these objectives.

For Venus, precise determination of atmospheric abundances, including phosphine and the $SO_2$ temporal variability (and possible relation to volcanic activity), explore the nature of the UV aerosol absorber and its variability, determine the long-term variability of the wind velocity, including tides contribution, and studies of the variety of wave phenomena at different altitudes and cloud levels. For Mars, measurements of trace species such as $H_2O_2$ and in particular gaseous $CH_4$, wind speeds from Doppler spectroscopy, cloud and dust distribution, including large-scale dynamical instabilities and planet encircling or Global Dust Storms. For Jupiter, establish periodicities and possible cycles of the long-term changes in the albedo and opacity of belts and zones, the vertical structure and properties of clouds and hazes in different places, including colour variability of long-lived anticyclones BA and GRS. Studies of the outbreaks of major planetary-scale disturbances, as the SEBD and NTBD, the changes their produce in aerosol properties, chemical abundances (in particular search for $NH_3$ and $H_2O$ ices to constrain vertical transport), and in their dynamical properties (temperatures, winds). Monitoring the impact rates of bolides (objects with a size approximately $\leq 40$ m) and survey of large impacts producing atmospheric debris bolides (objects with a size approximately $\geq 40$ m). For Saturn, monitor the potential development of Great White Spots, in particular in the southern hemisphere where they have never been observed, and in general of the degree of stormy convective activity. It would be also new to observe in the southern hemisphere, the development of planetary-scale waves like "the ribbon" and "hexagon". Look for the formation mechanism of large-scale and long-lived cyclones and anticyclones. In the equatorial area of both Jupiter and Saturn, follow the evolution of the thermal structure and the stratospheric oscillation, as well as changes in the haze opacity (visible and infrared) and in the structure of the wind profiles (meridional and vertical). Also for Jupiter and Saturn, determination of the number and nature of chromophore agents, their properties (particle size, imaginary index), and their relationship to dynamics, and in the case of Saturn, to the seasonal insolation. For Titan, the main objectives are the monitoring over time of chemical abundances of hydrocarbons, nitriles and other compounds, as well as of haze layers (opacity, particle properties), and their possible relationship including the study of their dependence with the seasonal cycle and solar activity. Survey of the methane cloud activity, frequency, horizontal extent, brightness and altitude and track their motions to retrieve winds in the lower atmosphere. Finally, for Uranus and Neptune the objectives are quite similar, and could include the study of the long-term evolution and possible seasonal dependence of hazes properties, $CH_4$ gas spatial distribution, survey of bright spot activity at different altitude levels, frequency and properties of dark ovals at short wavelengths, and detailed study of the wind profile in particular in the equatorial region to determine its temporal variability and vertical shear. An additional objective through the multi-wavelength observations will be to test the tiered structure of stacked circulation cells proposed by Fletcher et al (2020).

It is fair to state that in recent years the exoplanet community has achieved a basic understanding of the chemistry, energy balance and dynamics of the atmospheres of close-in gas giants. This knowledge has come from investigations that combine spectroscopy and photometry in different viewing geometries, and from their subsequent interpretation with increasingly sophisticated models. In spite of the work done, there remains much to be done. High-resolution spectroscopy in its multiple variants has been fundamental in the detection of many chemical species and at constraining temperatures and velocities. The ongoing construction of high-resolution spectrographs that will operate over broad spectral ranges at a new generation of ground-based telescopes ensures that the technique will remain relevant for decades. It is thus important to ensure that the information encoded in the spectra is optimally extracted. Reaching that point of maturity will help put better quantitative constraints on chemical abundances (probably with the aid of low-resolution spectra) that can be directly compared to models, and establish the connection of the abundances with the local temperature and velocities. Accepting this complexity in the interpretation of the measurements represents both a challenge and an opportunity to understand the main interactions and feedbacks in the atmosphere. Future work should also emphasize the connections between different atmospheric layers, each with possibly a different prevalent physics, and the connections of the atmospheres with the deep envelopes and the host stars. Exploring these connections is interesting in itself, and may help elucidate the main processes of planetary formation and evolution.

There is a clear interest in the community to understand the atmospheres of sub-Neptune and smaller planets. JWST will be the main facility to explore such worlds in the imminent future. The recent investigations with this telescope for planets transiting cool stars suggest that there may be many surprises in the coming years. Two questions that often arise in such investigations are whether the planets have atmospheres and what their dominant compositions are. These questions may be difficult to answer, and it is often the case that additional plausibility arguments based on, for example, atmospheric evolution are invoked. This demonstrates again how a challenge is turned into an opportunity to explore other aspects of an atmosphere.

The investigation of exoplanet atmospheres will advance rapidly in the coming years and decades, partly as a result of the new telescopes and instruments that are being commissioned or conceived. Solid progress in our understanding of the exoplanet atmospheres calls for a healthy balance between observations and theory. Observationally, it is desirable to continue maturing the techniques that are already available while developing others that remain under-developed. From the theoretical viewpoint, it seems recommendable to embrace the complexity of the atmospheres and the physical processes occurring in them.

Solar system planets have primarily served as the reference against which our understanding of exoplanets has been contrasted. The situation has evolved and the knowledge that we have been acquiring about exoplanet atmospheres, partly as a result of the variety expected in them, is offering new insight into the Solar System planets and their atmospheres. The feedback between both communities is not only at the conceptual level. It affects also the techniques that we are using in particular for the prediction of key atmospheric properties and the interpretation of the available observations (see e.g. Deeg and Belmonte, 2018). This dialogue will surely continue for a long time and will ideally merge into a higher-level understanding of planetary atmospheres and all the disciplines upon which it rests.

**Acknowledgements**

ASL was supported by Grupos Gobierno Vasco IT1742-22 and by Grant PID2019-109467GB-I00 funded by MCIN/AEI/10.13039/501100011033/.